\documentclass{mn2e}

\usepackage{graphicx}
\usepackage{amsmath}
\usepackage{amssymb}

\title[On the possibility of sub-TeV Gamma-ray emission from Cyg X-3]
{On the possibility of sub-TeV Gamma-ray emission from Cyg X-3}

\author[W. Bednarek]{W. Bednarek  \\
Department of Astrophysics, University of \L\'od\'z, ul. Pomorska 149/153, PL-90236 \L \'od\'z, Poland, bednar@astro.phys.uni.lodz.pl}

\begin{document}

\date{Accepted . Received ; in original form }

\pagerange{\pageref{firstpage}--\pageref{lastpage}} \pubyear{2007}

\maketitle

\label{firstpage}

\begin{abstract}
The compact X-ray binary system Cyg X-3 has been recently discovered as a source of GeV $\gamma$-rays by the AGILE and the {\it Fermi} satellites. It shows emission features in the GeV $\gamma$-rays similar to other $\gamma$-ray binaries which
were also observed in the TeV $\gamma$-rays (LS 5039 and LSI +61 303). The question appears whether Cyg X-3 can be also detected in the TeV $\gamma$-rays  by the Cherenkov telescopes.

Here we discuss this problem in detail based on the anisotropic inverse Compton (IC) $e^\pm$ pair cascade model successfully applied to TeV $\gamma$-ray binaries. 
We calculate the $\gamma$-ray light curves and $\gamma$-ray spectra expected from the cascade process occurring inside the Cyg X-3 binary system. It is found that the $\gamma$-ray light curves at GeV energies can be consistent with the $\gamma$-ray light curve observed by the {\it Fermi} for 
reasonable parameters of the orbit of the injection source of relativistic electrons.
Moreover, we show that in such a model the sub-TeV $\gamma$-ray emission (above 100 GeV) is expected to be below sensitivities of the present Cherenkov telescopes assuming that electrons are accelerated in Cyg X-3 to TeV energies. 
The next stage Cherenkov telescopes (MAGIC II, HESS II) should have the energy threshold in 
the range 20-30 GeV, in order to have a chance to detect the signal from Cyg X-3.
Otherwise, the positive detection of $\gamma$-rays at energies above a few tens of GeV requires a telescope with the sensitivity of $\sim 0.1\%$ of Crab Units. We conclude that detection of sub-TeV $\gamma$-rays from Cyg X-3 by on-ground telescopes has to probably wait for the construction of the Cherenkov Telescope Array (CTA). 

\end{abstract}

\section{Introduction}

The enigmatic binary system, Cyg X-3, has been claimed as a powerful $\gamma$-ray source at different
energies since the first satellite and on-ground observations starting from 70-80-ties.
The reports on the MeV-GeV $\gamma$-ray emission were diverse. A marginal detection has been claimed by the SAS-2
and EGRET (Lamb et al.~1977, Mori et al.~1997). However, the negative result has been reported by the COS B (Hermsen et al.~1987). Also early optimistic results reported by the Cherenkov telescopes
were not confirmed (e.g. Weekes~1988, 1992; Chadwick et al.~1990). Only recently, the AGILE and {\it Fermi} satellites
reported positive transient detection of Cyg X-3 at GeV energies (Tavani et al.~2009, Abdo et al.~2009a).
The $\gamma$-ray emission from Cyg X-3 has been discovered before the major radio flares (Tavani et al.~2009). It shows a modulation with the period of the binary system (Abdo et al.~2009a). The emission above 100 MeV is well described by a single power law with differential spectral index 
$-2.7\pm 0.05(stat)\pm 0.20(syst)$ and the peak flux $\sim 2\times 10^{-6}$ ph cm$^{-2}$ s$^{-1}$
(Abdo et al.~2009a). Interestingly, the $\gamma$-ray light curve at GeV energies from Cyg X-3 
shows general features that are quite similar to those observed recently from the TeV $\gamma$-ray binaries
LS 5039 and LSI 303 +61, which were also observed by the {\it Fermi-LAT} at GeV energies (Abdo et al.~2009b,c).
Therefore, future observations of Cyg X-3 with the modern Cherenkov telescopes seem to be well motivated. In fact, during last years the MAGIC telescope has observed Cyg X-3 for $\sim 70$ hours in different emission states. No positive signal has been reported up to now (Saito et al.~2009). The upper limits are on the level of $1\%$ of the Crab Unit (C.U.) above $\sim 250$ GeV.

The observed modulation of GeV $\gamma$-ray emission from Cyg X-3 strongly suggests that photons has to originate in the radiation process which occurs inside the binary system.
The most likely scenario is the interaction of electrons with the anisotropic radiation of the WR type companion star. Electrons are accelerated at some physical process either close to accreting object or in the jet or in the vicinity of a pulsar or at the shock wave between the jet and the stellar wind (as postulated by models proposed for TeV $\gamma$-ray binaries). However, due to very strong radiation field created by such luminous star inside compact binary Cyg X-3 (surface temperature $T_{\star}\sim 10^5$ K, stellar radius $R_{\star}\sim 2\times 10^{11}$ cm, and separation of the compact object $D\sim 2R_{\star}$), $\gamma$-rays should initiate IC $e^\pm$ pair cascades as considered for TeV $\gamma$-ray binaries (e.g. Bednarek~2000, Orellana et al.~2007, Khangulyan et al.~2008).
Such IC e$^\pm$ pair cascade scenario has been already investigated also for the Cyg X-3 binary system (Bednarek~1997 and Sierpowska \& Bednarek~2005). It has been shown in those papers that the optical depths for $\gamma$-rays created inside the Cyg X-3 binary system are huge (they can reach values up to a few hundred in specific directions). Therefore, escape of $\gamma$-rays with energies above a few tens of GeV is not very likely. In this paper, we perform detailed calculations of the $\gamma$-ray spectra escaping towards the observer located at different directions in respect to the orbital plane of the Cyg X-3 binary system for different
locations of the source of energetic leptons within the binary. We calculate the expected $\gamma$-ray light curves at energies suitable for the satellite and Cherenkov telescopes.

\section{Phase dependent Gamma-ray optical depths}

The optical depths of $\gamma$-ray photons in the anisotropic radiation of the massive star has
been calculated in general case (arbitrary injection place within the binary) for the star with the parameters of the WR star in Cyg X-3 by Bednarek~(1997, see Fig.~2 in that paper). Here we calculate the $\gamma$-ray optical
depth light curves for the case of $\gamma$-rays injected from a point source on a circular orbit
around the star with a radius, $R_\gamma$. Since the inclination angle of the Cyg X-3 binary system is expected in the range $i = 30^{\rm o}-60^{\rm o}$, we show how the optical depths change with the phase of the binary system for the limiting values of the inclination angle.
We apply the following parameters for the WR star in Cyg X-3: its radius $R_\star = 2\times 10^{11}$ cm,
surface temperature $T_\star = 9\times 10^4$ K and the radius of the orbit of the compact object $a = 2.25R_\star$. These parameters are consistent with the limits put by Cherepashchuk \& Moffat~(1994).  Note that the injection place of $\gamma$-rays does not need to be identical with the location of the compact object in this binary system. 

We consider two general geometrical situations for the location of the injection source of $\gamma$-rays:

\begin{enumerate}

\item It is in the plane of the binary system but at different distances from the companion star (in the range $R_\gamma = 1.5 - 20R_\star$).
Such geometry can give us an idea about the $\gamma$-ray emission in the case of acceleration regions extended in the plane of the binary system (e.g. the shock structure created in the collision of the stellar and pulsar winds).

\item It is above the plane of the binary system for the distance of the compact object from the companion star
(in the range $D = 1 - 20R_\star$).
Such calculations are suitable in the case of the acceleration region extended in the direction of the jet which is perpendicular to the plane of the binary. 

\end{enumerate}

The optical depths for $\gamma$-rays propagating in the anisotropic radiation of the companion star
from the injection place up to the infinity for these two situations are shown in Figs.~1 and 2.
In the case of a circular orbit with different radii, we note clear variability of the optical
depths for $\gamma$-rays with different energies. The general tendency is that the level of variability 
of the optical depth decreases with energy of $\gamma$-ray photons in the range of $\sim$33 GeV to $\sim$1 TeV.
The $\gamma$-ray optical depths change more drastically with the phase of the binary system for larger inclination angles. As expected, the optical depths reach the lowest values when the injection source is
in front of the companion star (phases close to 0.5). Optical depths can reach values below unity at certain
region around the phase 0.5. This region becomes less extended for larger energies of $\gamma$-ray photons.
In the case of the orbit with the radius $R_\gamma = 1.5r_\star$, the optical depths decrease for small and large phases due to the eclipse of the $\gamma$-ray source by the companion star (in this case the optical depths are calculated only to the stellar surface).

\begin{figure*}
\vskip 18.5truecm
\includegraphics{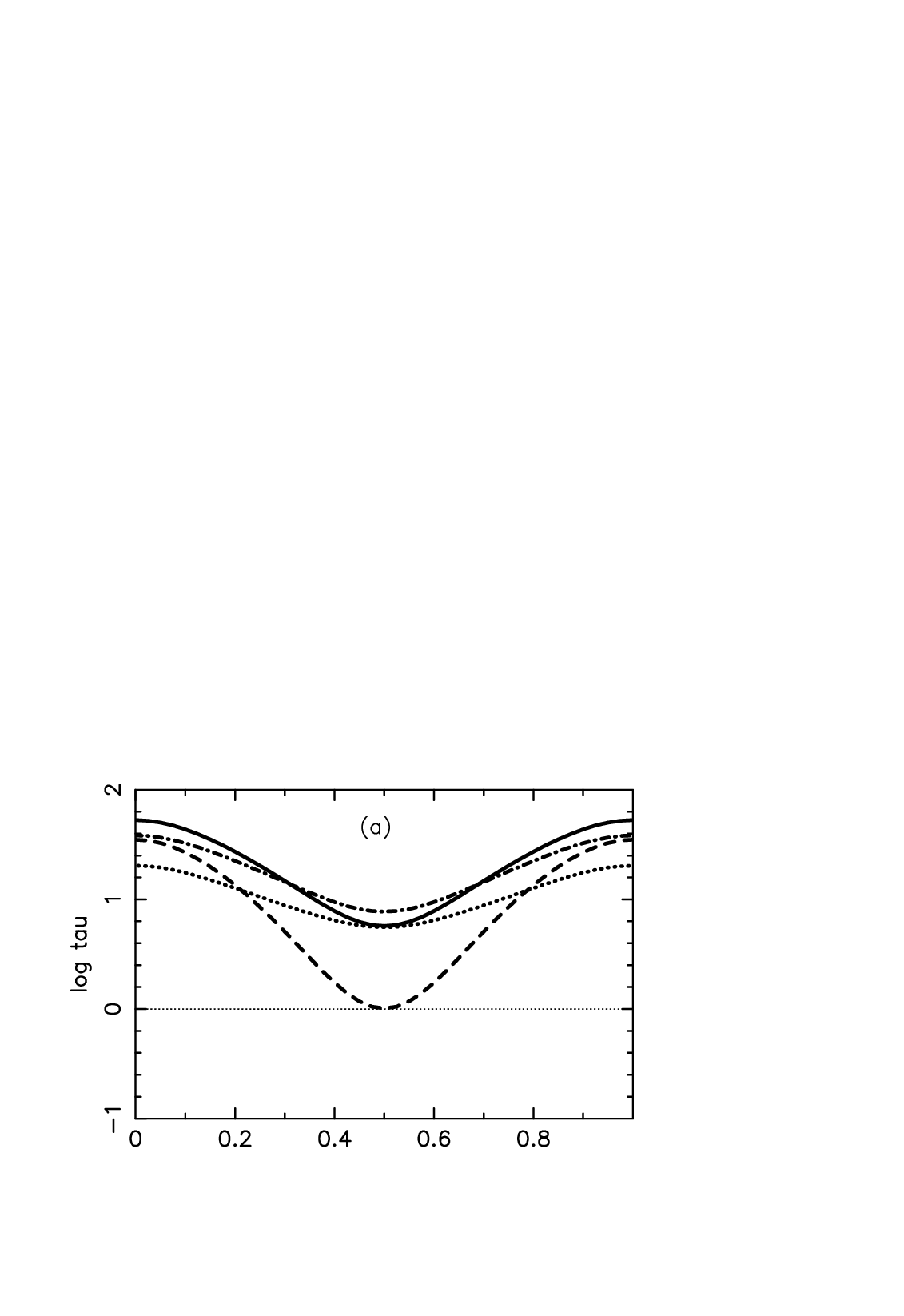}
\includegraphics{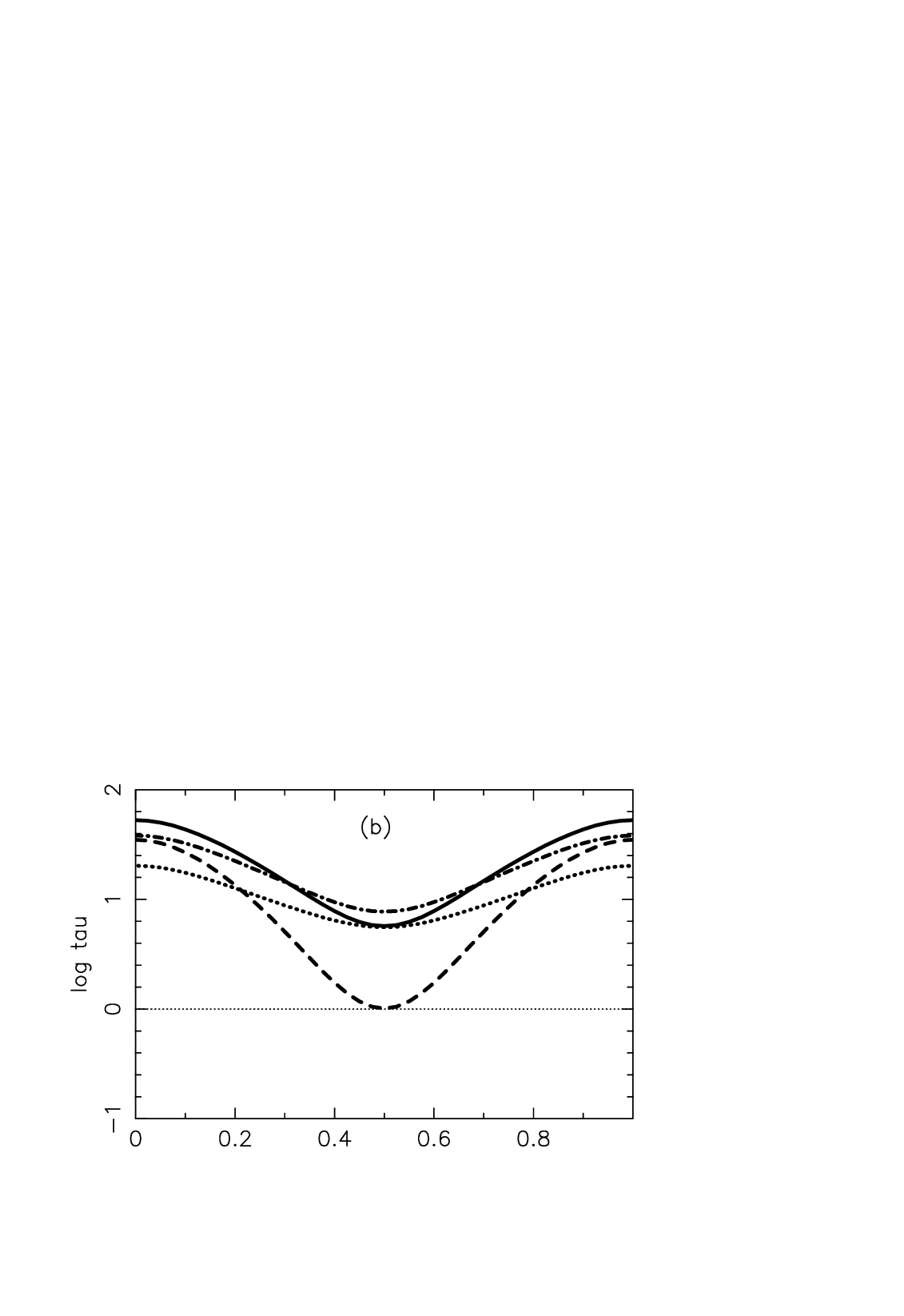}
\includegraphics{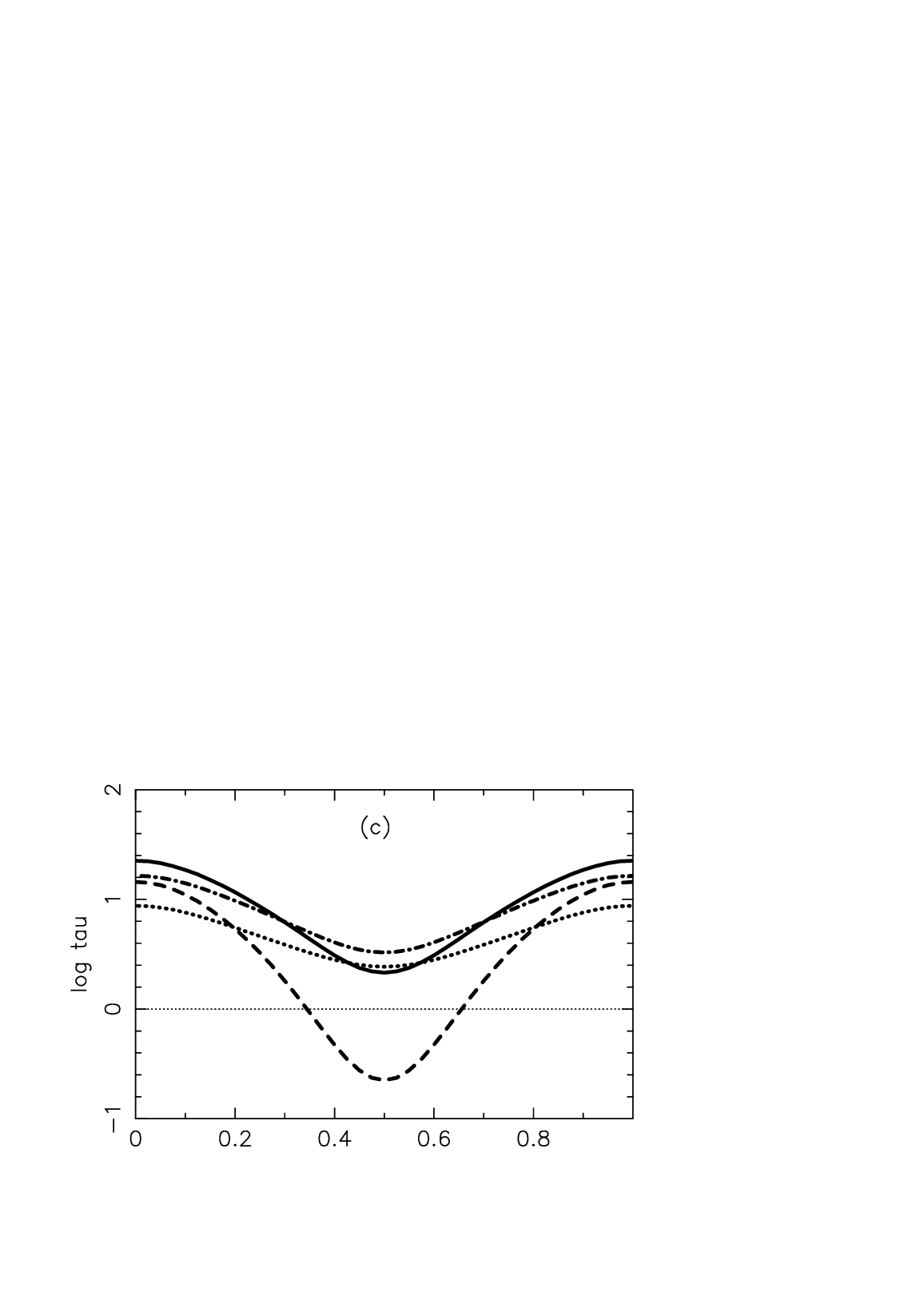}
\includegraphics{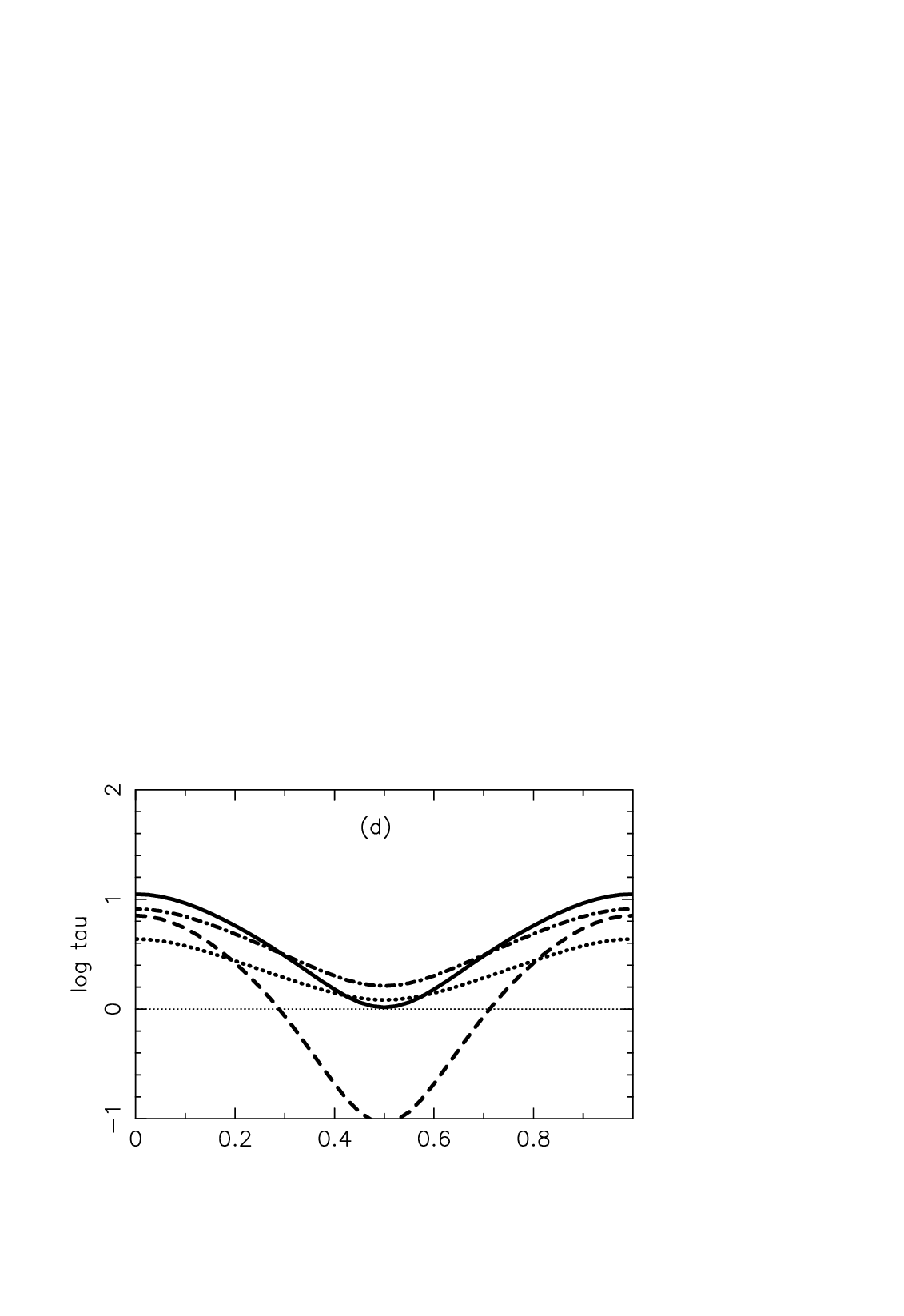}
\includegraphics{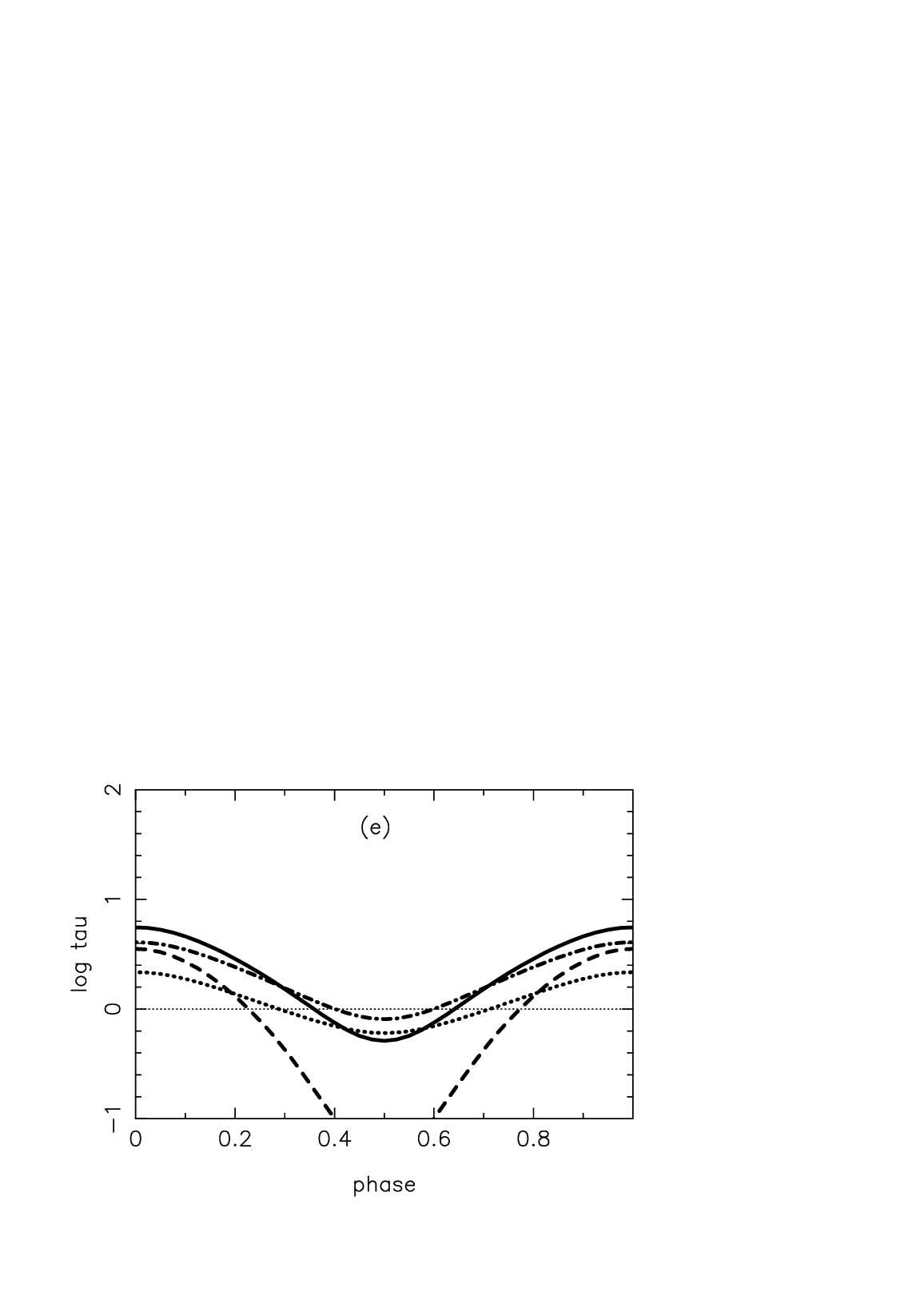}
\includegraphics{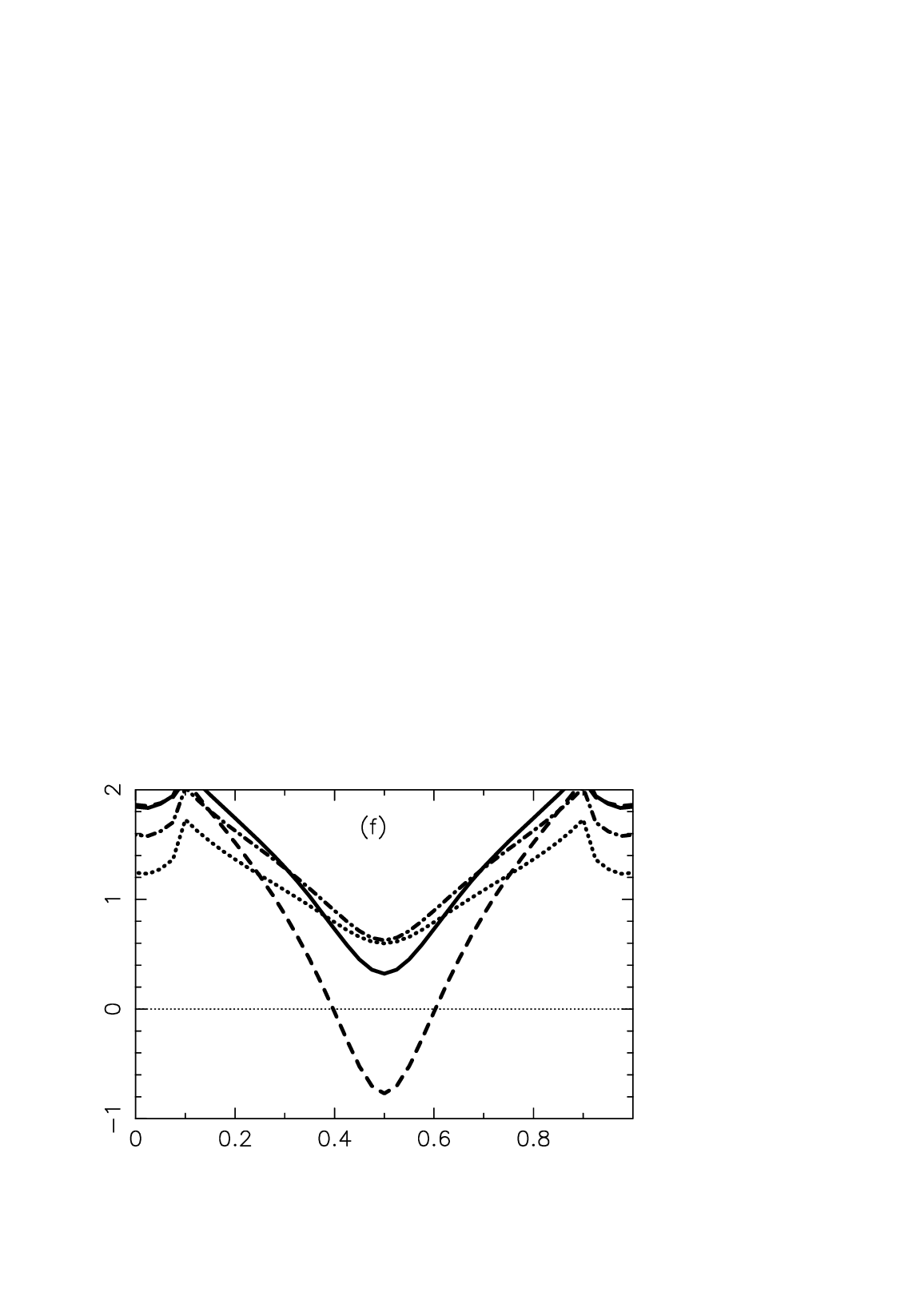}
\includegraphics{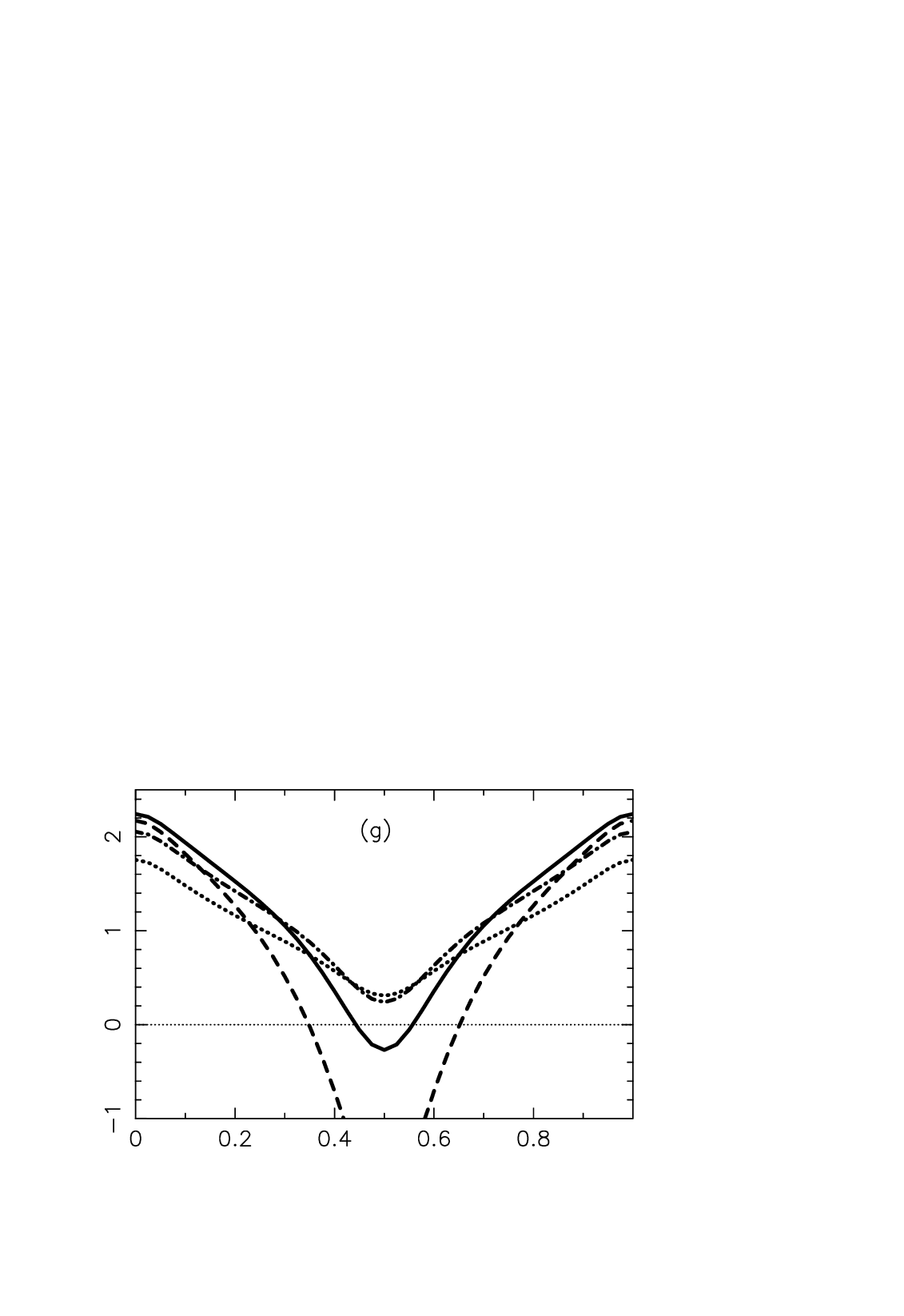}
\includegraphics{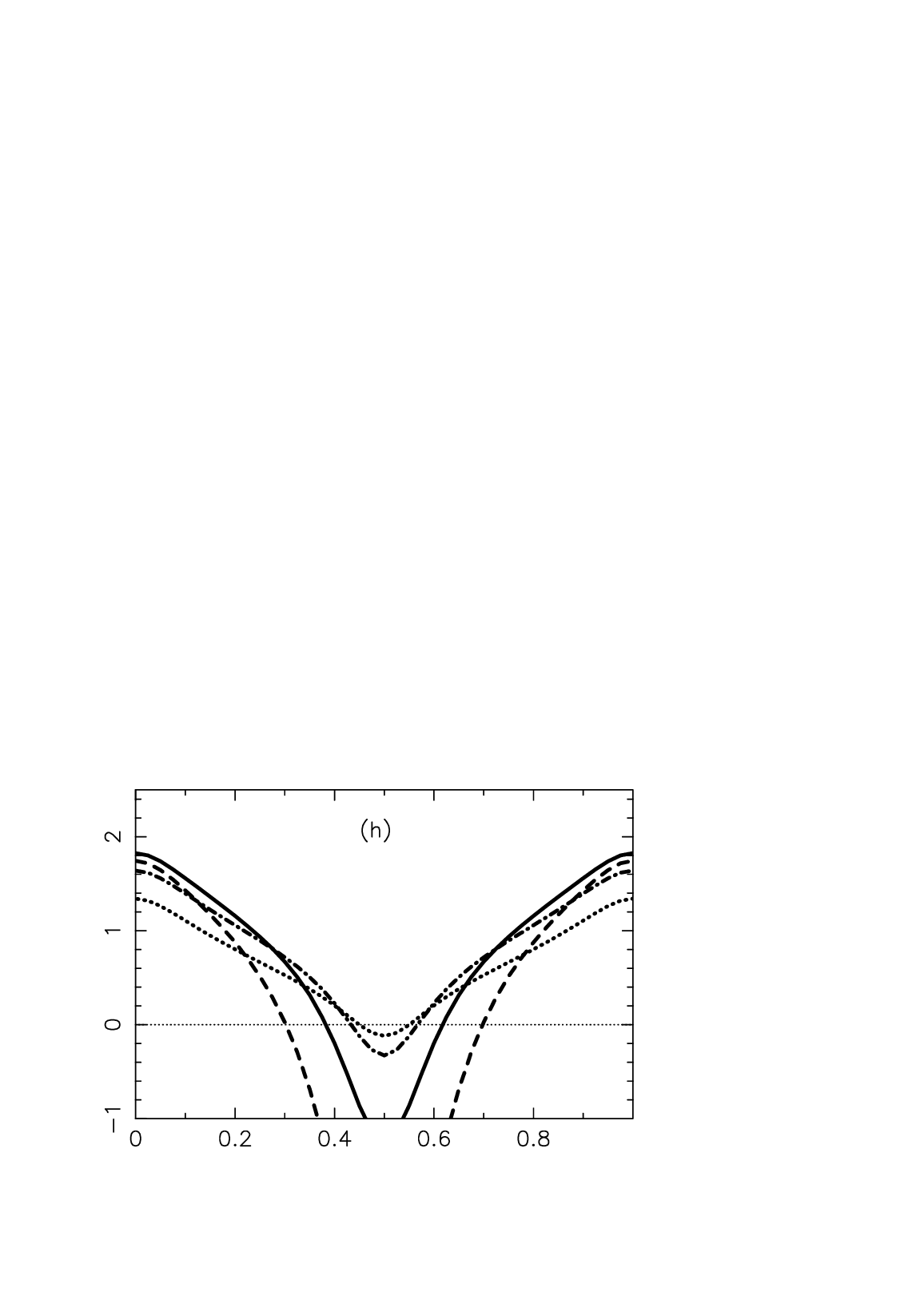}
\includegraphics{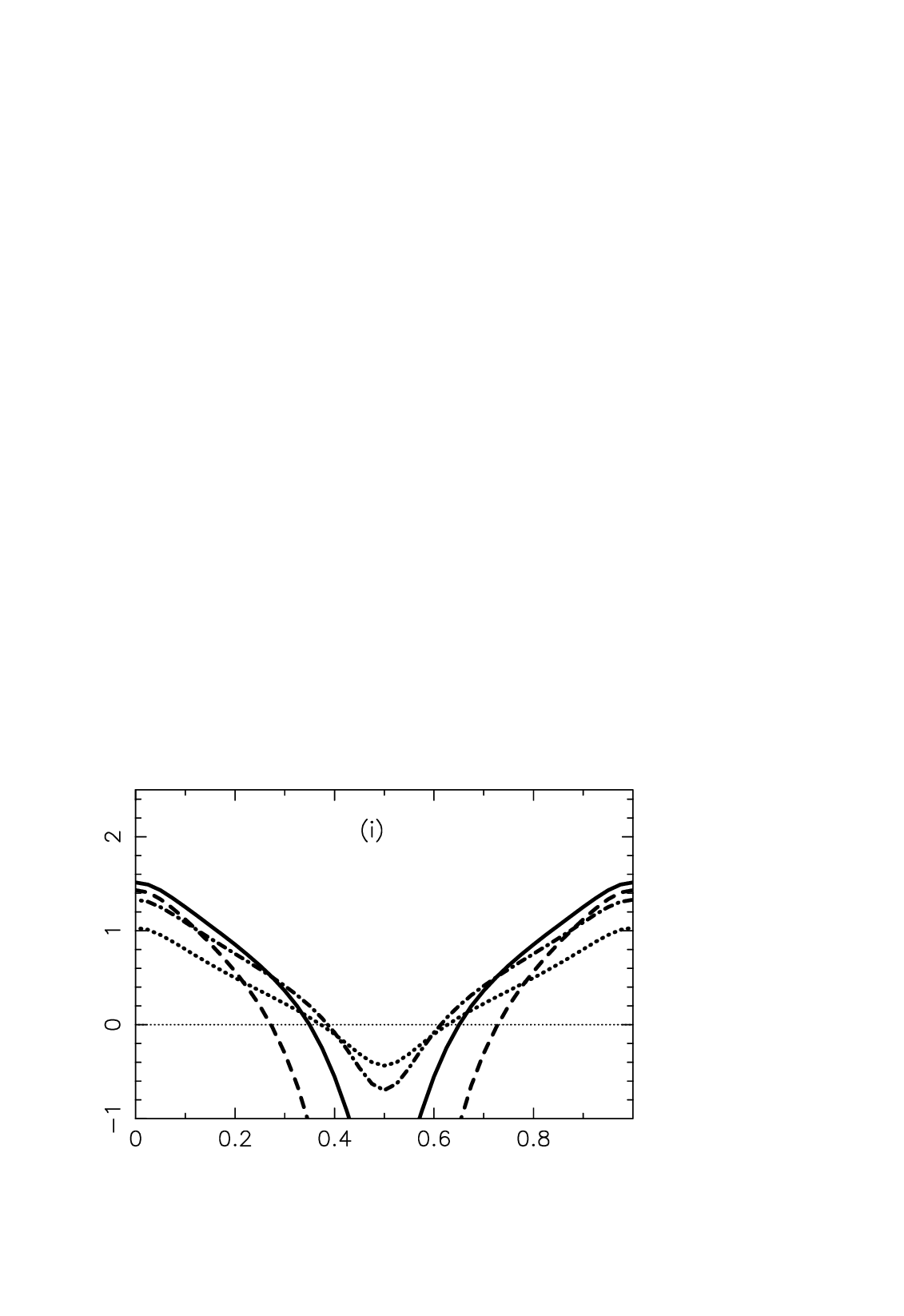}
\includegraphics{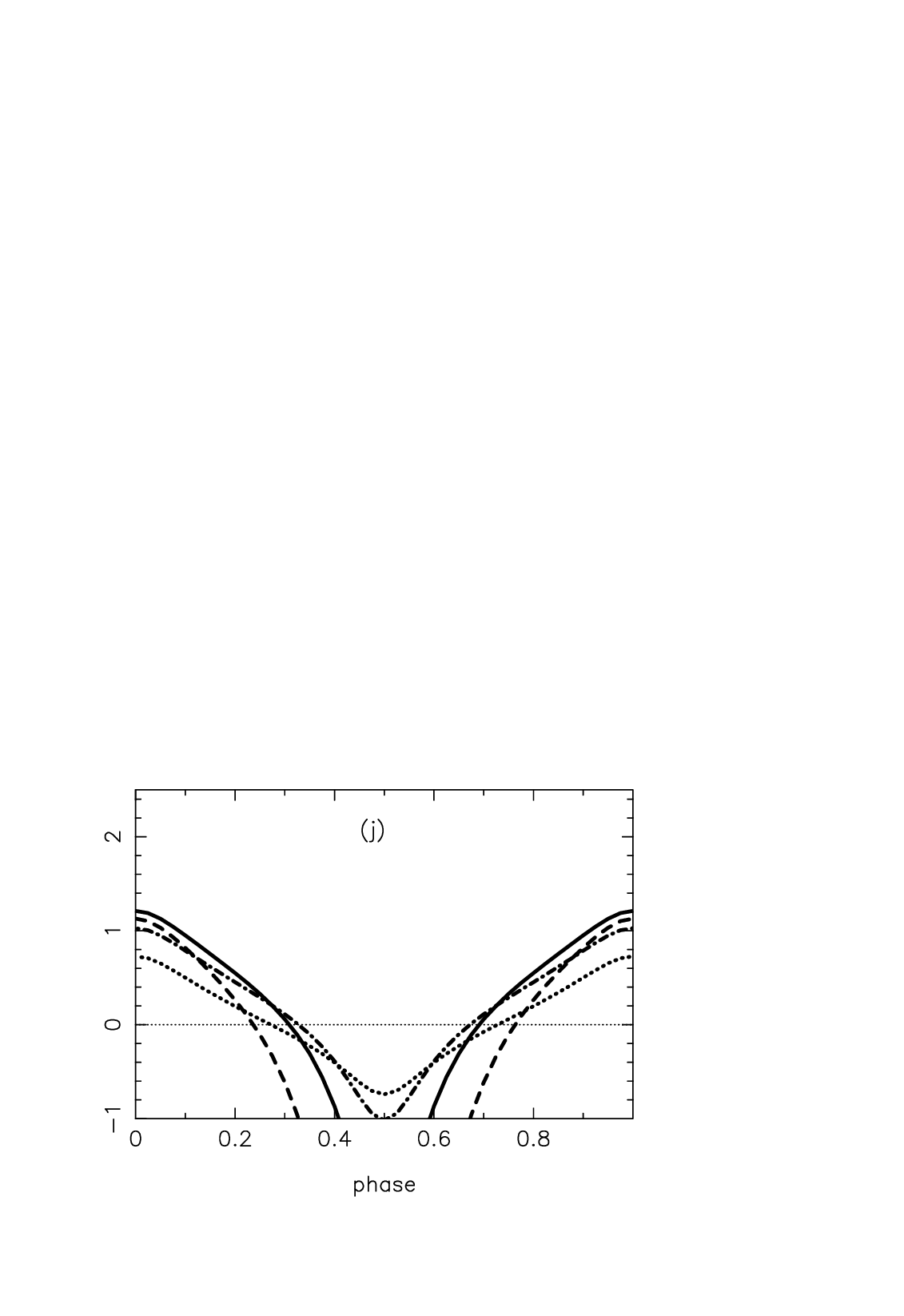}
\caption{The optical depths for $\gamma$-ray photons injected at a specific place inside the binary system for energies $E_\gamma = 33$ GeV (dashed curves), 100 GeV (solid), 330 GeV (dot-dashed), and 1 TeV (dotted) as a function of the phase of the binary system Cyg X-3. The phase zero corresponds to the location of the injection source behind the massive star. Figures on the left are obtained for the inclination of the binary system equal to $i = 30^o$ and on the right for $i = 60^o$.
The injection source is on the circular orbit within the plane of the binary system but at different distances from the companion star: $R_\gamma = 1.5R_\star$ (a and f), $2.25R_\star$ (b,g), $5R_\star$ (c,h), $10R_\star$ (d,i),$20R_\star$ (e,j). Thin  dotted line marks the optical depth equal to unity. The companion star has the radius $R_\star = 2\times 10^{11}$ cm and the surface temperature $T_\star = 9\times 10^4$ K (as expected for the companion star in the Cyg X-3 binary system).}
\label{fig1}
\end{figure*}

The $\gamma$-ray optical depth light curves calculated for different distances above the plane of the binary system show also local minima at phases corresponding to the location of the injection place in front of the companion star (see Fig.~2). These minima become at first deeper with increasing distance of the injection place from the orbital plane. The appearance of these features is related to the decrease of the angle between direction towards the observer and the direction towards the injection place of $\gamma$-rays (counted from the stellar center) for larger distance, $D$, from the orbital plane. However, for large distances, 
$D$, the dips in the $\gamma$-ray light curves become shallower and the general level of the optical depth becomes lower. As a result, modulation of the optical depths with the period of the binary system becomes relatively low. Based on these calculations, we conclude that the largest contribution to the modulation of the $\gamma$-ray emission with the period of the binary system comes from the intermediate distances from the orbital plane. However, the distance, from which the maximum contribution to the modulation of the $\gamma$-ray signal is seen, depends also on the inclination angle of the binary system.

\begin{figure*}
\vskip 18.5truecm
\includegraphics{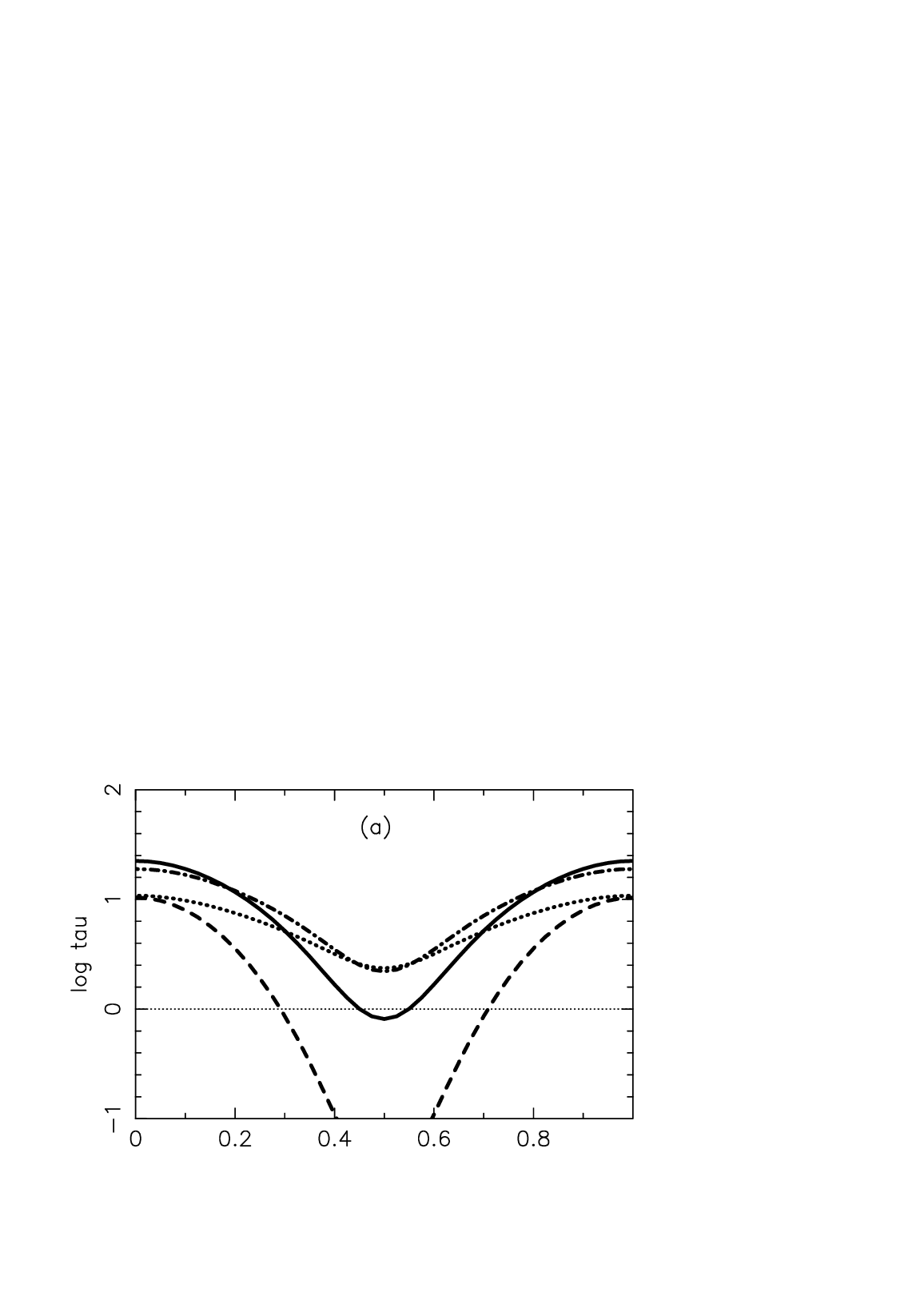}
\includegraphics{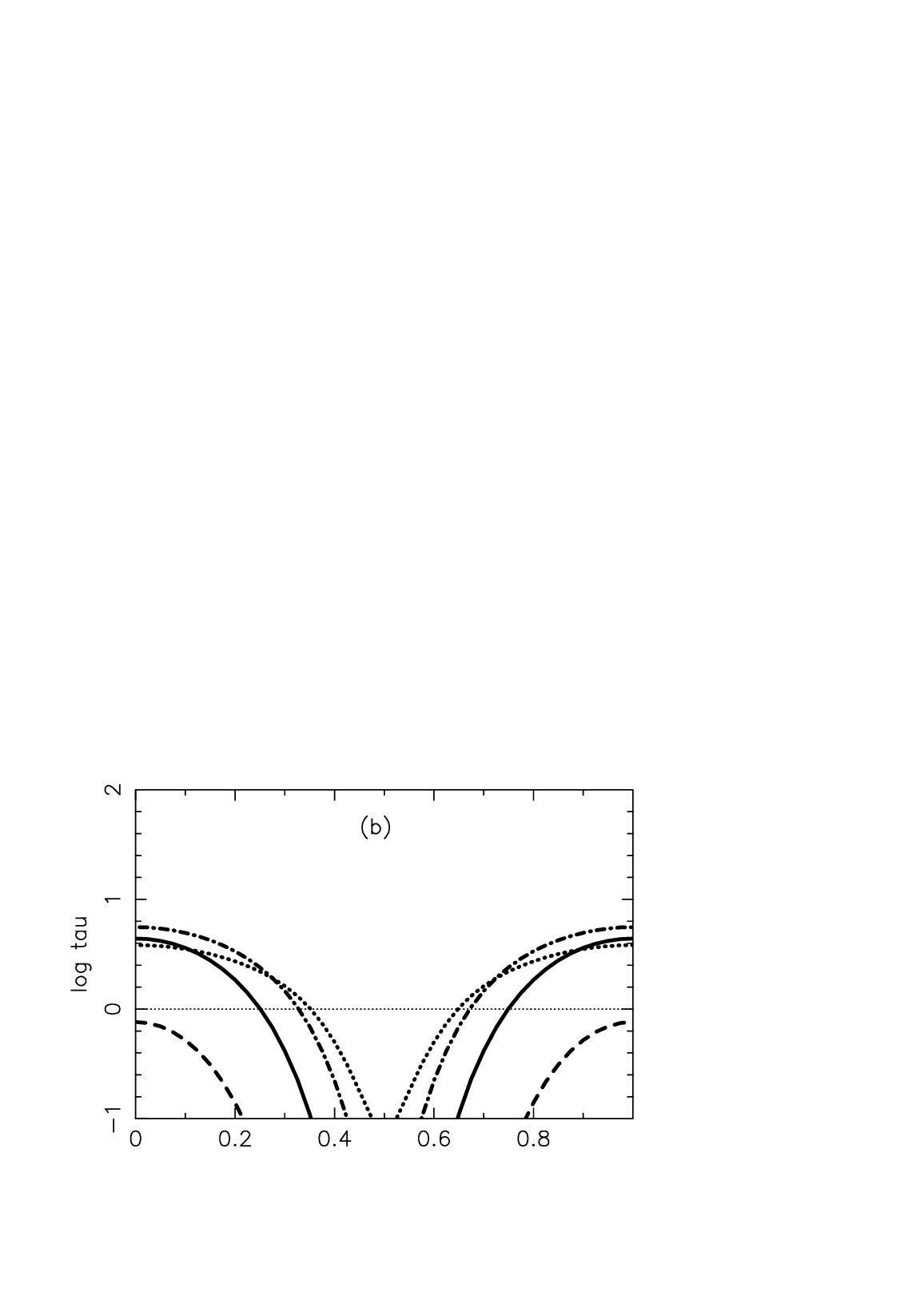}
\includegraphics{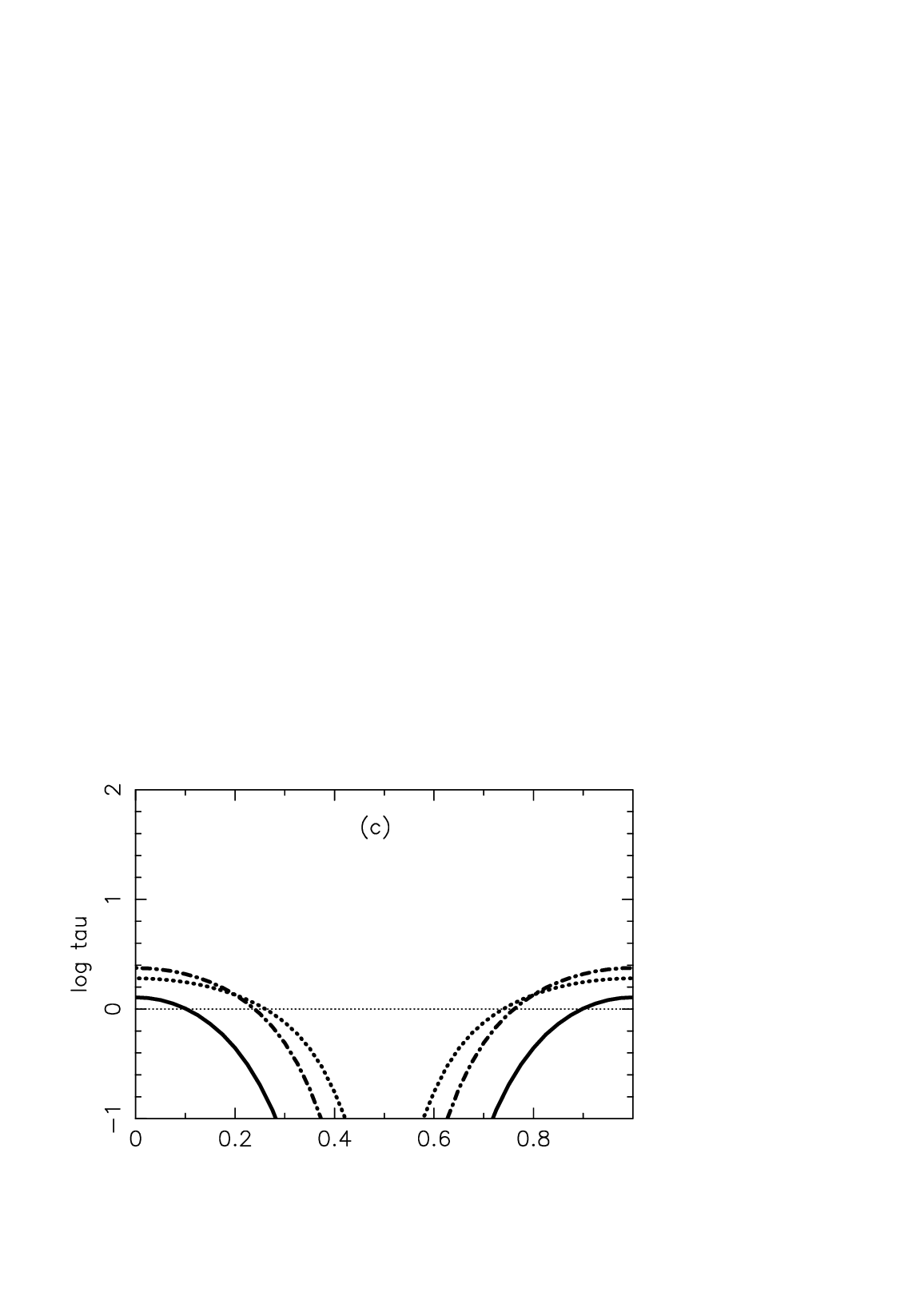}
\includegraphics{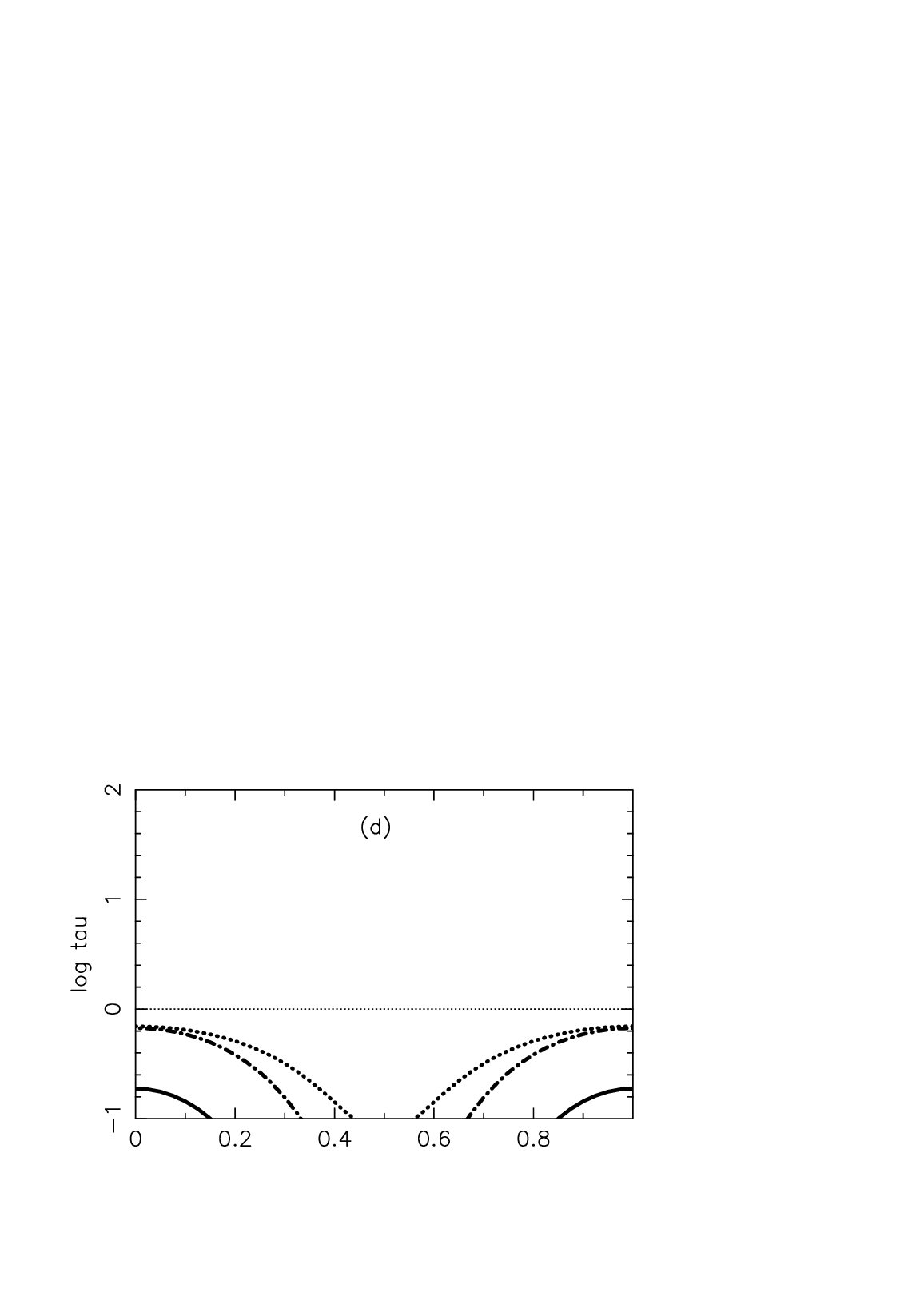}
\includegraphics{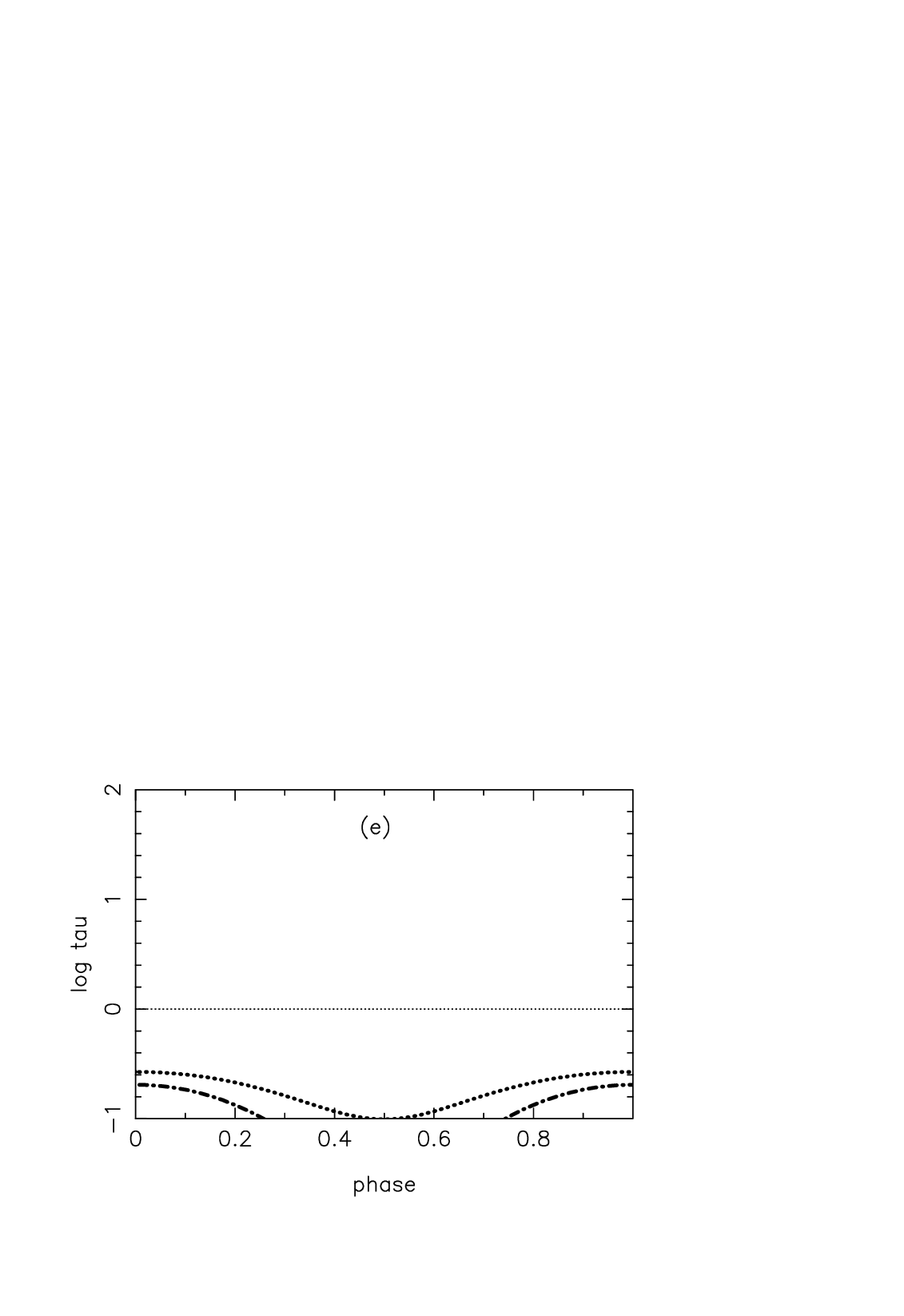}
\includegraphics{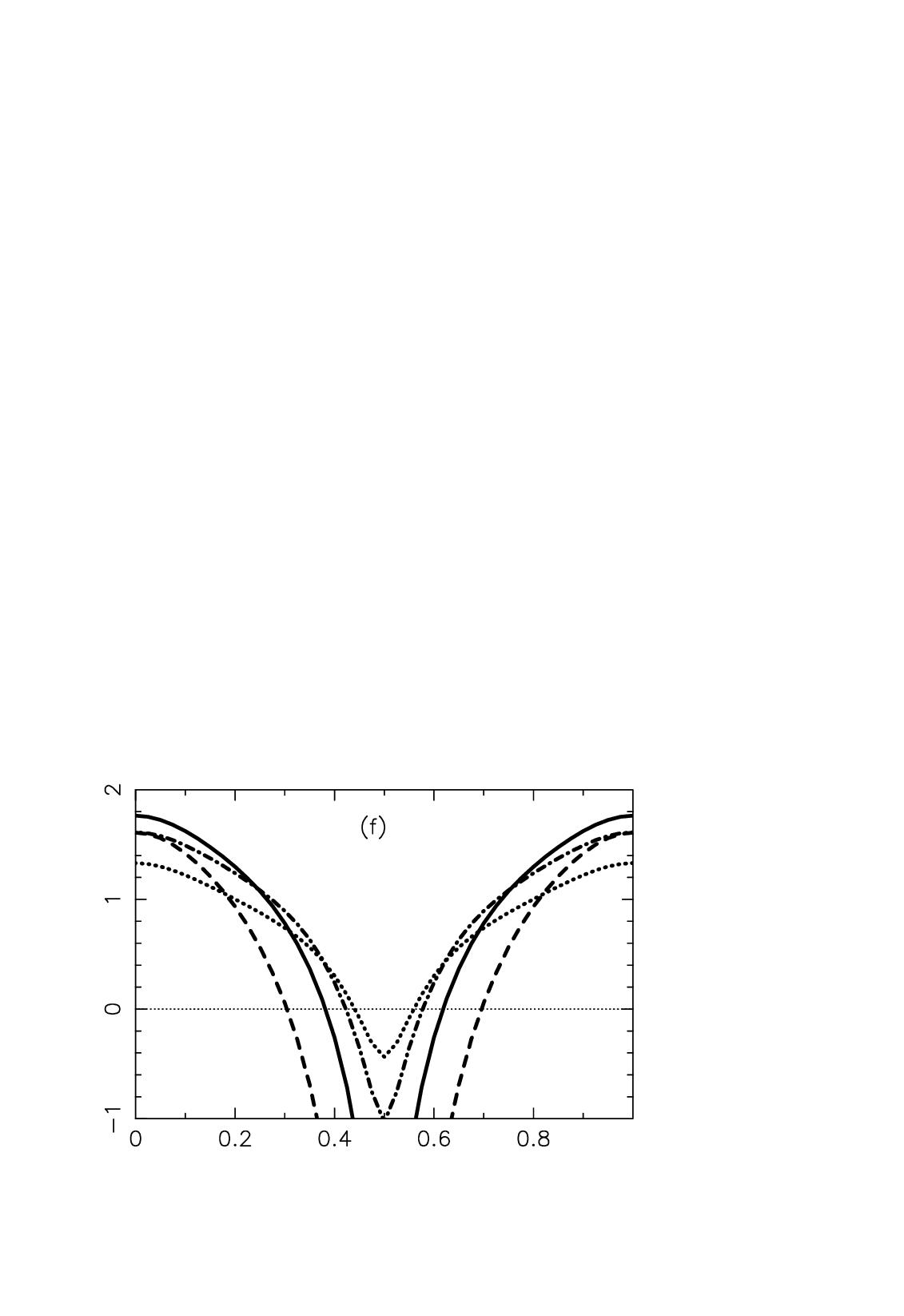}
\includegraphics{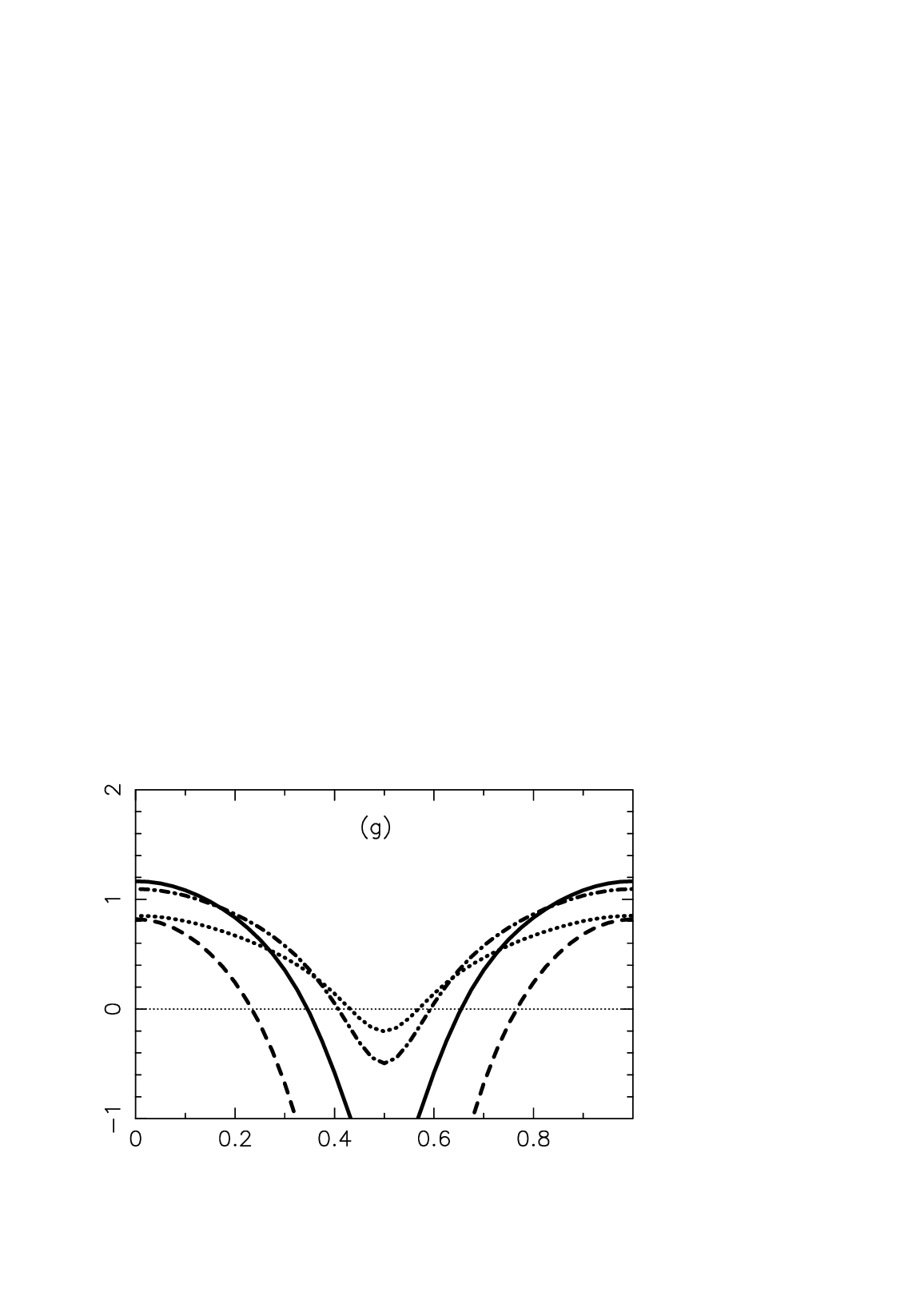}
\includegraphics{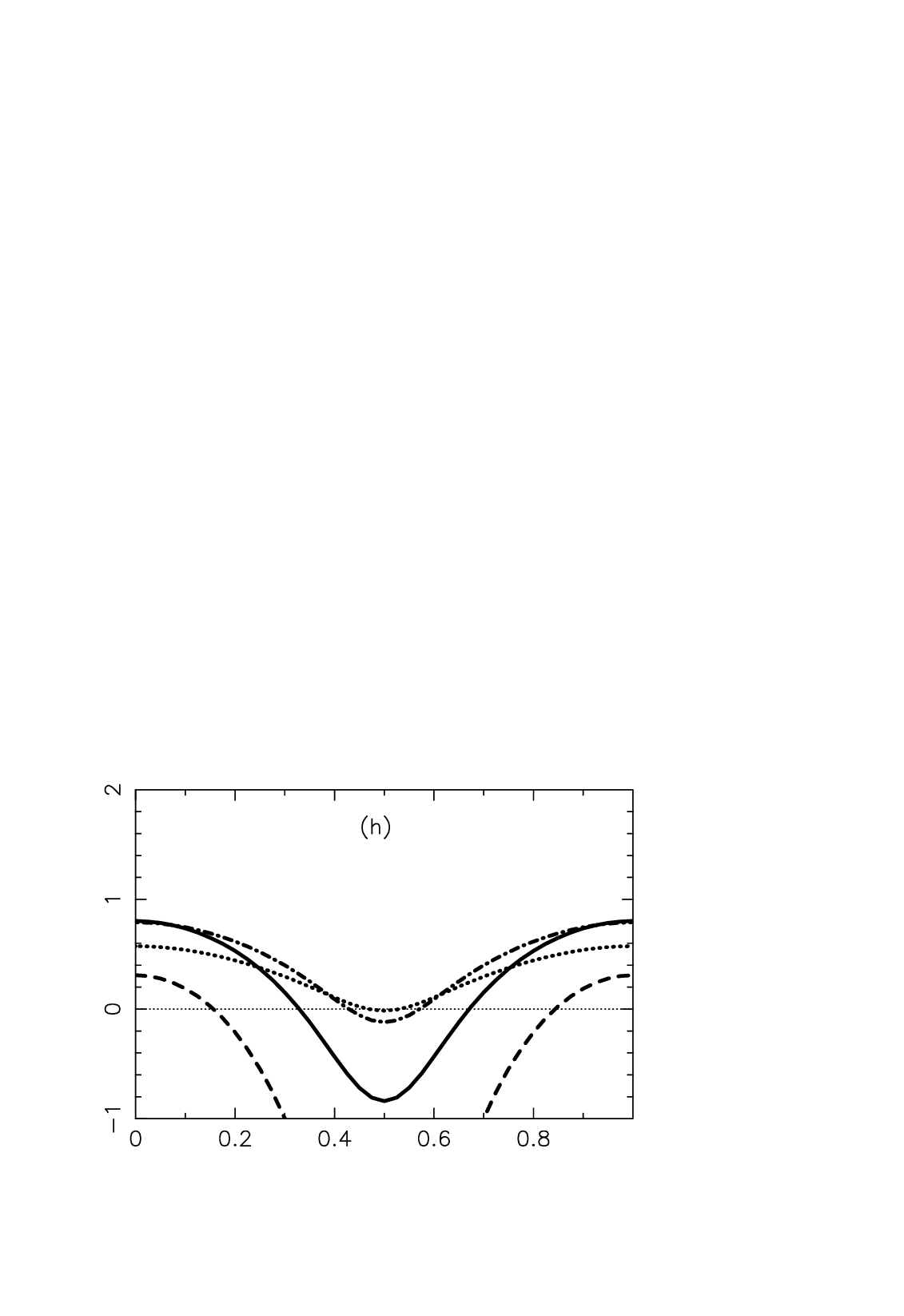}
\includegraphics{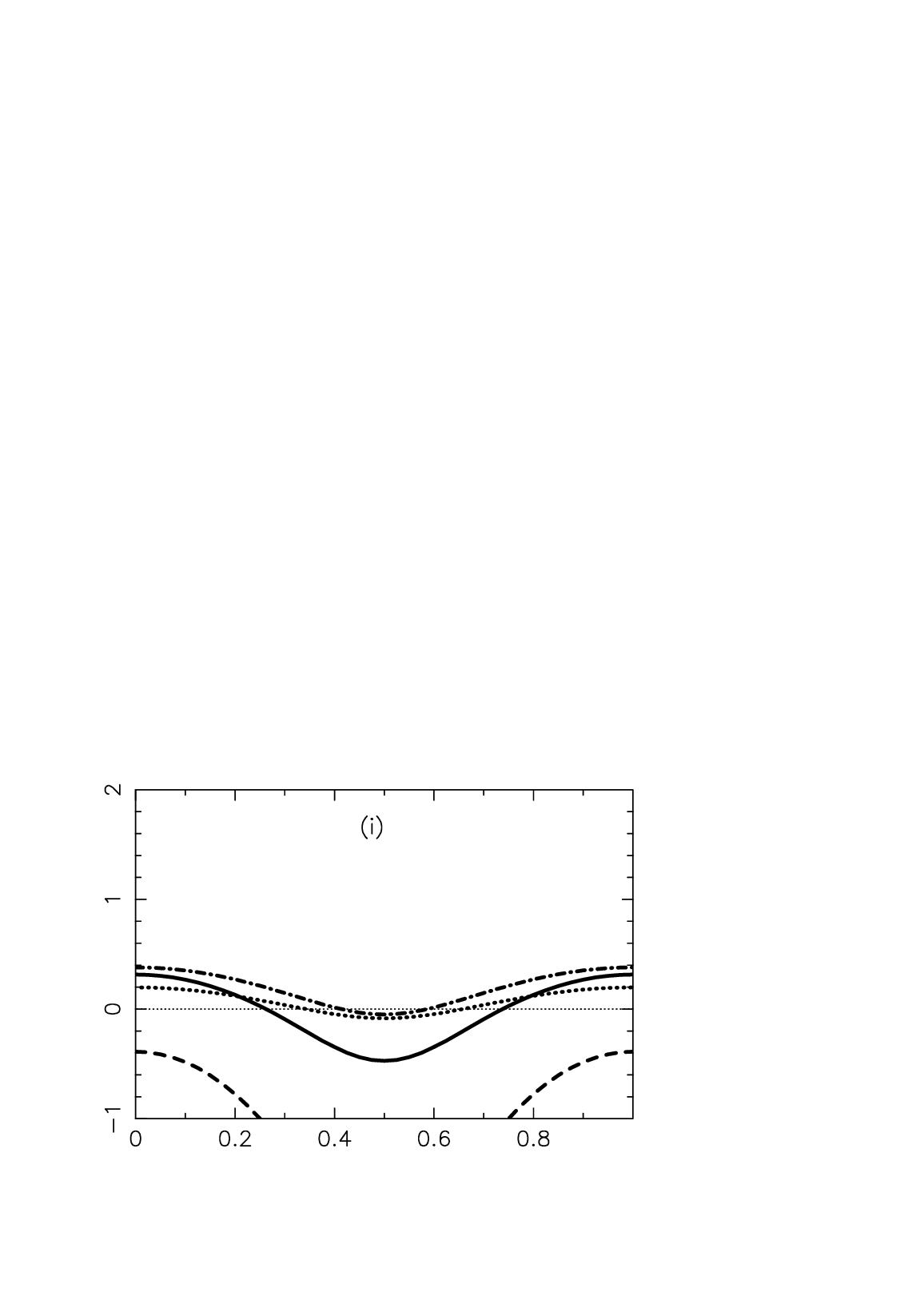}
\includegraphics{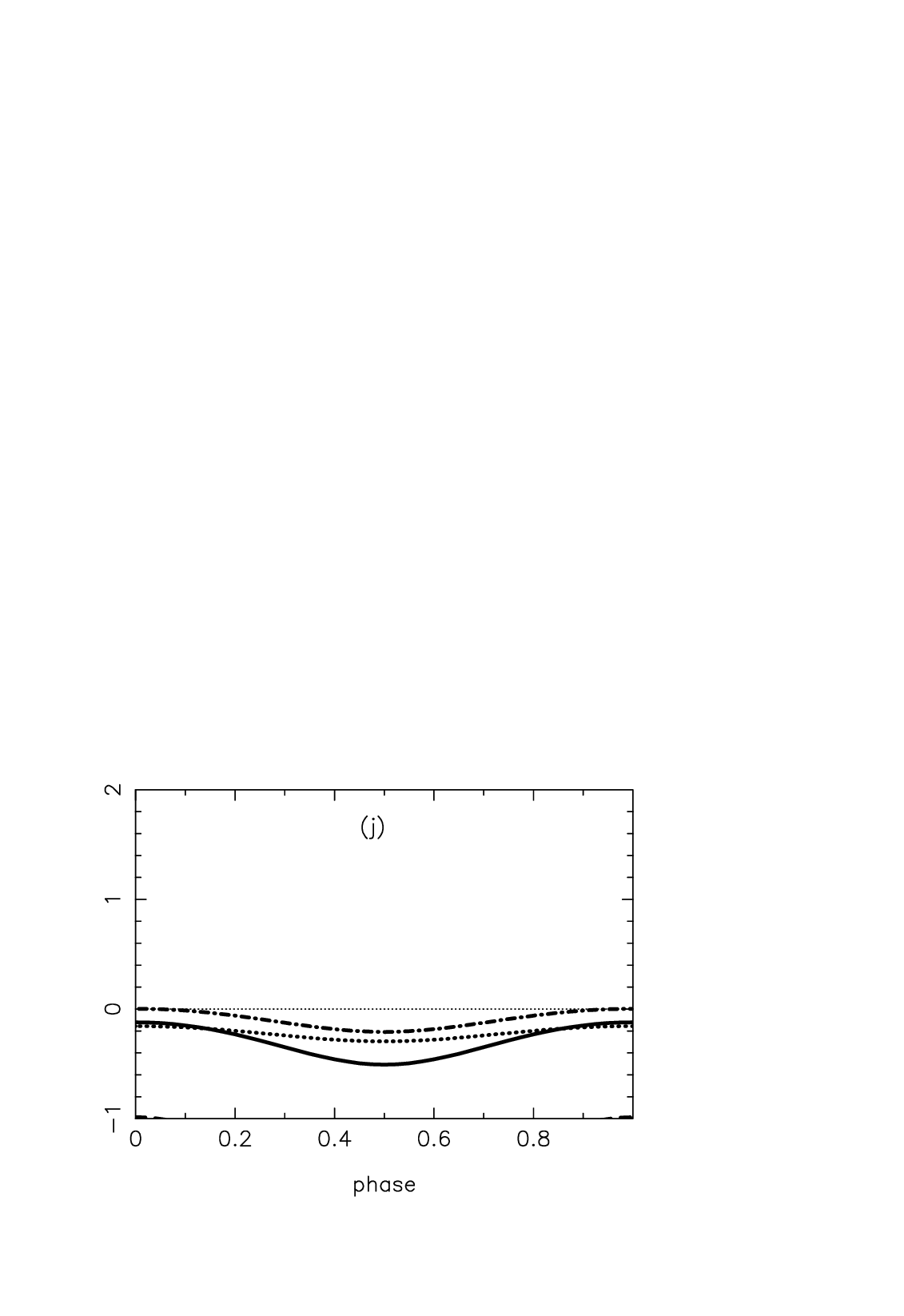}
\caption{As in Fig.~1 but for the case of the $\gamma$-ray injection source located in the jet 
launched from the compact object perpendicularly to the plane of the binary system. 
The compact object is at the distance of $a = 2.25R_\star$ from the companion star. 
Specific figures show the optical depths for distances along the jet equal to: 
$D = 1R_\star$ (a,f), $3R_\star$ (b,g), $5R_\star$ (c,h), $10R_\star$ (d,i), $20R_\star$ (e,j).}
\label{fig2}
\end{figure*}

Based on the calculations of the phase dependent optical depths for $\gamma$-ray photons with different energies, we can conclude that the cascading effects should be very important in the case of Cyg X-3 binary system for both considered inclination angles. However, the $\gamma$-ray spectra escaping to the observer have a chance to extend up to the sub-TeV energies only for a range of phases around $\varphi = 0.5$, corresponding to the location of the injection region in front of 
the companion star. On the other hand, when the injection source is behind the companion star, the optical depths are clearly above unity for most of the considered locations of the injection source and energies of $\gamma$-ray photons. For these phases, $\gamma$-ray emission should be limited only to energies below a few
tens of GeV. In the next section we discuss the $\gamma$-ray emission features in terms of a simple scenario in which relativistic electrons are injected isotropically from a point  like source located within the binary system.

\section{General scenario for gamma-ray production}

The observed modulation of GeV-TeV $\gamma$-ray emission with the period of the binary systems 
(LS 5039, LSI 303 +61, Cyg X-3) indicates that the production of $\gamma$-rays occurs inside the volume of the binary system. The companion stars in these binaries are very luminous which again suggest that emission process is related to strong radiation field
created by these stars. At the conditions 
envisaged above, the production of $\gamma$-rays likely occurs in the cascade process since the cross sections for the inverse Compton process and the  $e^\pm$ pair creation in collisions
of $\gamma$-rays with the stellar low energy photons are comparable. In fact, relativistic hadrons  
might also contribute to the $\gamma$-ray flux provided that they find enough target
provided by the dense stellar winds or the accretion disk around a compact object. 
Due to the observed modulation of the $\gamma$-ray signal, the radiation processes likely occur
within the binary system.
The specific radiation mechanism which is consistent with the observed features of the $\gamma$-ray emission from the binary systems has been already studied before the discovery of TeV $\gamma$-ray emission from massive binary systems (see e.g. Bednarek~2000, Sierpowska \& Bednarek~2005). It assumes that relativistic electrons are accelerated inside the binary system in one of the popular general picture, i.e either by the pulsar or the pulsar wind shock or the shock in the jet or at the jet base.  
In the case of Cyg X-3 binary system, the jet scenario is more likely due to clear evidences of the existence of jets in this source (Marti et al.~2001, Miller-Jones et al.~2004).

Let us consider the conditions in the jet which allow acceleration of electrons to a few TeV.
The maximum energies of particles are determined by the balance between the energy
gain from the acceleration mechanism and energy losses on the radiation processes.
In the case of the shock acceleration scenario, the energy gain rate can be parametrized by,
\begin{eqnarray}
{\dot P}_{\rm acc} = \xi c E/R_{\rm L}\approx 10^{13}\xi B~~~~{\rm eV~s^{-1}},
\label{eq1}
\end{eqnarray}
\noindent
where $E$ is the energy of electron, $\xi$ is the acceleration efficiency, $R_{\rm L} = E/eB$ is its Larmor radius, $B$ is the magnetic field inside the acceleration region (in Gauss), $e$ is the electron charge, and $c$ is the velocity of light. The synchrotron energy loss rate of these electrons during acceleration process is given by,
\begin{eqnarray}
{\dot P}_{\rm syn}(\gamma) = 4\sigma_{\rm T}cU_{\rm B}\gamma^2/3\approx 7.4\times 10^{-4}B^2\gamma^2~~~{\rm eV~s^{-1}},
\label{eq2}
\end{eqnarray}
\noindent
where $U_{\rm B} = B^2/8\pi$ is the energy density of the magnetic field, $\gamma$ is the Lorentz factor of electrons, and $\sigma_{\rm T}$ is the Thomson cross section. By comparing Eq.~\ref{eq1} and Eq.~\ref{eq2}, we obtain the limit on the maximum energies of electrons accelerated at the shock,
\begin{eqnarray}
E_{\rm max}^{\rm syn}\approx 60(\xi/B)^{1/2}~{\rm TeV}.
\label{eq2b}
\end{eqnarray}
\noindent
Electrons can reach TeV energies (e.g. $>3$ TeV), provided that the magnetic field in the acceleration site fulfills the condition (obtained by reversing Eq.~\ref{eq2b}), 
\begin{eqnarray}
B < 400\xi~{\rm G}.
\label{eq3}
\end{eqnarray}
On the other hand, acceleration process of electrons can be also saturated by their IC energy losses in the strong radiation from the stellar companion (i.e. the hot WR type star). The IC energy losses in the Thomson (T) regime increases fast with energy. However, in the Klein Nishina (KN) regime, they increase slowly, depending only logarithmically with energy.
The critical electron Lorentz factor, at the border these two regimes is,
\begin{eqnarray}
\gamma_{T/KN} = mc^2/3k_{\rm B}T\approx 2\times 10^4/T_5,
\label{eq4}
\end{eqnarray}
\noindent
where $m$ is the electron rest mass, $k_{\rm B}$ is the Boltzmann constant, $T = 10^5T_5$ K is the surface temperature of the companion star.
Unless the acceleration process is saturated in the T regime, it is also not likely saturated in the KN regime since the energy losses in the KN regime depend weakly (only logarithmically) with electron energy. We evaluate the conditions under which acceleration of electrons can be saturated by IC energy losses in the T regime. The energy loss rate on this process is given by
\begin{eqnarray}
{\dot P}^{\rm T}_{\rm IC}(\gamma) = 4\sigma_{\rm T}cU_{\rm rad}\gamma^2/3\approx 1.3\times 10^4T_5^4\gamma^2/r^2~~~{\rm eV~s^{-1}},
\label{eq5}
\end{eqnarray}
\noindent
where the energy density of radiation is $U_{\rm rad} = 4.7\times 10^{17}T_5^4/r^2$ eV cm$^{-3}$,
and $r$ is the distance from the companion star in units of stellar radius.
The IC energy losses at the border between T and KN regimes can be approximately estimated by putting $\gamma_{T/KN}$ into the above formula,

\begin{eqnarray}
{\dot P}^{\rm T}_{\rm IC}(\gamma_{T/KN}) \approx 5\times 10^{12}T_5^2/r^2~~~{\rm eV~s^{-1}}.
\label{eq6}
\end{eqnarray}
Electrons can reach energies above $mc^2\gamma_{T/KN}$, when their energy gain rate is clearly above the IC energy loss rate at the border between the T and KN regimes. By comparing Eq.~\ref{eq1} and Eq.~\ref{eq6}, we obtain the lower limit on the magnetic field strength at the acceleration region for which electrons can reach TeV energies,
\begin{eqnarray}
B > T_5^2/(2\xi r^2).
\label{eq7}
\end{eqnarray}
\noindent
This limit is $\sim 1/(10\xi)$~G for the case of the Cyg X-3 binary system in which the temperature of the companion star is estimated on $\sim 10^5$ K and the distance of the compact object from the star is $r = 2.25$.
The above limits on the magnetic field strength in the acceleration region which allow acceleration of  electrons to 3 TeV, can be converted to the lower limit on the acceleration efficiency, $\xi > 0.016$ (from comparison of Eq.~ \ref{eq3} and Eq.~\ref{eq7}). We conclude that in the case of the shock acceleration scenario (operating within the jet), electrons can be in principle accelerated to a few TeV, provided that the acceleration mechanism is efficient enough. Note, that $\xi$ in the Crab Nebula is expected to be not far from unity and the theoretical estimates give the values of $\xi$ close to $\sim 0.1$ in the case of relativistic shocks (e.g. Malkov \& Drury~2001). Applying the limiting case $\xi\sim 1$, we estimate the magnetic field in the acceleration region of the jet on $B_{\rm j}\sim 0.1 - 400$ G (able to accelerate electrons to a few TeV). Such magnetic fields in the jet of microquasar seem to be characteristic at distances from its base comparable to the dimensions of the companion star, i.e. acceleration process still occurs inside the binary system 
(see e.g. Bednarek~2006a, Bosch-Ramon, Romero \& Paredes~2006).   
The high energy electrons and $\gamma$-rays can be also injected within the binary system in other acceleration mechanism, e.g. as a result of electron acceleration in the reconnection regions in which case synchrotron losses may become negligible (e.g. Schopper, Lesch \& Birk~1998), or as a secondary products of hadronic interactions occurring between relativistic hadrons and the background matter inside the binary system or at the vicinity of the compact object (e.g. Romero et al.~2003).

TeV electrons produce first generation of $\gamma$-rays in the IC process by scattering anisotropic radiation from a companion star. These $\gamma$-rays are absorbed in the interaction with this same radiation field. As a result, $\gamma$-ray spectrum is formed in the IC $e^\pm$ pair cascade process occurring in the whole volume of the binary system. The efficiency of the cascade process (the final $\gamma$-ray spectra escaping towards the observer) strongly depends on the location of the observer in respect to the source of primary electrons and the companion star.  
A basic prediction of such model is the anticorrelation of the GeV and TeV $\gamma$-ray emission with the orbital period of the binary system (see Bednarek~2000, Bednarek~2006b). Such features have been recently observed in the GeV-TeV $\gamma$-rays from  LS 5039 and LSI +61 303 (Abdo et al.~2009b,c).

We apply this successful general cascade scenario to the binary system Cyg X-3, in order to investigate its likely behaviour at sub-TeV $\gamma$-ray energies. Such predictions can be very useful for planning future observations of Cyg X-3 with the Cherenkov telescopes. In order to control the number of free parameters of the model, we rather prefer to discuss
general picture (without specifying any concrete model for particle acceleration).
Therefore, we assume that somewhere inside the binary system a point like isotropic source of relativistic electrons appears. The location of the source is described by its distance from the companion star, $R_\gamma$ (measured in the plane of the binary system), and the distance, $D$ (above the plane of the binary system). The source is located at the phase, $\varphi$ (phase zero corresponds to the source behind the companion star in respect to the observer), and the observer is located at the inclination angle $i$. The source accelerates electrons with the power law spectrum extending up to TeV energies. In our calculations, we apply the single power law differential spectrum for electrons,
$dN/dE\propto E^{-3.4}$, with the cut-off at 3 TeV. The spectral index has been chosen in order to be consistent with the differential spectral index of the GeV $\gamma$-ray emission observed by the {\it Fermi} satellite
(Abdo et al.~2009a). With these assumptions on the source of relativistic electrons, we analyse the "pure" IC $e^\pm$ pair cascade in the radiation of a massive star characterised by the parameters of the WR star observed in Cyg X-3. We perform cascade calculations under specific assumptions. At first, the local isotropization of secondary cascade $e^\pm$ pairs has been assumed. Such assumption is valid provided that the Larmor radius of electrons is much smaller than the mean free path, $\lambda_{\rm IC}$, for their IC energy losses (Bednarek~1997), i.e. 
\begin{eqnarray}
2\pi R_{\rm L} < \lambda_{\rm IC}.  
\label{eq8}
\end{eqnarray}
\noindent
The above condition is the most restrictive for electrons with energies at the border between the T and KN regimes (in the KN regime $R_{\rm L}/\lambda_{\rm IC}^{\rm KN}\approx const$).
We estimate $\lambda_{\rm IC}^{\rm T/KN} = cE/{\dot P}^{\rm T}_{\rm IC}(\gamma_{T/KN})\approx 6\times 10^7 r^2/T_5^3$ cm, for electrons with energy $E_{\rm T/KN} = m\gamma_{\rm T/KN}$. Then, the condition given by Eq.~9 can be transferred to the limit on the magnetic field in the region of cascade development,
\begin{eqnarray}
B_{\rm cas} > 3.6T_5^2/r^2~{\rm G}.
\label{eq9}
\end{eqnarray}
For other energies of electrons than $E_{\rm T/KN}$, the limit is less restrictive.

It is assumed that primary electrons and secondary leptons cool completely in the IC process up to 500 MeV. Note that more realistic scenario (which include a number of possible additional processes) can be very complicated. It will need many free parameters for correct description.
Unfortunately, such parameters are at present not well known preventing detailed analysis of the complite cascade picture at this stage. For example, we neglected possible effects
due to the strong magnetic field in the region of the cascade development such as synchrotron energy losses and partial re-distribution of directions of leptons by the ordered magnetic field (particles following local magnetic field). The importance of these effects were
already considered in (Bednarek~1997, see Eq.~3 and Fig.~4). It is possible to 
estimate the magnetic field strength in the cascade region (i.e. the region around the companion star) for which the synchrotron energy losses of secondary cascade $e^\pm$ pairs can be neglected in respect to IC energy losses. In the T regime, it is enough to compare the energy density of the magnetic field with the energy density of stellar radiation ($U_{\rm B} = U_{\rm rad}$). Then, the upper limit on the magnetic field strength in the cascade region is, 
\begin{eqnarray}  
B_{\rm cas}^{\rm T} < 4.3\times 10^3 T_5^2/r~{\rm G}.
\label{eq10}
\end{eqnarray}
However, the limit on the magnetic field is much more restrictive for electrons scattering radiation in the KN regime. By analysing dependence of the energy losses of electrons on these two radiation processes, we conclude, that with good approximation the limit on B in the KN regime should scale as $B_{\rm cas}^{\rm KN}\approx
B_{\rm cas}^{\rm T}(E_{\rm T/KN}/E)$. Therefore, for $e^\pm$ pairs with energies of 3 TeV, 
the limit on the magnetic field strength in the cascade region is of the order of $\sim 10$ G.
Although the magnetic field around the companion star drops fast from its surface (close to the surface in the dipole region as $\propto r^{-3}$ and in the region dominated by the wind as $\propto r^{-2}$), it may happen that synchrotron  process may extract some additional energy from the cascade $e^\pm$ pairs. However, this will not influence the final conclusions of our paper on the sub-TeV $\gamma$-ray emission escaping from the binary system. The synchrotron losses will result in even lower level of the $\gamma$-ray fluxes than predicted by our "pure" IC $e^\pm$ pair cascade scenario.
Note, that synchrotron energy losses have been taken into account in some versions of the IC $e^\pm$ pair cascade scenario (e.g. Sierpowska \& Bednarek~2005, Bednarek \& Giovannelli~2007 and also Bosch Ramon et al.~2008). 

We also neglected the adiabatic energy losses of leptons and non-local production of secondary $\gamma$-rays by secondary leptons (i.e. efficient escape of leptons from their place of origin). These two processes are difficult to treat reasonably with a limited number of free parameters. We note that all the effects mentioned above can be considered in detail when  precise multiwavelength results are available allowing realistic constraints of the free parameters describing these processes in the case of specific binary system.

In the following subsections we calculate the $\gamma$-ray light curves at energies above 1 GeV, 100 GeV, and 330 GeV. We show the $\gamma$-ray spectra at specific phases of the injection source. The results are discussed for the circular and eccentric orbits of the source of relativistic electrons.

\subsection{Gamma-ray light curves}

We calculate the cascade $\gamma$-ray light curves expected for two inclination angles ($i = 30^{\rm o}$ and $60^{\rm o}$) of 
the binary system Cyg X-3 assuming the point-like isotropic source of relativistic electrons.
Electrons have the power law spectrum as described above. 
The fluxes of $\gamma$-rays are obtained above three energies (1 GeV, 100 GeV, and 330 GeV), characteristic for the satellite observations and Cherenkov telescopes.
In the first set of calculations (shown in Fig. 3), the injection source is at a circular orbit 
but at different distances from the companion star. As expected, the GeV $\gamma$-ray light curves show clear modulation with the orbital period of the binary system with the maximum at phases when the injection source is behind the companion star. For small inclination angles ($i = 30^{\rm o}$), the $\gamma$-ray light curves are very similar. Note that the angle between the injection source and the observer does not change with the phase of the binary system for different radii of the orbit.  
This almost independence of GeV $\gamma$-ray light curve on the distance from the 
companion star is a consequence of a type of the IC $e^\pm$ pair cascade considered in this paper (see Sect. above). Note that electrons interact mainly with soft photons coming from the direction of the companion star. When electrons are located farther from the star then the radiation field becomes more anisotropic and the first generation of $\gamma$-rays is more focused towards the companion star. These cascade $\gamma$-rays interact with the soft photons on average at similar distances from the companion star. 
As a result, most of the GeV cascade emission comes from the vicinity of the companion star. This effect is also responsible for the so called {\it focussing effect of $\gamma$-rays by a strong radiation of the companion star} noted in Bednarek~(2000).
$\gamma$-ray modulation pattern changes with distance from the companion star for large inclination angles (i.e. $60^{\rm o}$, compare the two upper figures in Fig.~3). Differences are clearly seen for the phases when the injection source is in front of the companion star (phases close to 0.5). 
This is a consequence of the inefficiency of the cascade in the outward direction from the companion star. Note that the injection source is hidden behind the companion star in the case of an orbit with the radius of $1.5R_\star$ and the phases close to zero (see dashed curves in the upper Fig. 3). That's why, $\gamma$-ray flux drops drastically. But, still some $\gamma$-ray photons emerge from the opposite site of the star due to the focussing effect mentioned above.

\begin{figure*}
\vskip 11.5truecm
\includegraphics{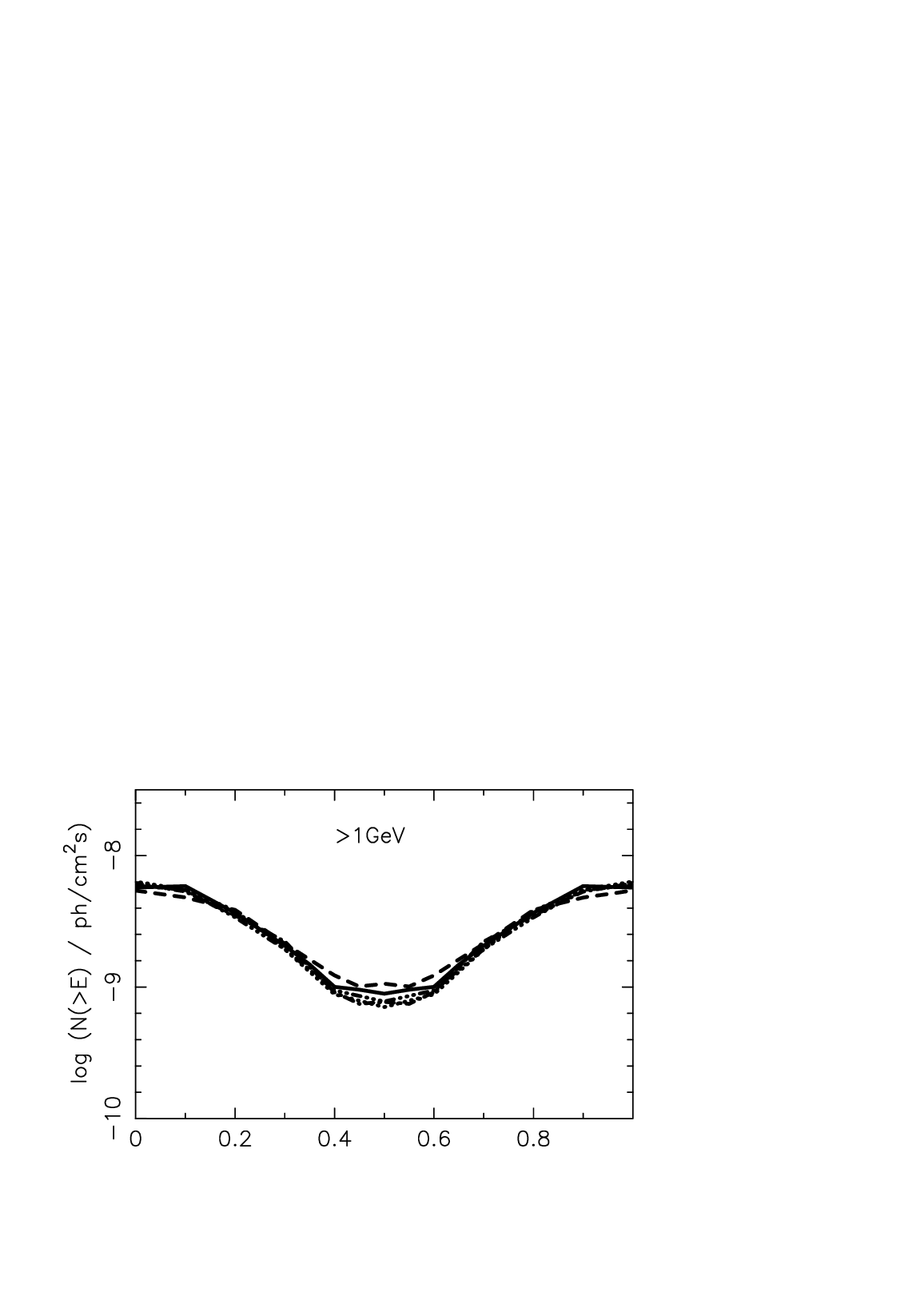}
\includegraphics{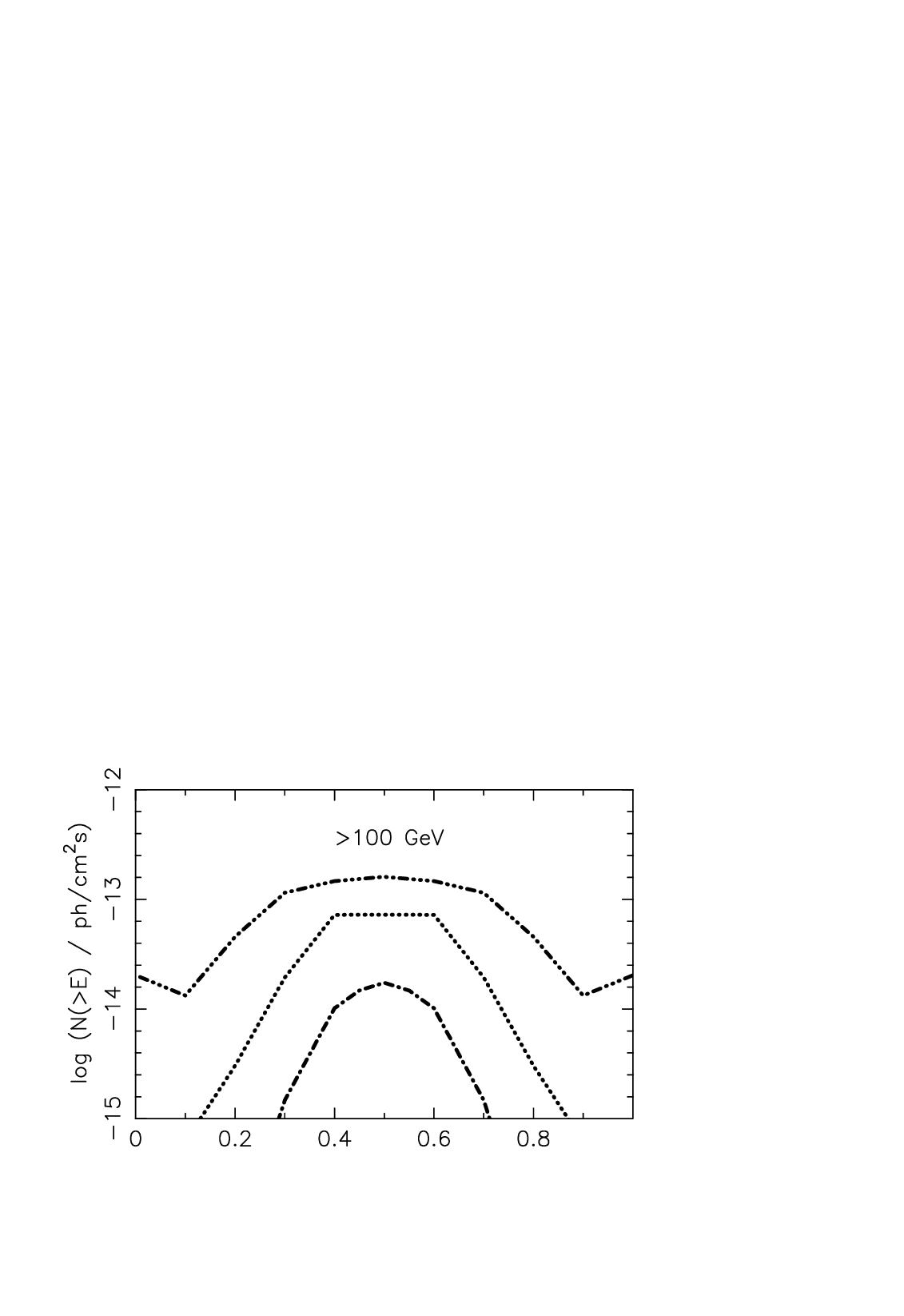}
\includegraphics{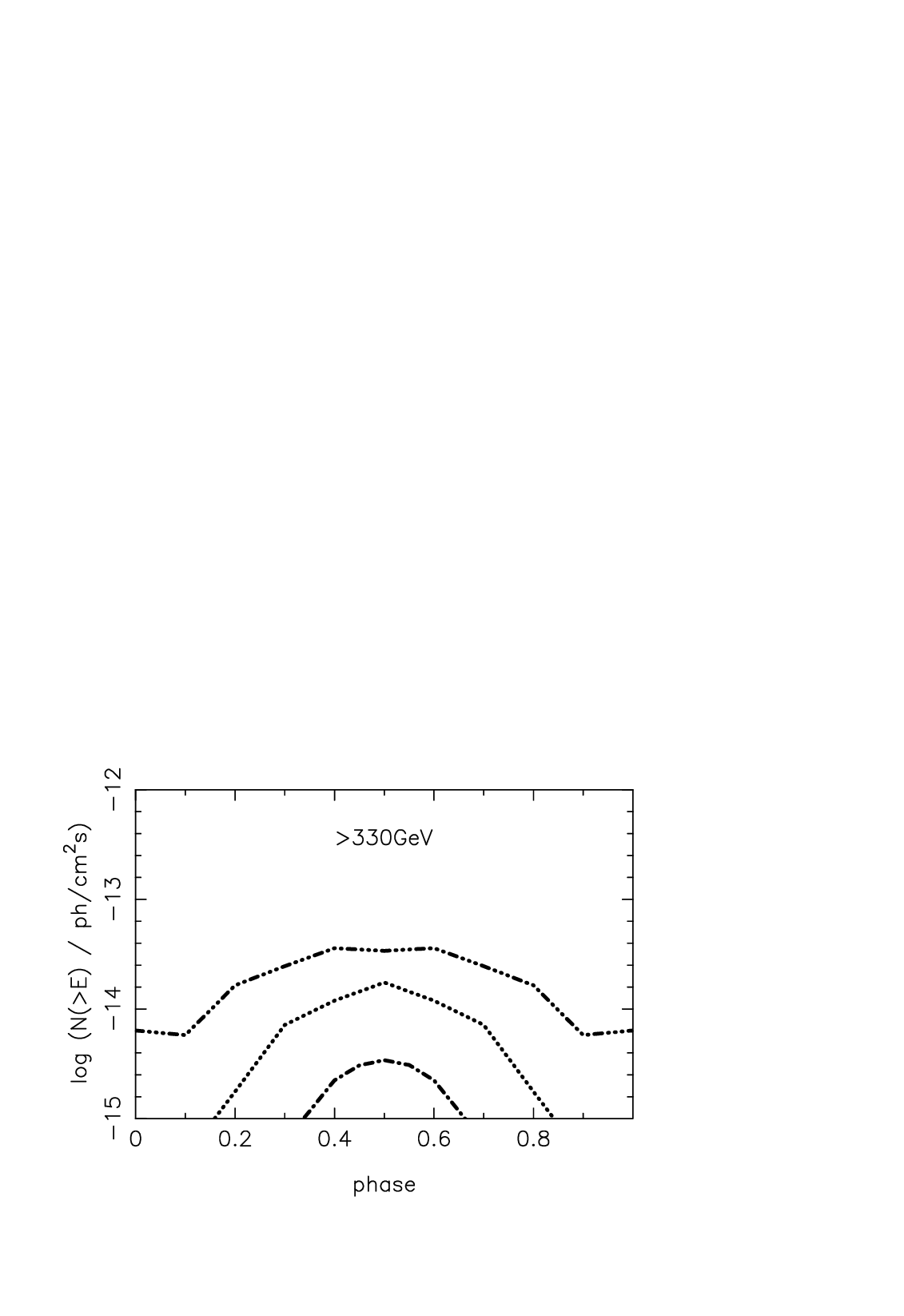}
\includegraphics{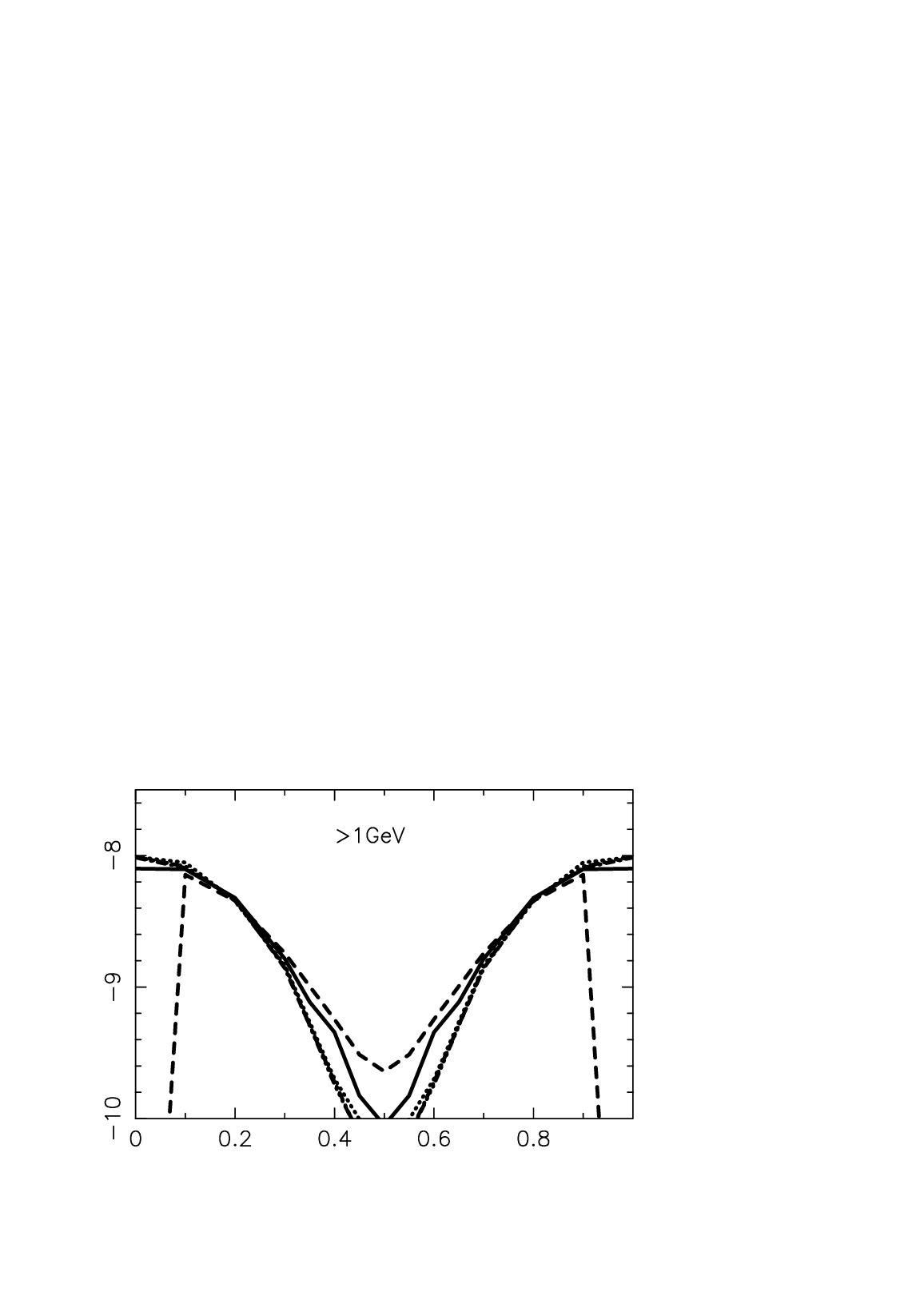}
\includegraphics{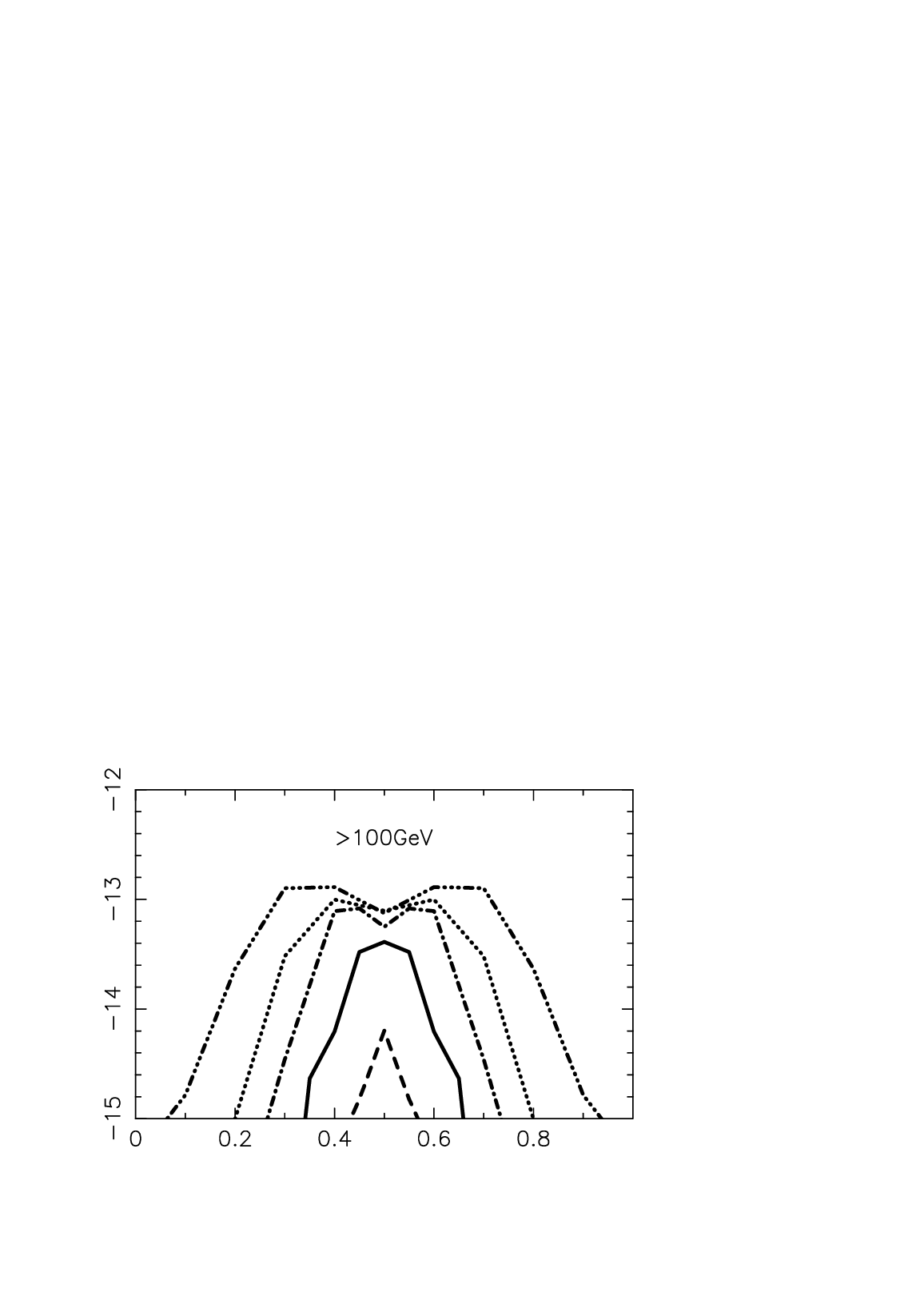}
\includegraphics{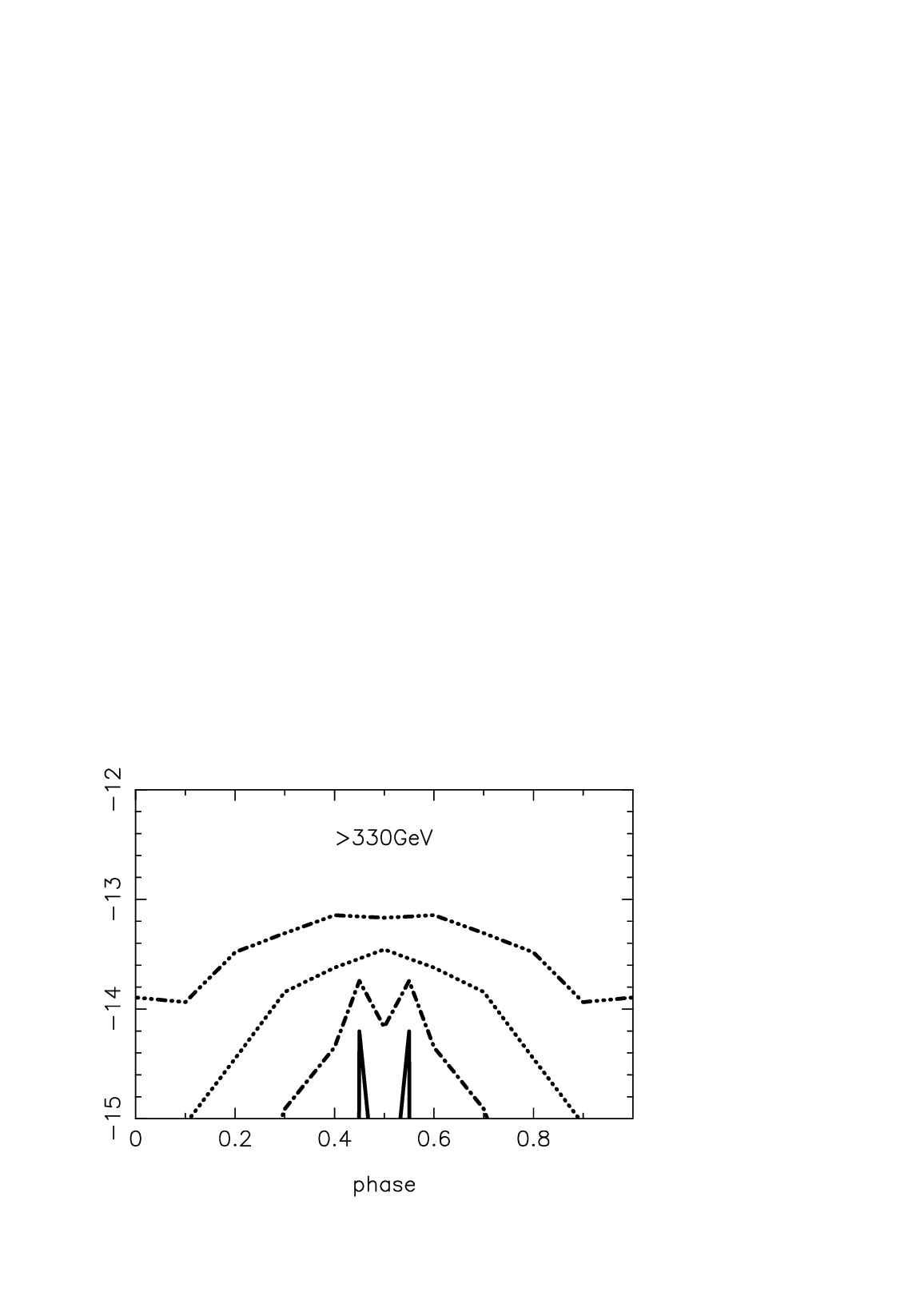}
\caption{The cascade $\gamma$-ray light curves expected for a point like isotropic source of relativistic electrons located on a circular orbit around the companion star in the plane of Cyg X-3 binary system 
for energies: $E_\gamma > 1$ GeV (upper panel), $>100$ GeV (middle), and $>300$ GeV (bottom). 
The phase zero corresponds to the location of the injection source of electrons behind the massive star. Electrons are injected with the differential power law spectrum, $\propto E^{-3.4}$, up to 3 TeV.
The injection rate of electrons is independent on the phase of the binary system.
Figures on the left are obtained for the inclination of the binary system equal to $i = 30^o$ and on the right for $i = 60^o$. The radius of the orbit around the companion star is equal to: $R_\gamma = 1.5R_\star$ (dashed curves), $2.25R_\star$ (solid), $5R_\star$ (dot-dashed), $10R_\star$ (dotted),$20R_\star$ (triple-dot-dashed). The companion star has the radius $R_\star = 2\times 10^{11}$ cm and the surface temperature $T_\star = 9\times 10^4$ K as expected for the case of Cyg X-3 binary system.}
\label{fig3}
\end{figure*}

The sub-TeV $\gamma$-ray light curves show clear anticorrelation with the GeV $\gamma$-ray light curves. They peak at phases when the injection source is in front of the companion star. Sub-TeV $\gamma$-ray light curves show strong dependence on the radius of the orbit since on average this emission originates in the cascade process occurring at larger distances from the companion star.
Moreover, the range of phases, at which sub-TeV emission is significant, decrease with the radius of the orbit of the electron injection source. For example, at energies above 100 GeV, the emission is concentrated in the range of phases between $\sim 0.3-0.7$ for the distance $R_\gamma = 20R_\star$ but only in the range $0.4-0.6$, for the distance $R_\gamma = 2.25 r_\star$. We conclude that interesting fluxes of sub-TeV $\gamma$-rays can only have a chance to emerge from the binary system for the phases when the optical depths are below unity (compare Fig.~3 with Fig.~1). Stronger sub-TeV emission is expected for larger inclination angles of the binary system.

\begin{figure*}
\vskip 11.5truecm
\includegraphics{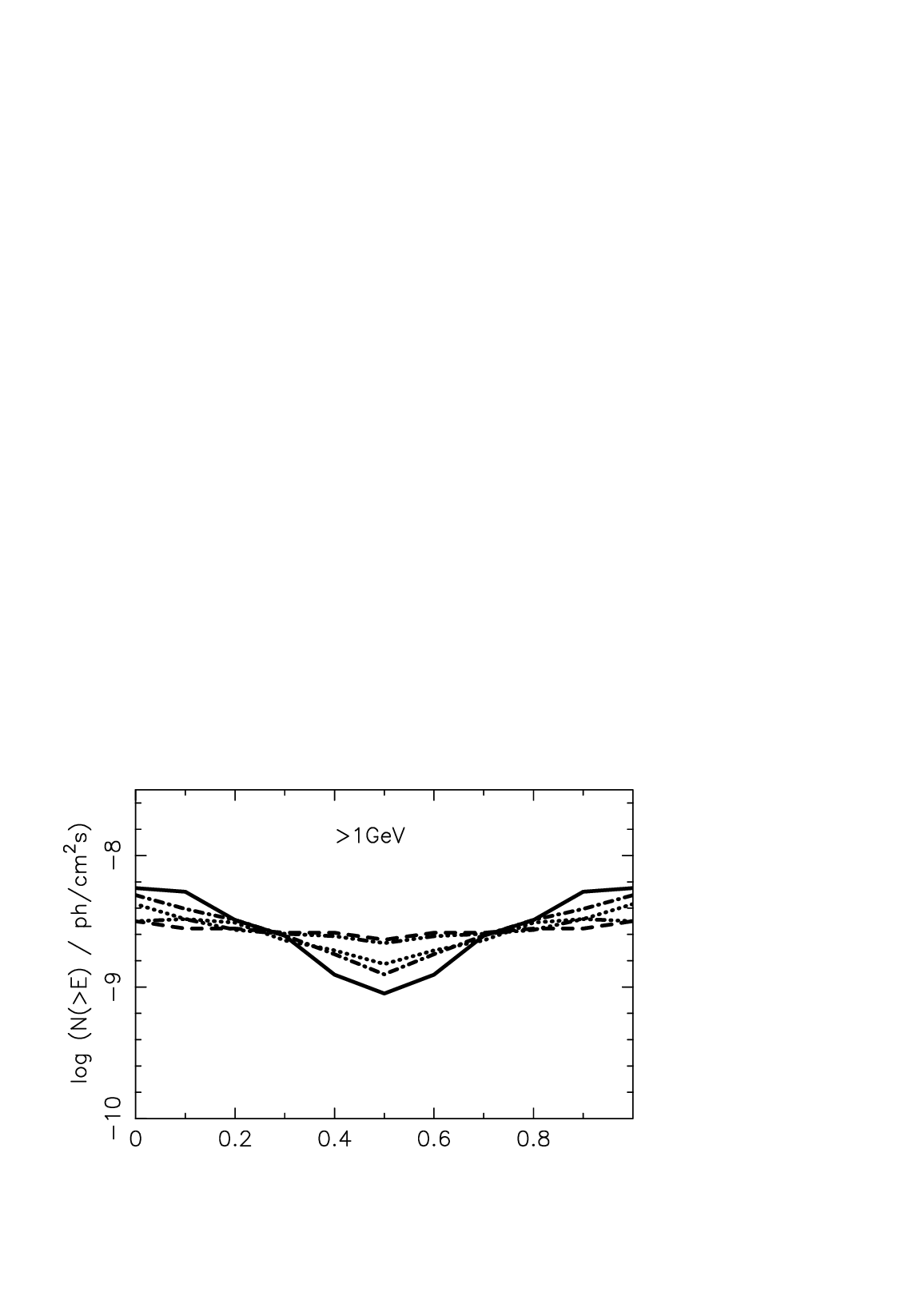}
\includegraphics{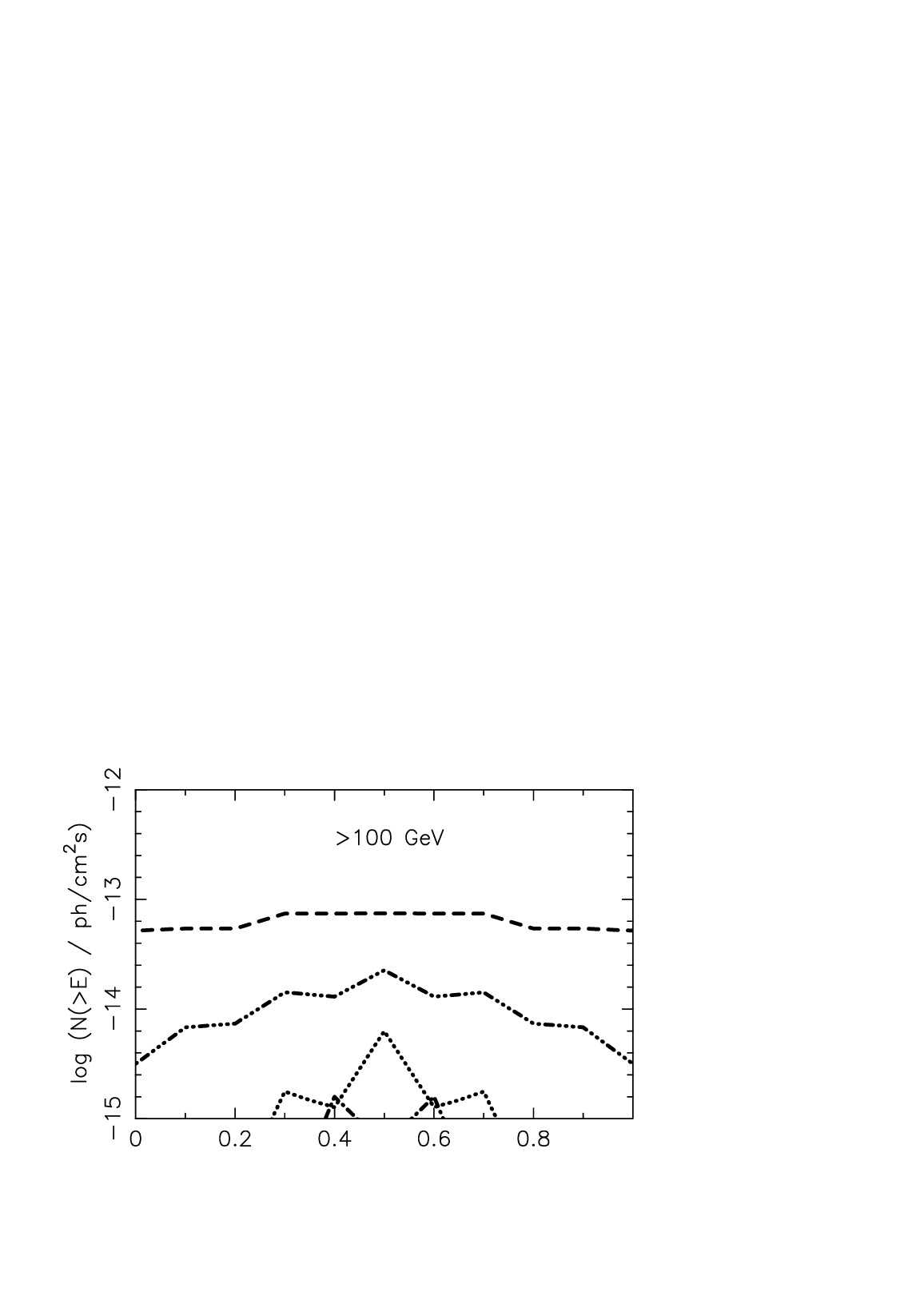}
\includegraphics{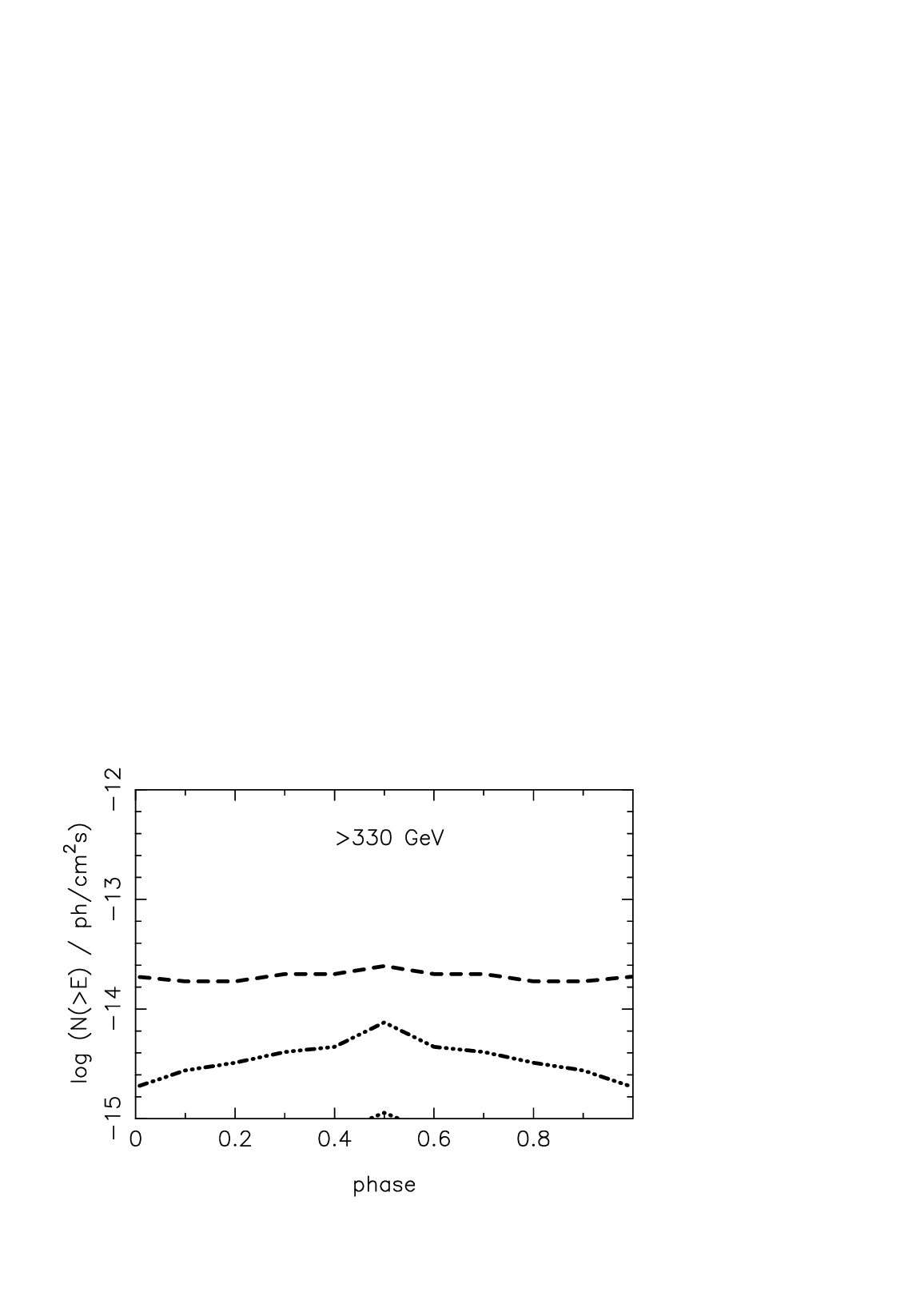}
\includegraphics{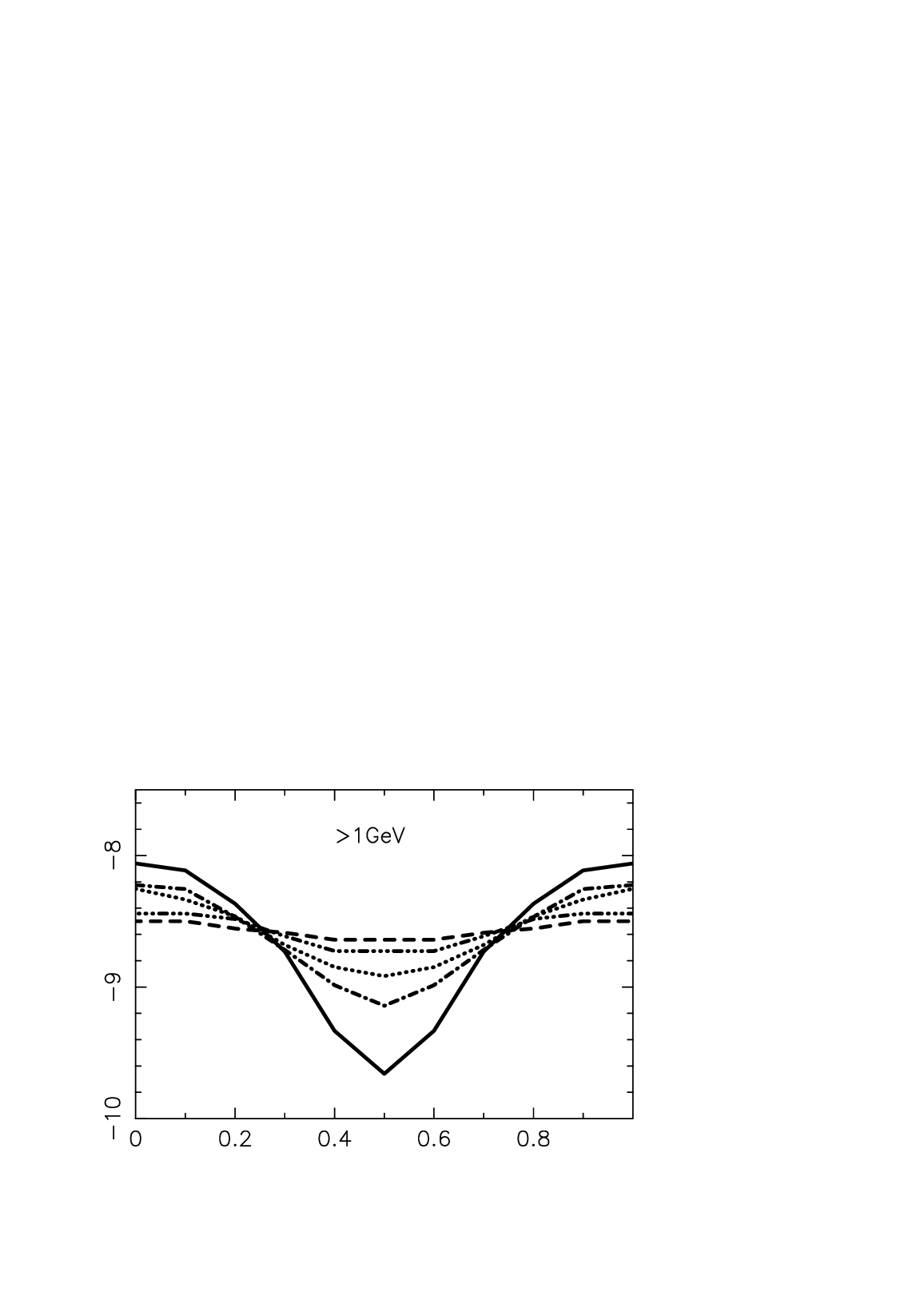}
\includegraphics{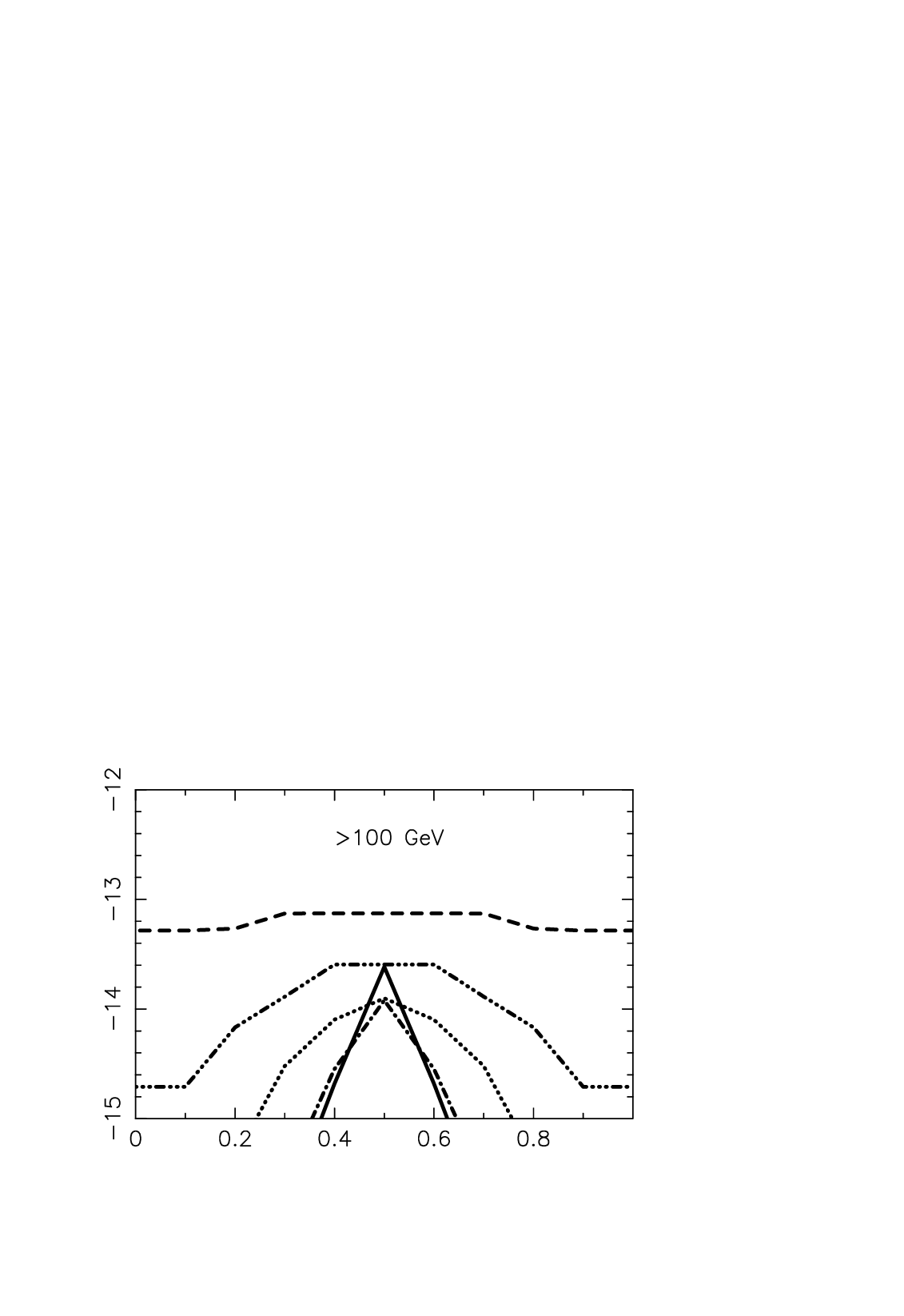}
\includegraphics{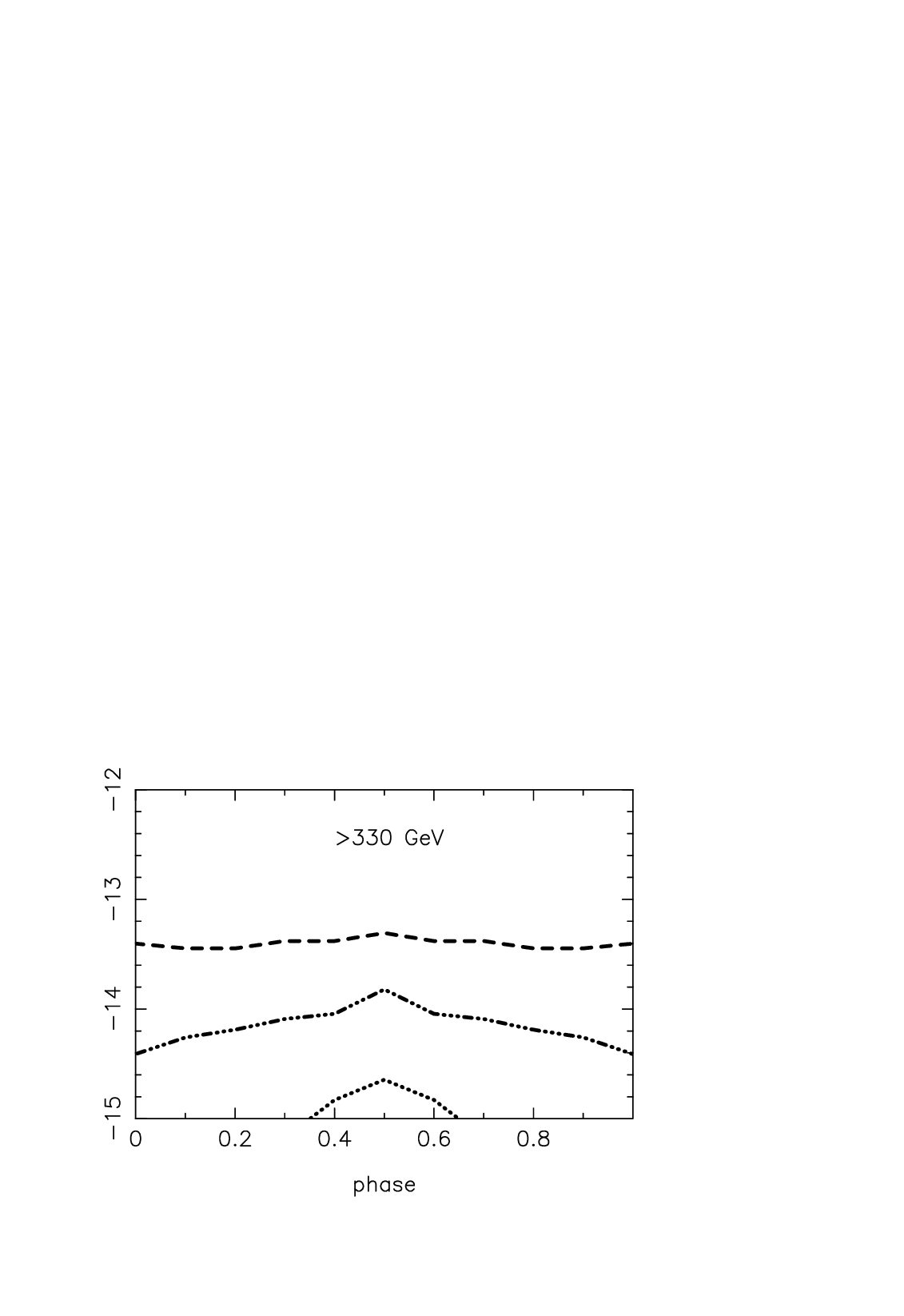}
\caption{As in Fig.~3 but for the injection source of electrons located at the distance of the compact object ($a = 2.25R_\star$) but above the plane of the binary system: $D = 1R_\star$ (solid), $3R_\star$ (dot-dashed), $5R_\star$ (dotted), $10R_\star$ (triple-dot-dashed), and $20R_\star$ (dashed).}
\label{fig4}
\end{figure*}

The dependence of the cascade $\gamma$-ray light curves on the distance from the plane of the binary system, for the radius of the orbit of the injection source of electrons equal to $R_\gamma = 2.25R_\star$ (also the distance of the compact object) is shown in Fig.~4. Note that the level of modulation of the GeV $\gamma$-ray flux with the period of the binary system decreases with the distance from the orbital plane being negligible for distances 
$>10R_\star$. For the inclination angle of the binary system equal to $i = 30^{\rm o}$, also 
the sub-TeV $\gamma$-ray flux shows weak modulation at large distances above the plane of the binary system.
The level of modulation of the GeV-TeV $\gamma$-ray signal increases significantly for larger inclination angles (see the case for $i = 60^{\rm o}$). As in the previous cases, the GeV and sub-TeV emission is clearly anticorrelated. Again, for compact orbits of the injection source, the sub-TeV $\gamma$-ray flux is limitted to the range of phases when the injection source is in front of the companion star. This range of phases increases with the distance above the plane of the binary system.

In summary, our cascade calculations show that the GeV and sub-TeV $\gamma$-ray emission is expected to be strongly modulated with the period of the binary system for the injection source of electrons within a few stellar radii
from the companion star. The GeV and sub-TeV $\gamma$-ray light curves show clear 
anticorrelation, with the maximum of GeV emission when the injection source is behind the companion 
star. The level of modulation of the GeV and sub-TeV $\gamma$-ray fluxes significantly drops with the distance from the companion star. We conclude that detection of the strong modulation of the GeV $\gamma$-ray flux indicates that $\gamma$-ray emission comes from within the binary system.
In such a case, the sub-TeV $\gamma$-ray emission is expected on a rather low level, strongly modulated with the period of the binary system, but with the maximum flux anticorrelated  with the GeV $\gamma$-ray flux.

\subsection{Gamma-ray spectra}

\begin{figure*}
\vskip 18.5truecm
\includegraphics{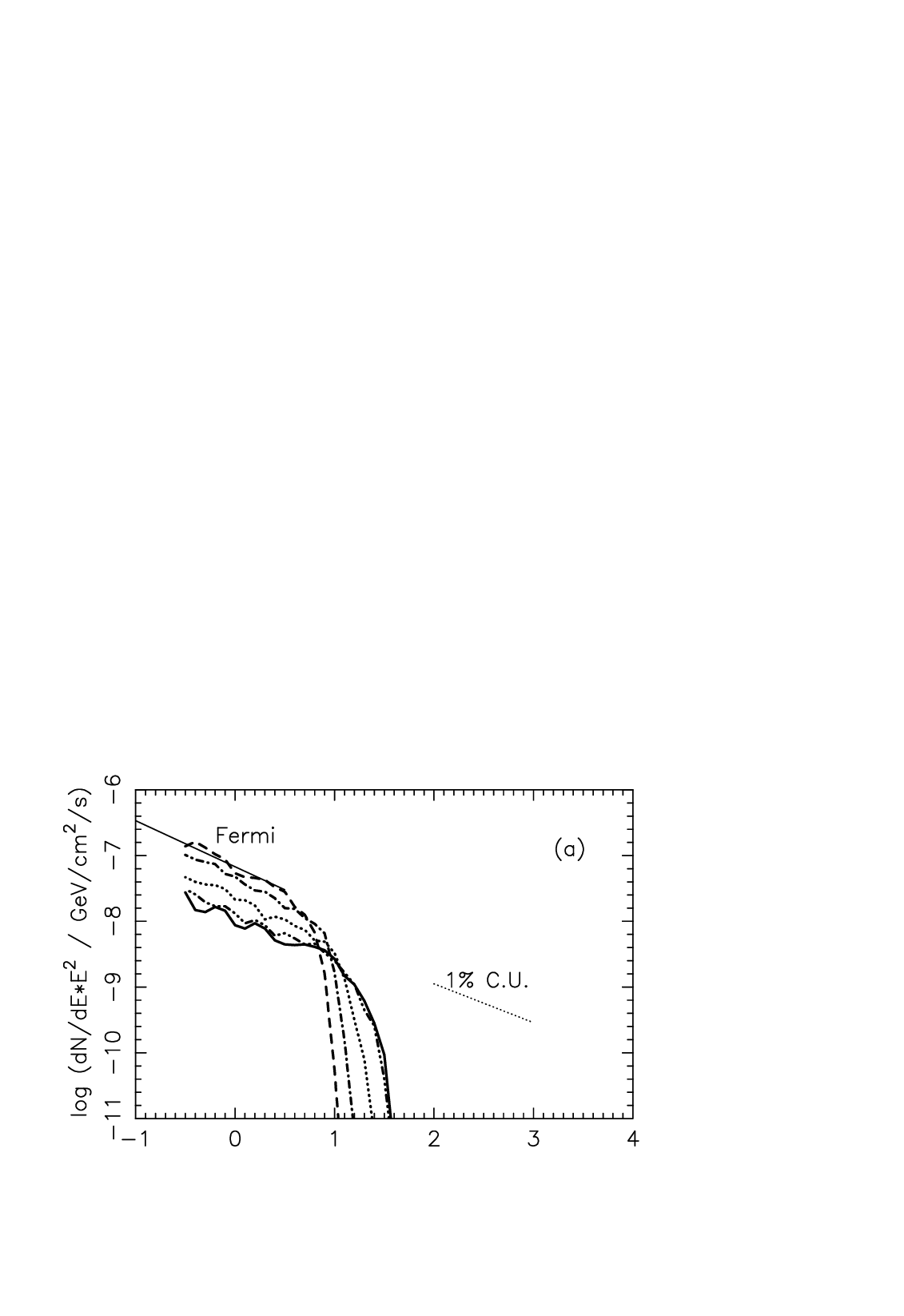}
\includegraphics{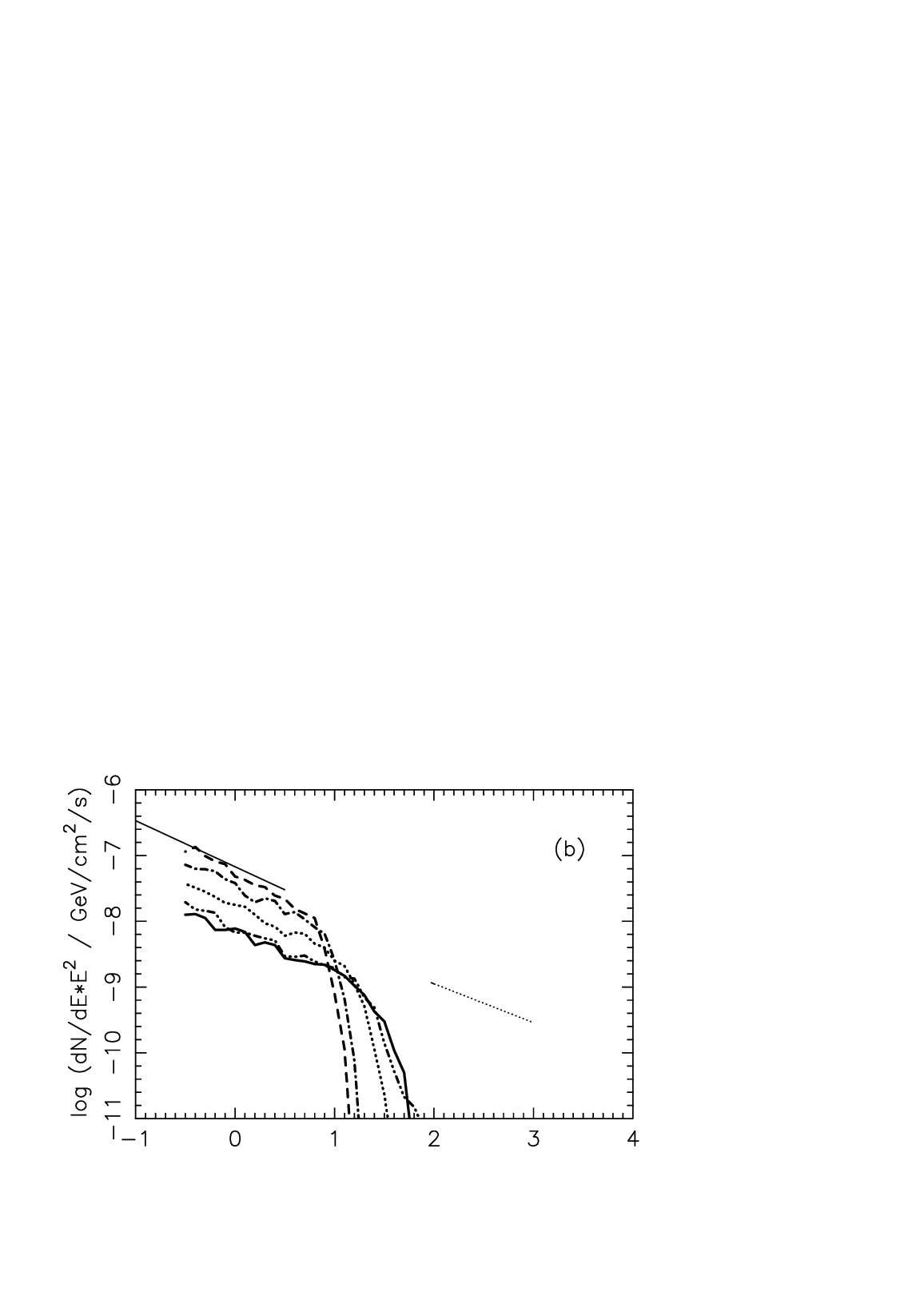}
\includegraphics{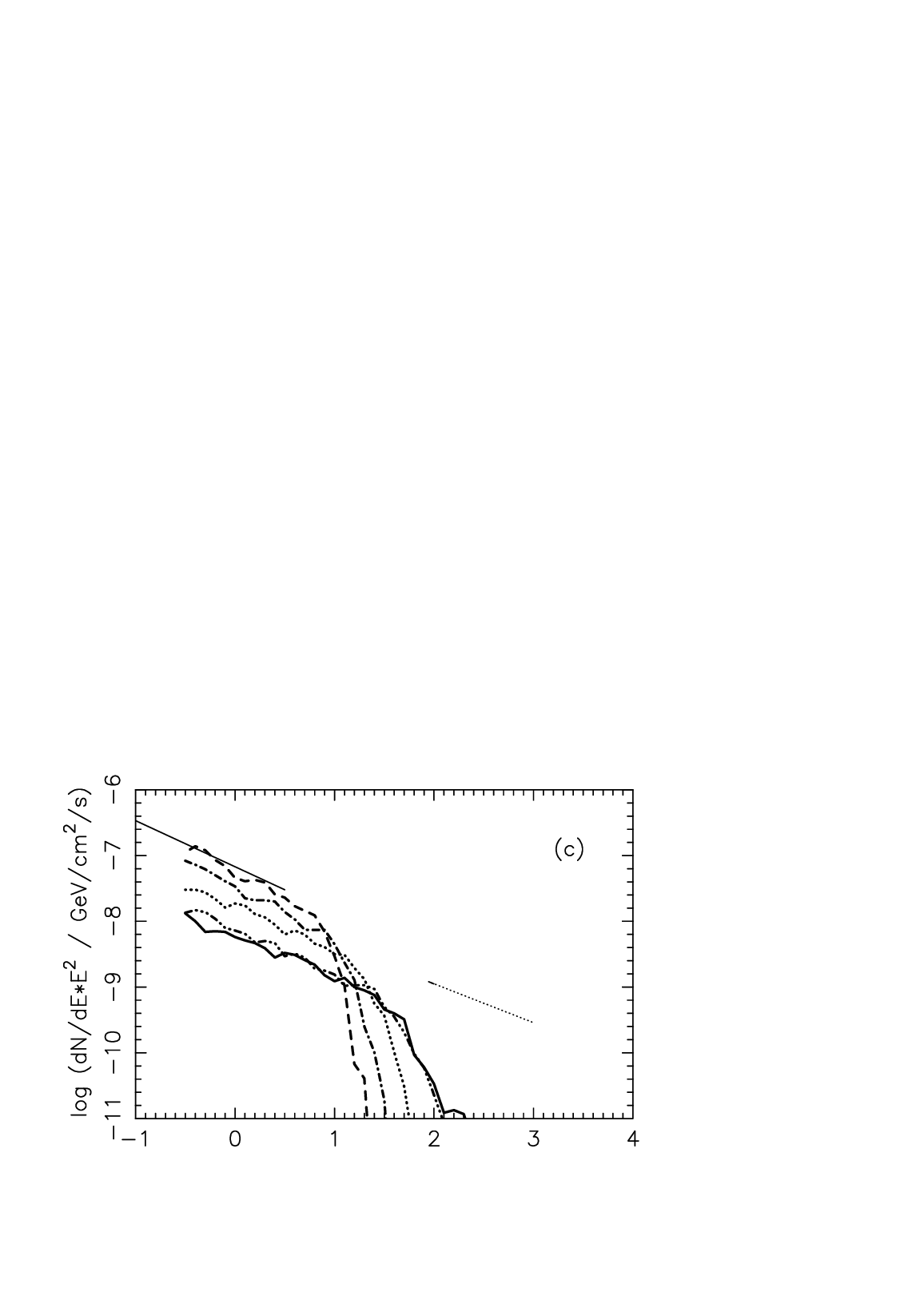}
\includegraphics{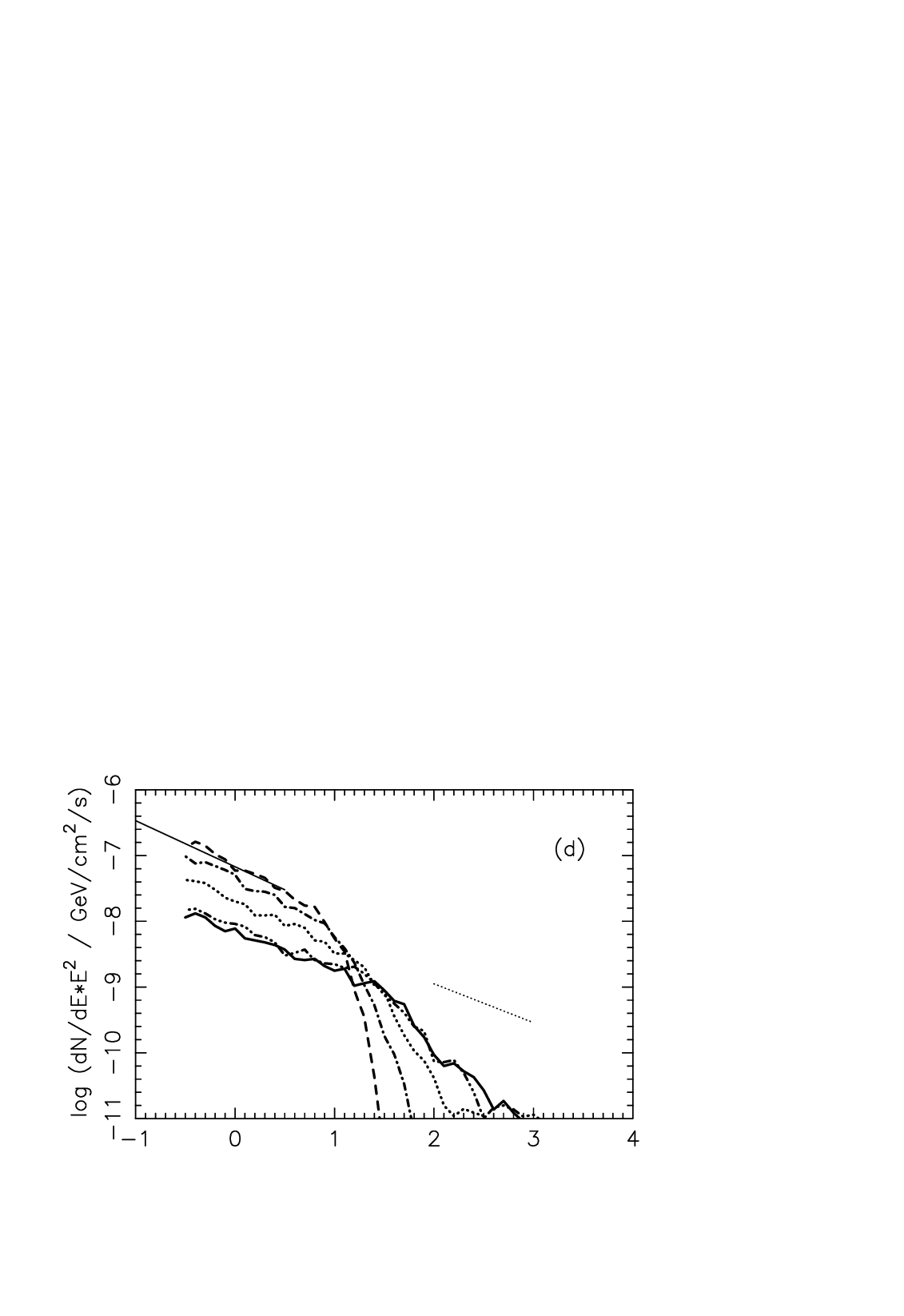}
\includegraphics{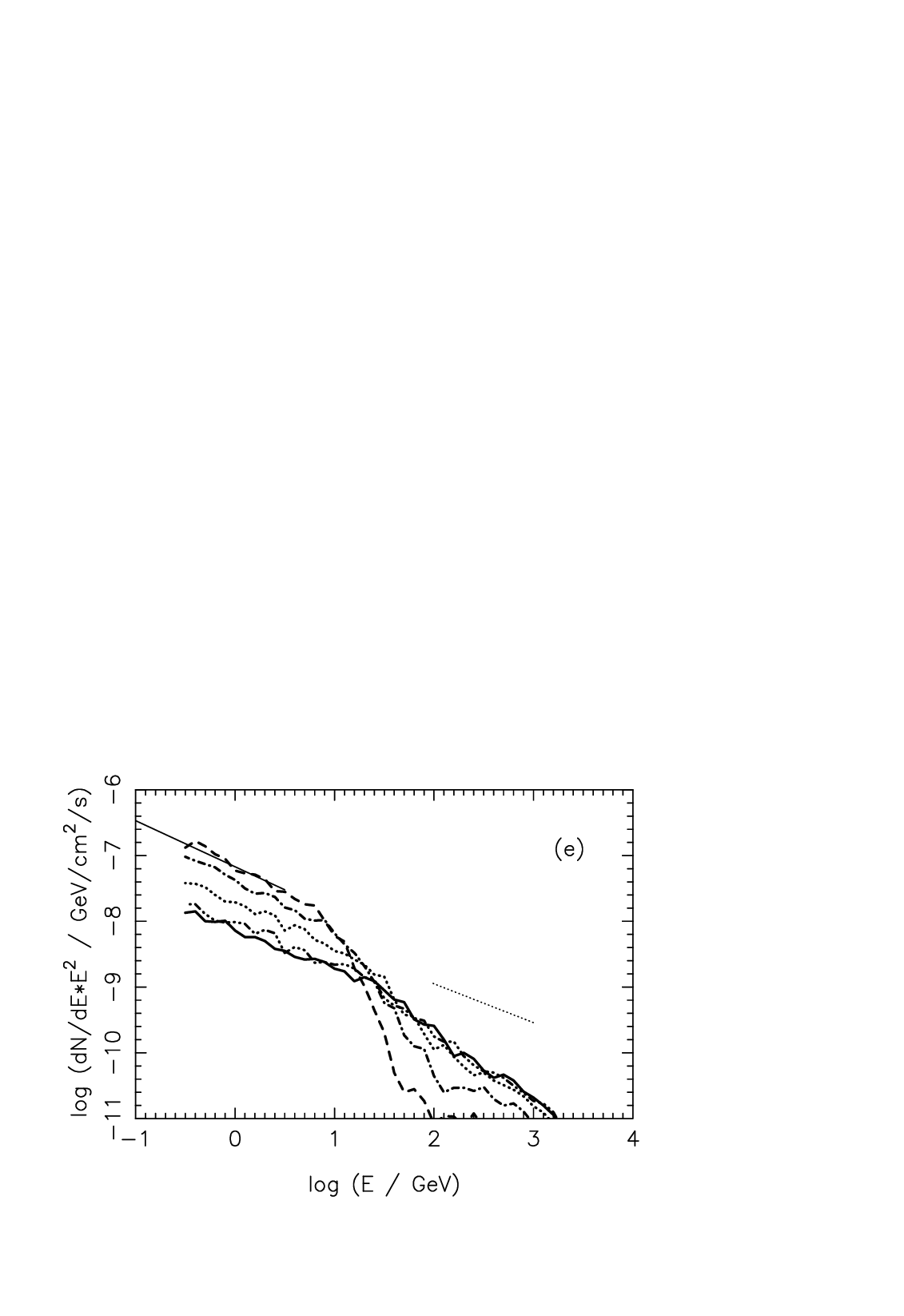}
\includegraphics{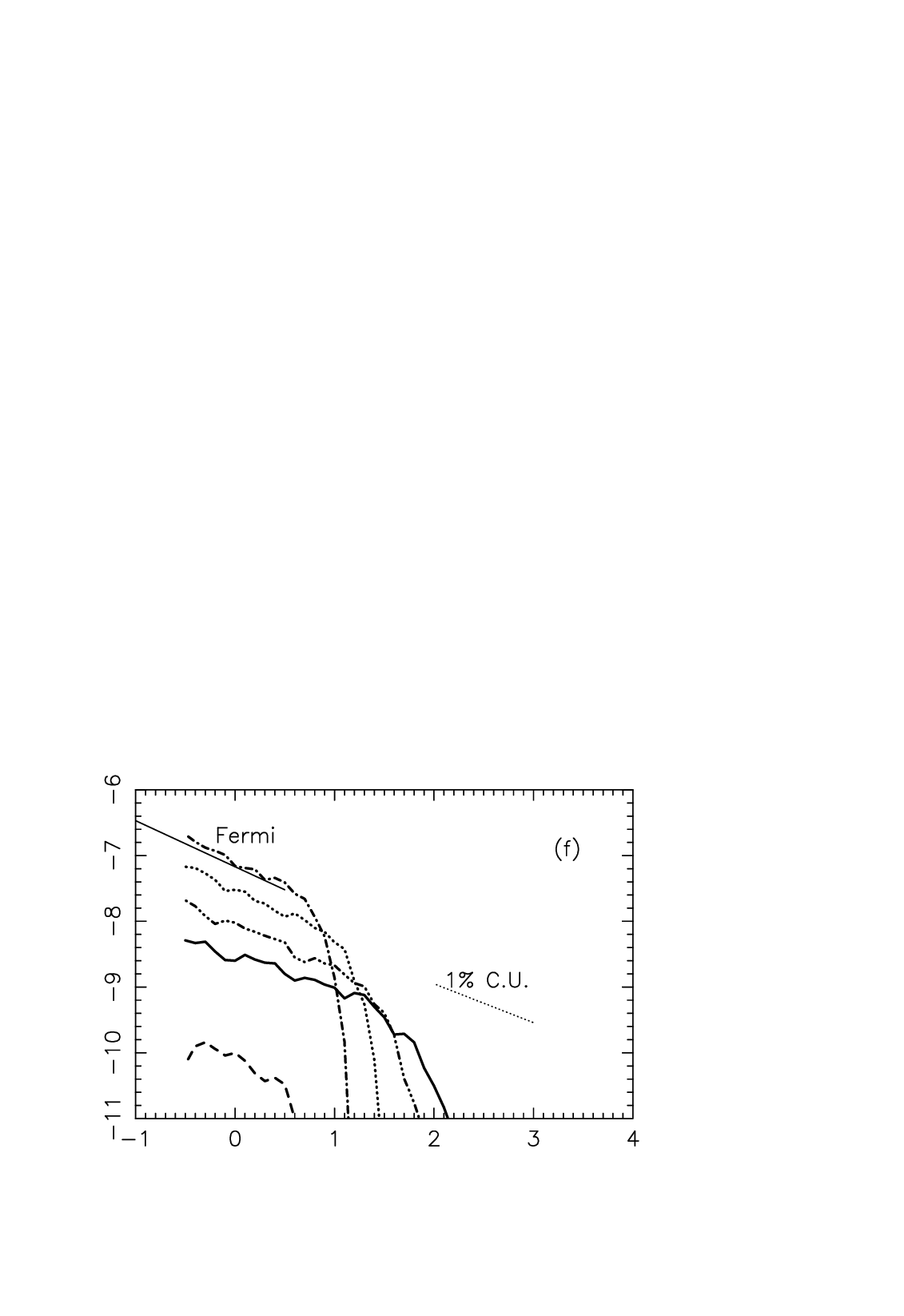}
\includegraphics{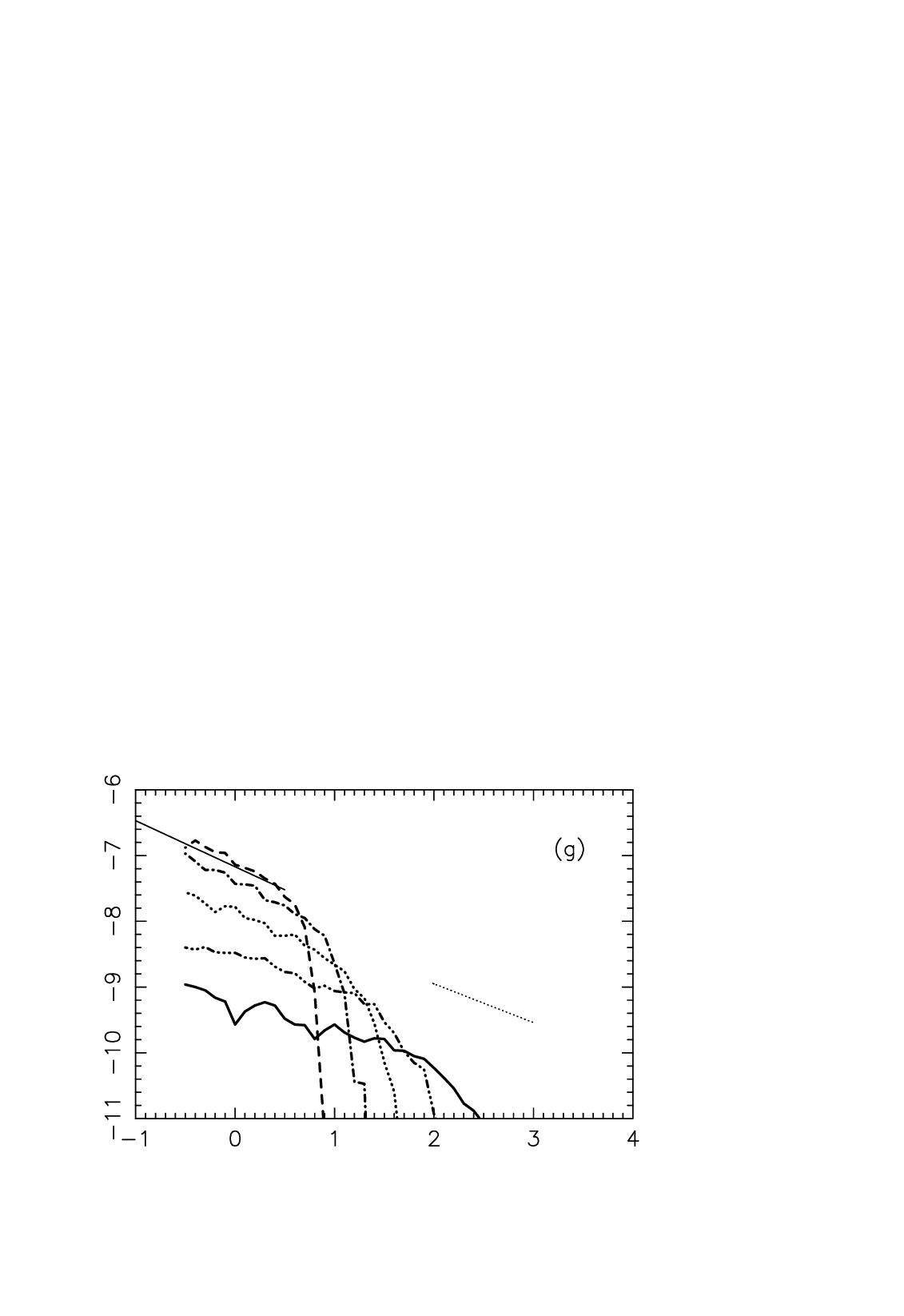}
\includegraphics{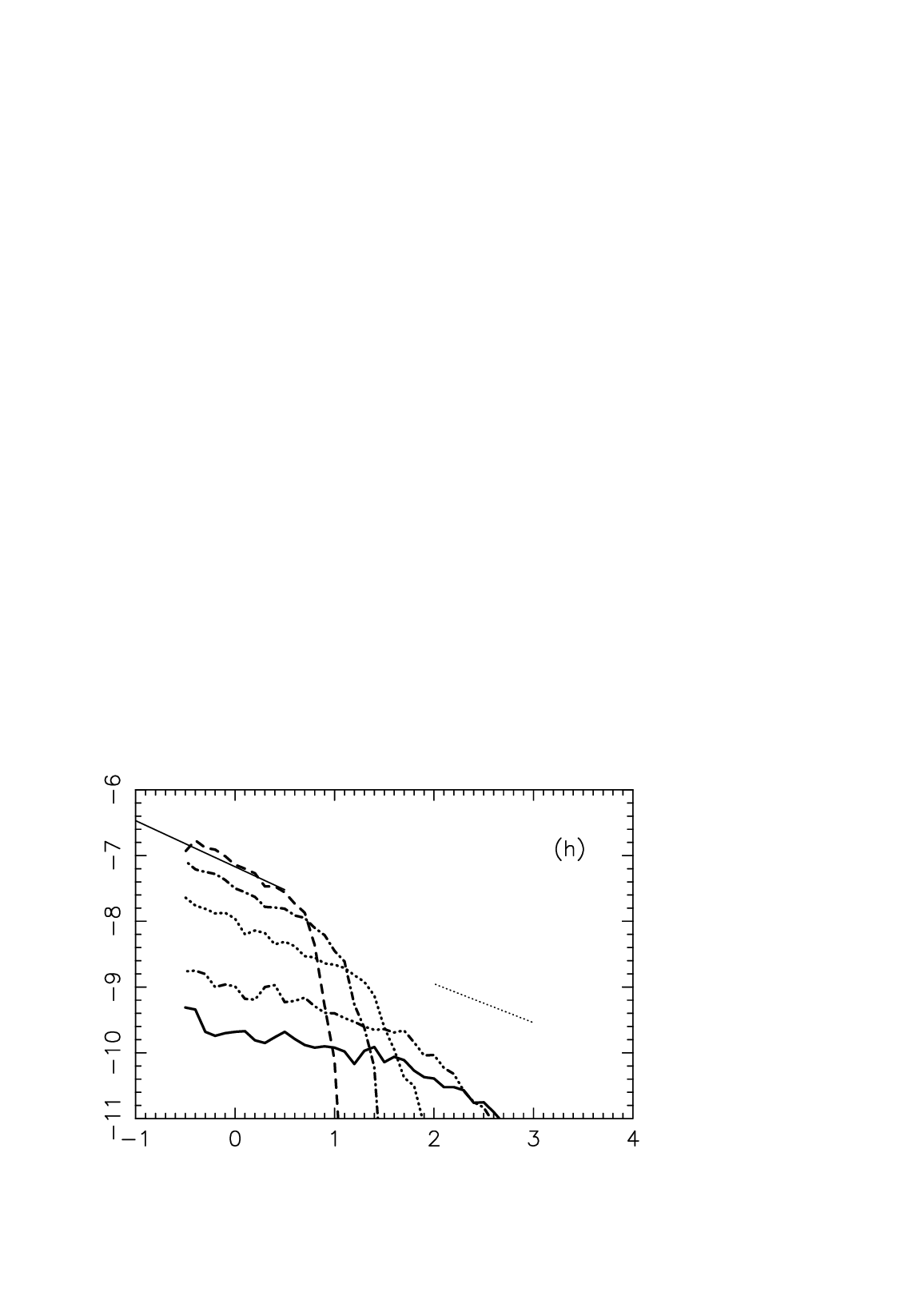}
\includegraphics{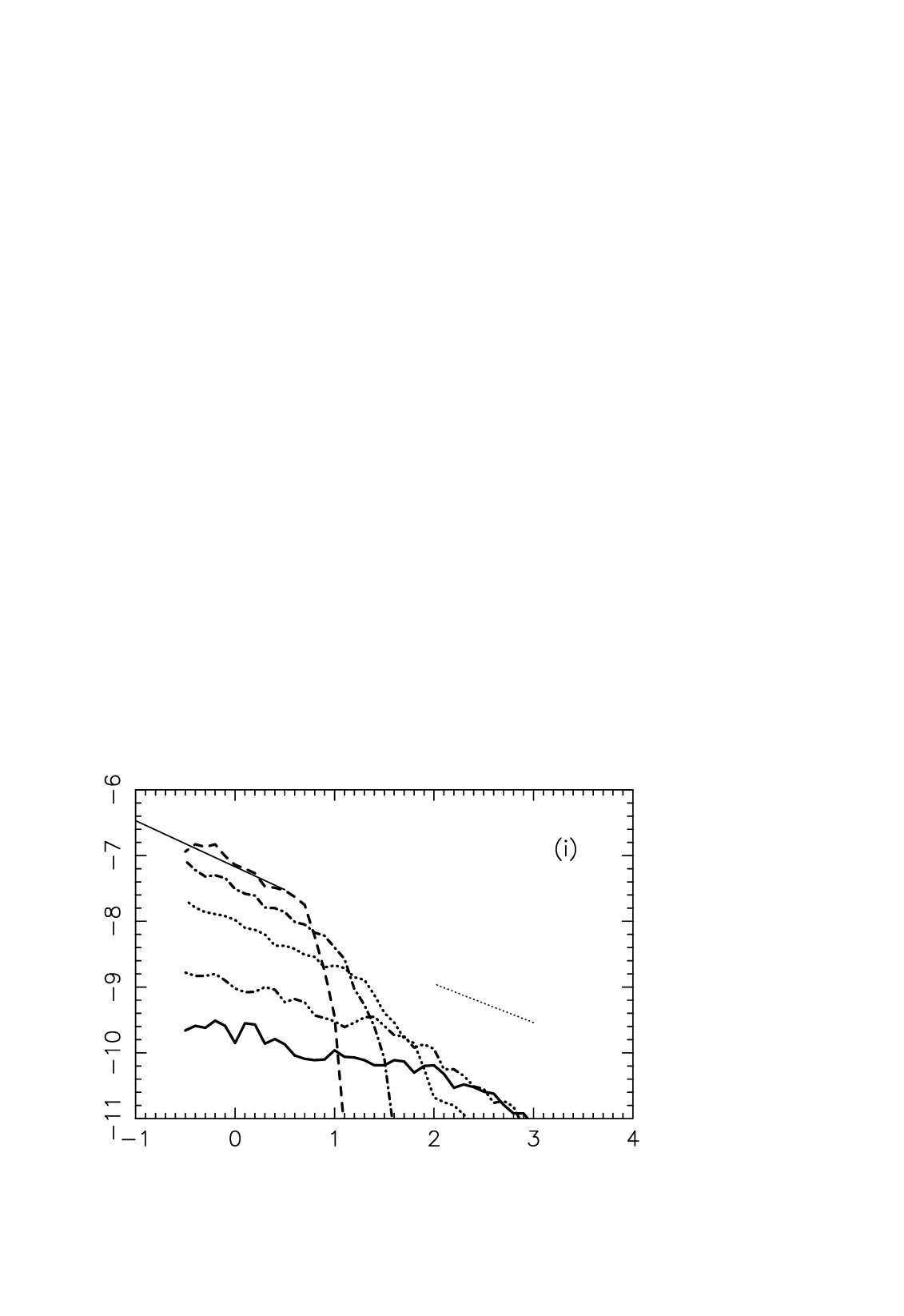}
\includegraphics{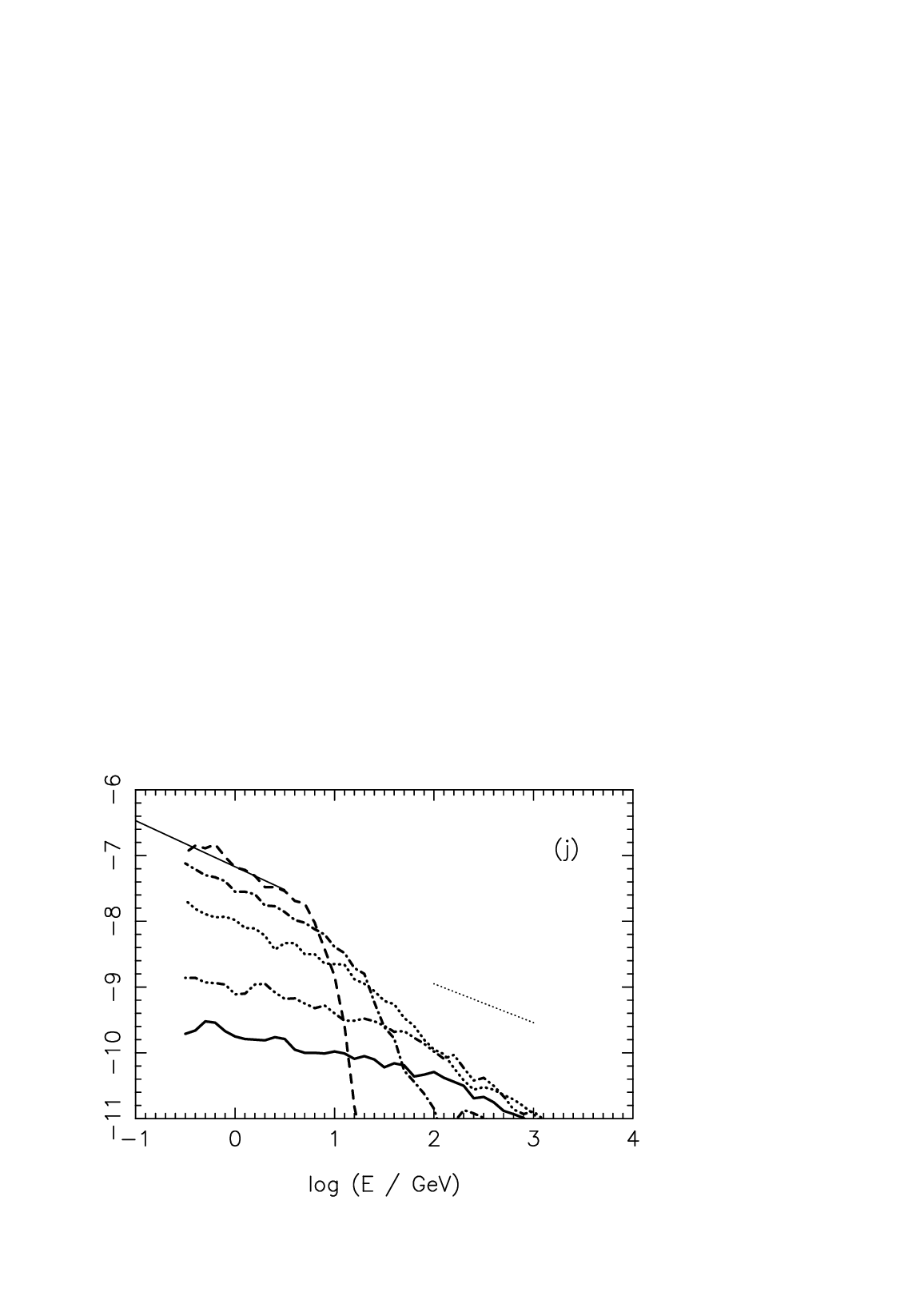}
\caption{The $\gamma$-ray spectra produced by electrons injected from a point-like isotropic source in the orbital plane of the binary system but at different distances from the companion star: $R_\gamma = 1.5R_\star$ (a and f), 
$2.25R_\star$ (b,g), $5R_\star$ (c,h), $10R_\star$ (d,i), and $20R_\star$ (e,j).
Primary electrons have a power law spectrum with differential spectral index $-3.4$ extending
up to 3 TeV.
Specific curves show the spectra for different phases of the injection source:
$\varphi = 0.$ (dashed curves), 0.1 (dot-dashed), 0.3 (dotted), 0.4 (triple-dot-dashed),
and 0.5 (solid). The phase zero 
corresponds to the location of the injection place behind the massive star. Figures on the left 
are obtained for the inclination of the binary system equal to $i = 30^o$ and on the right for 
$i = 60^o$. The thin solid line shows the GeV peak emission observed by the ${\it Fermi}$ and the thin  dotted line marks the sensitivity of the modern Cherenkov telescopes which is on the level of $1\%$ Crab Unit. The companion star has 
the radius $R_\star = 2\times 10^{11}$ cm and the surface temperature $T_\star = 9\times 10^4$ K, as expected for the case of Cyg X-3 binary system.}
\label{fig5}
\end{figure*}
\begin{figure*}
\vskip 18.5truecm
\includegraphics{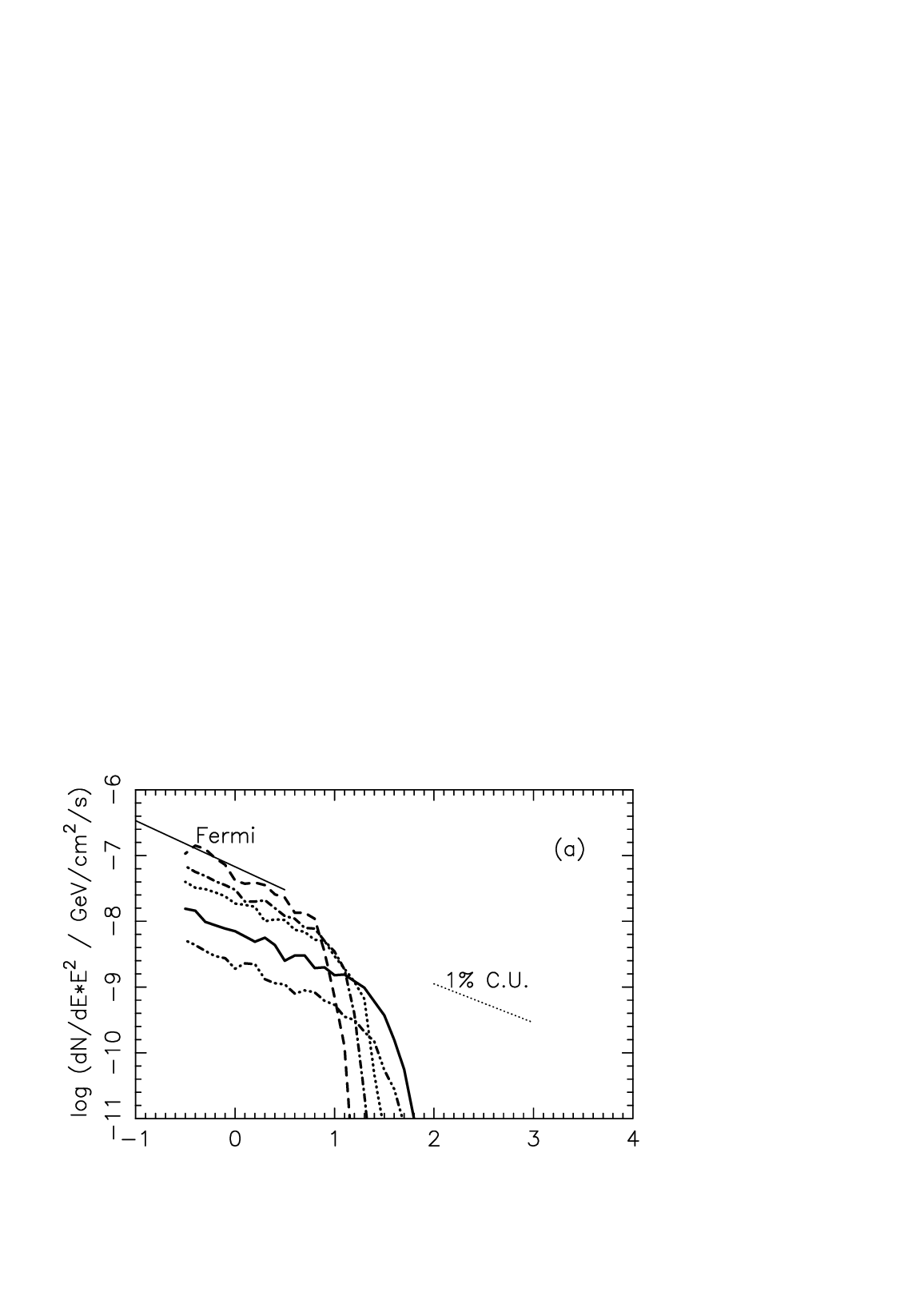}
\includegraphics{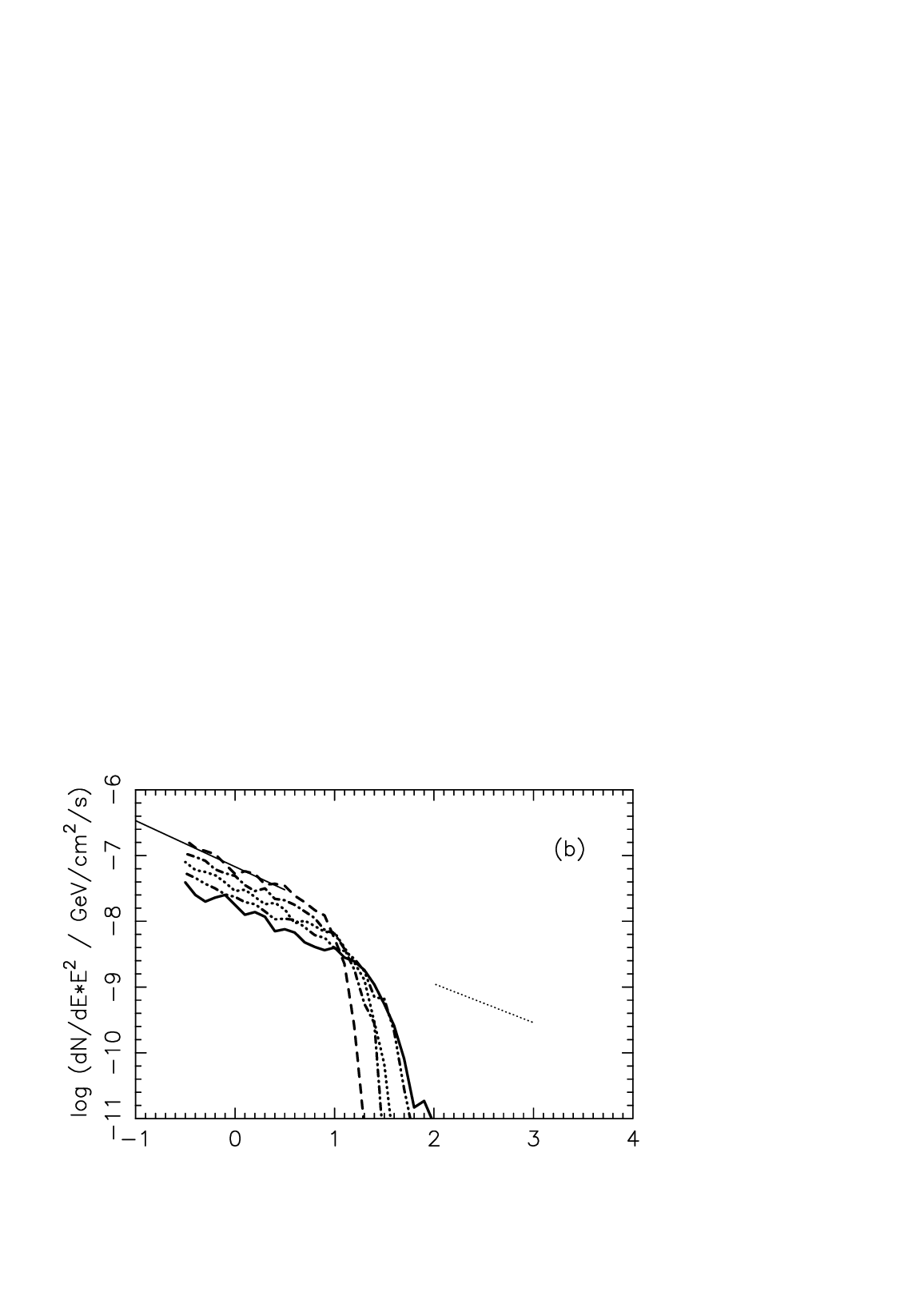}
\includegraphics{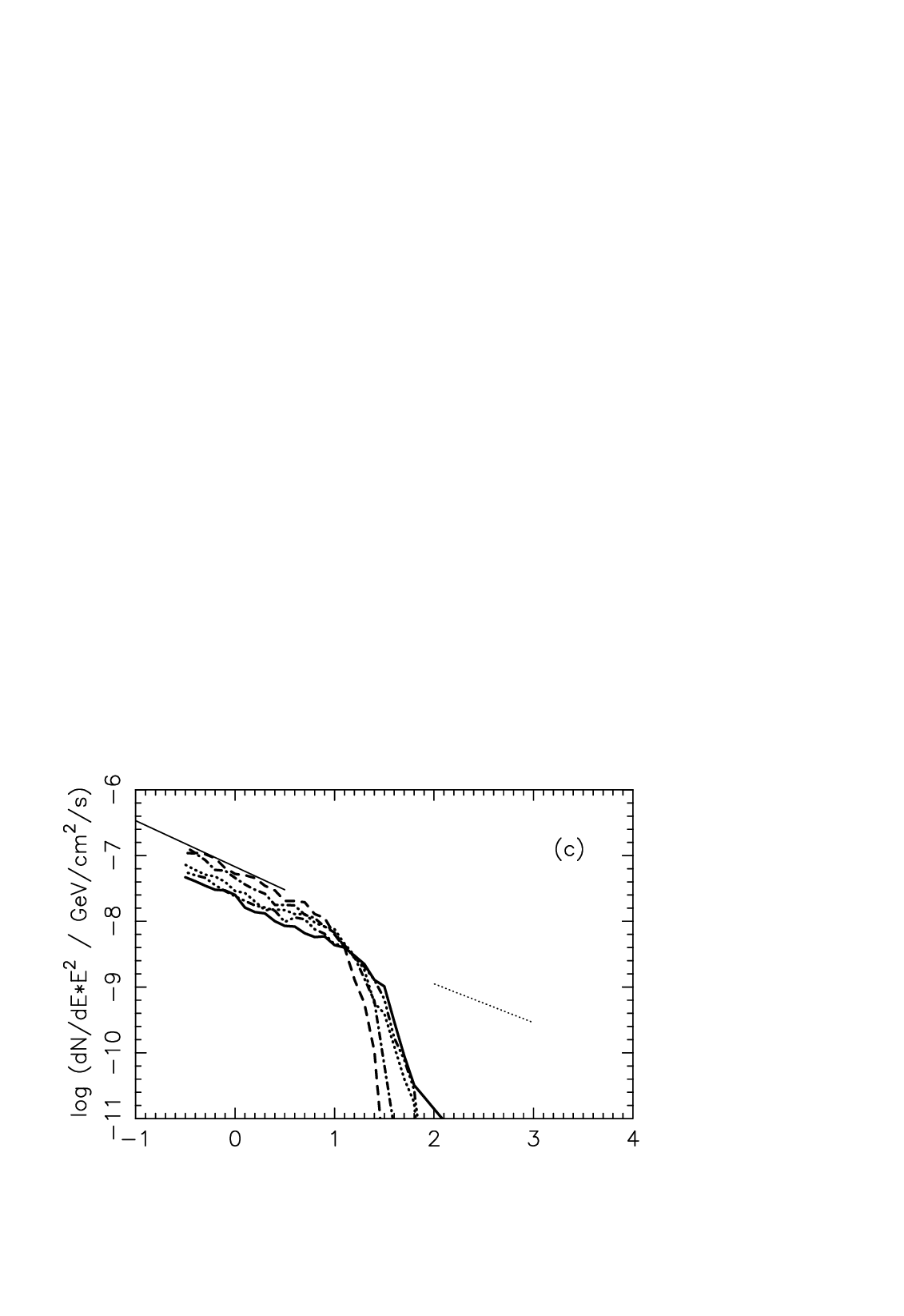}
\includegraphics{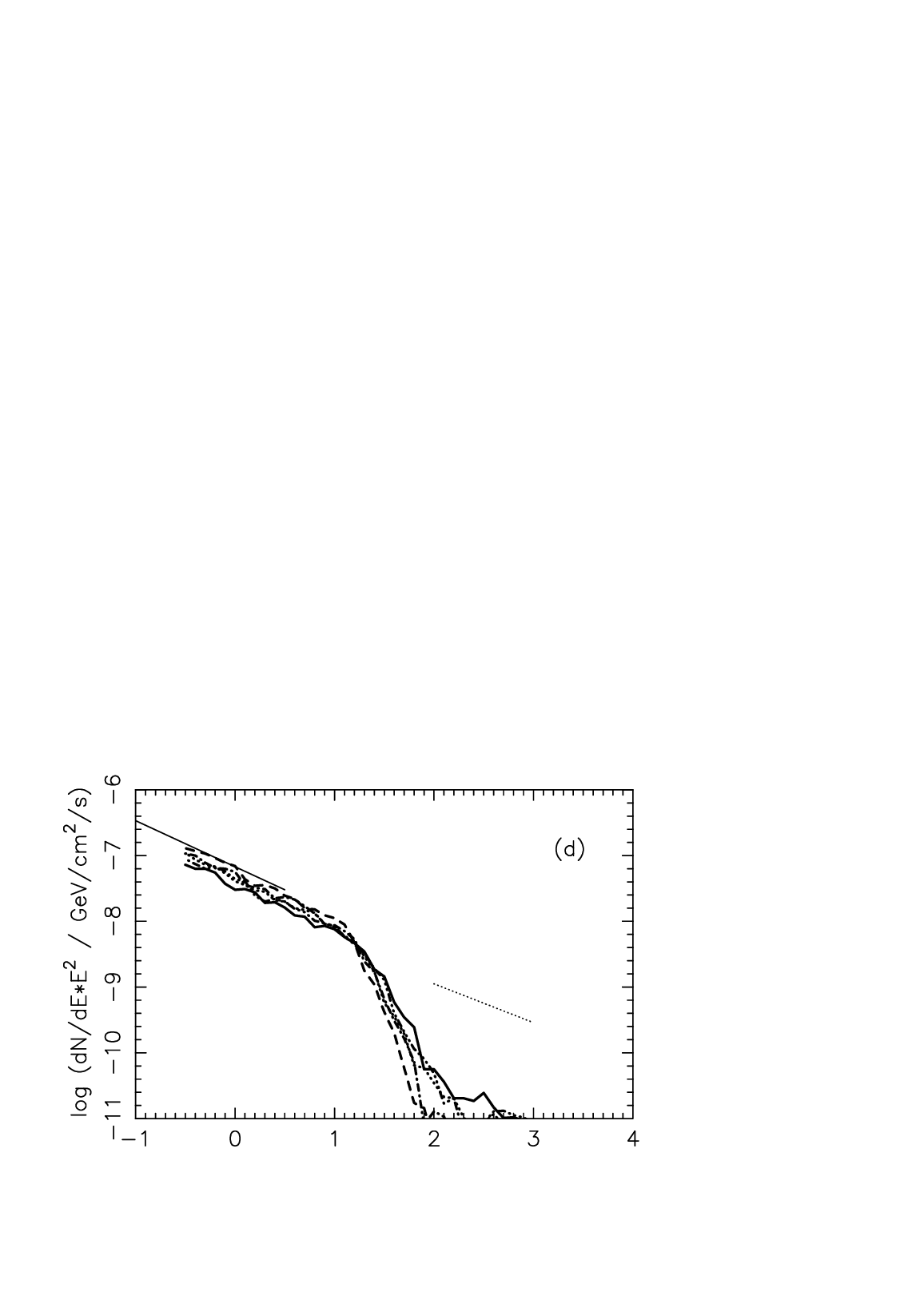}
\includegraphics{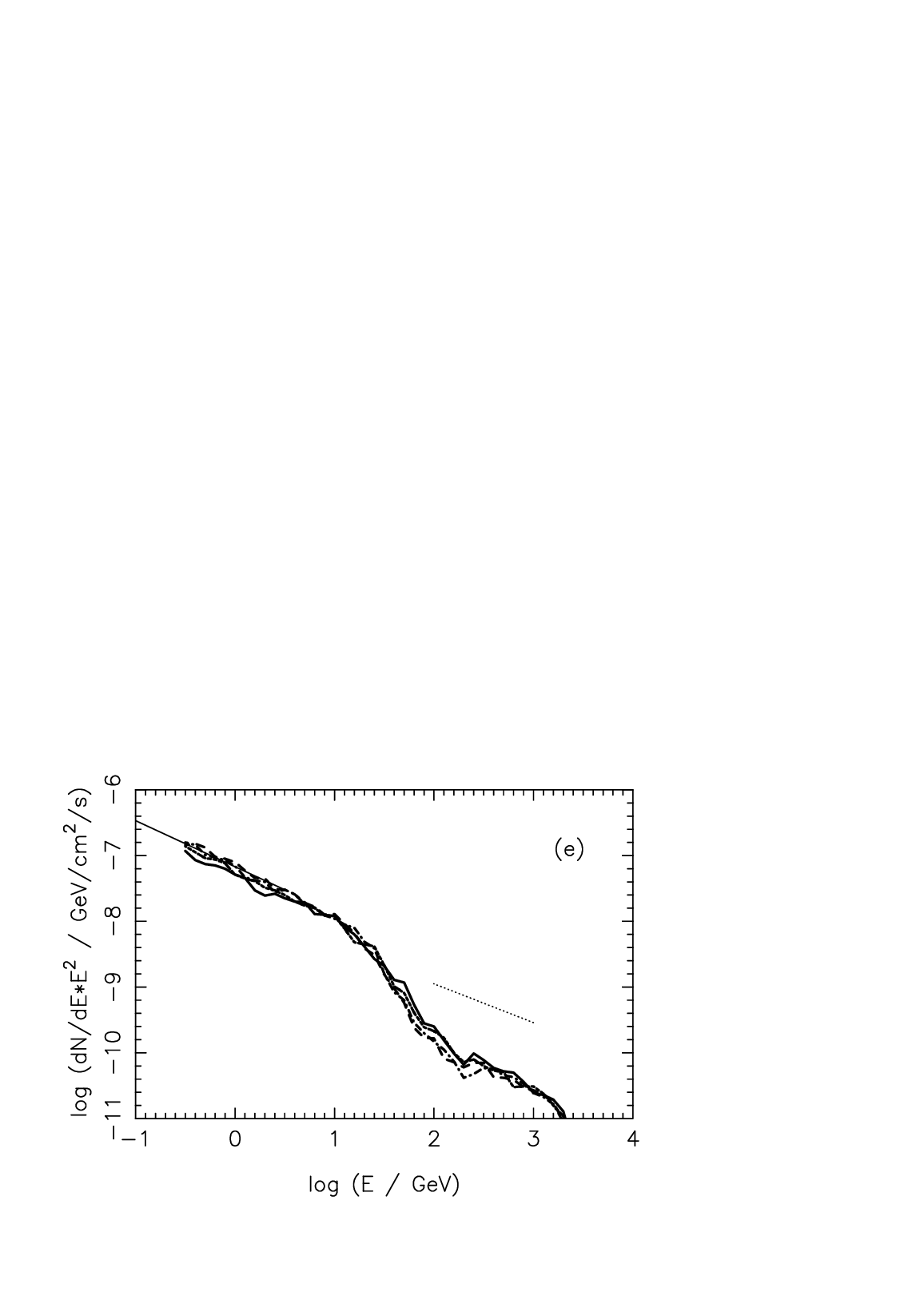}
\includegraphics{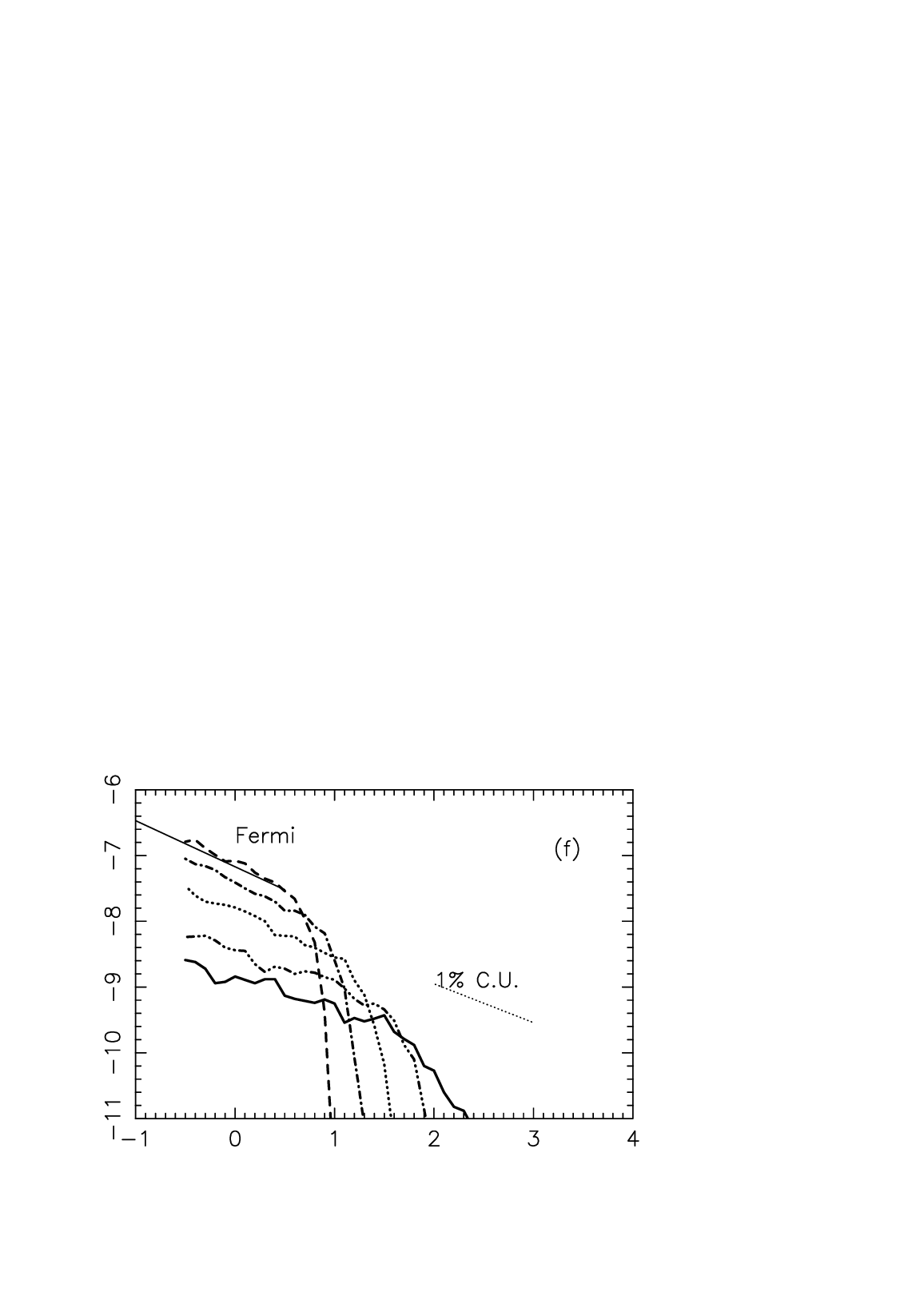}
\includegraphics{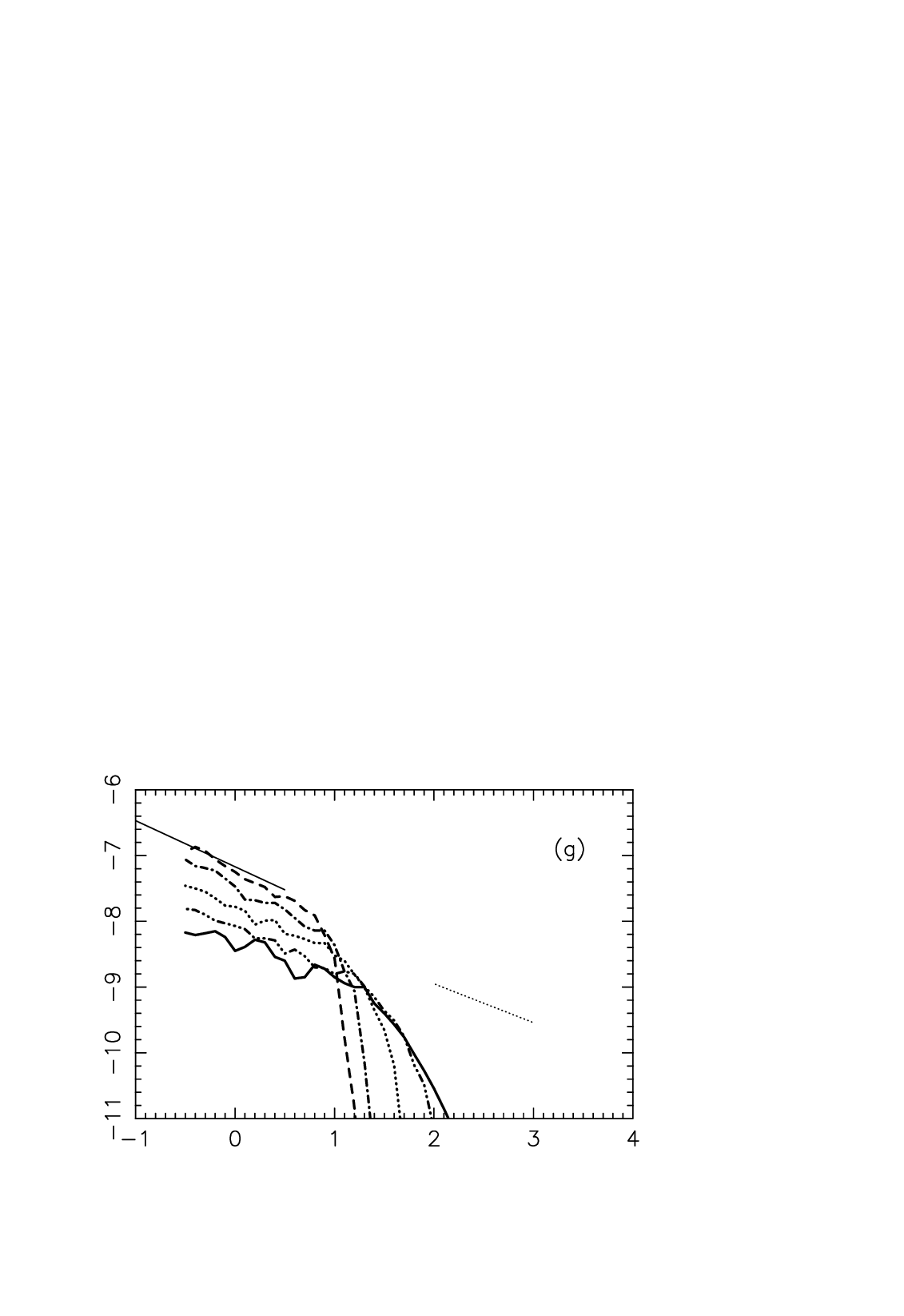}
\includegraphics{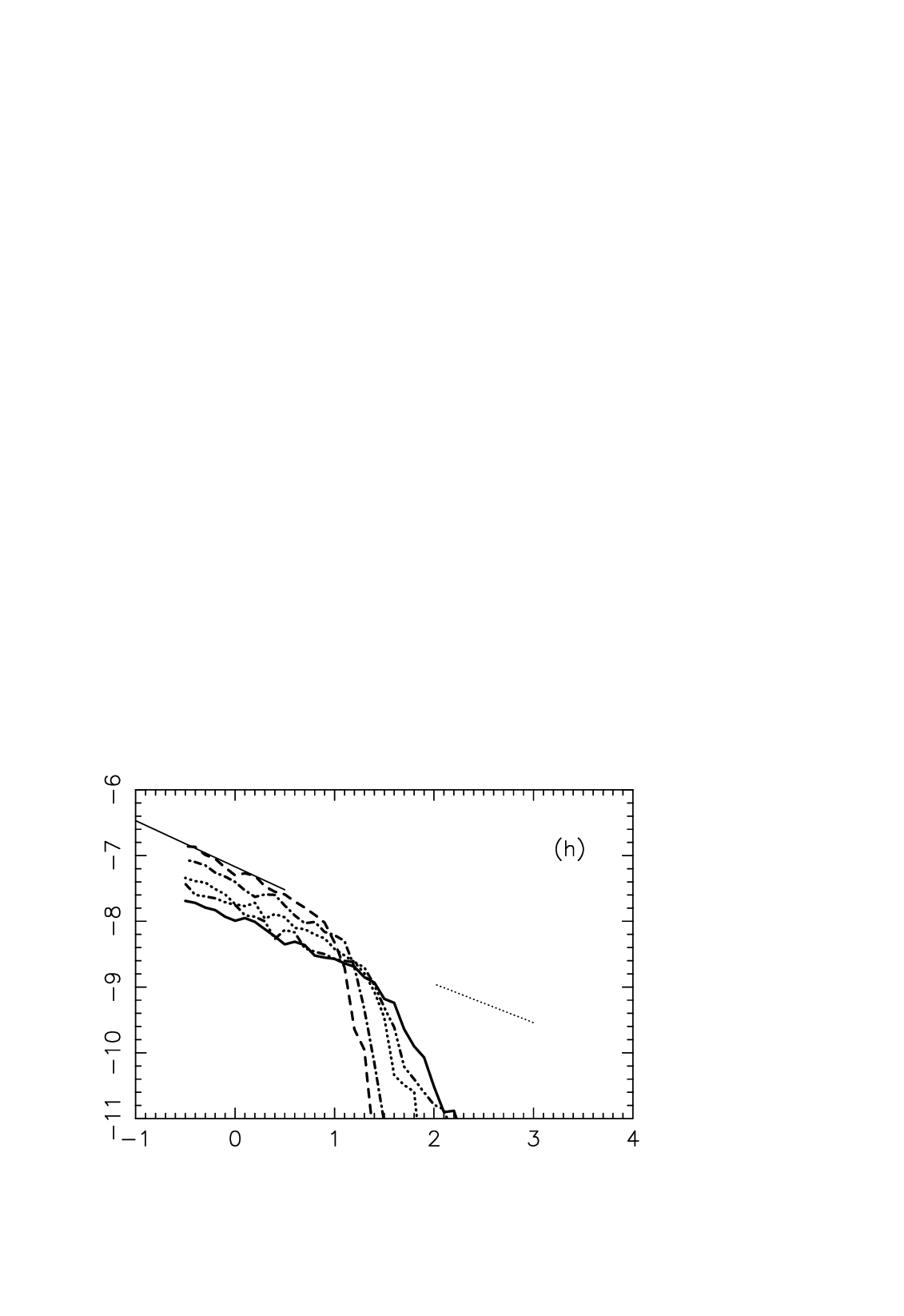}
\includegraphics{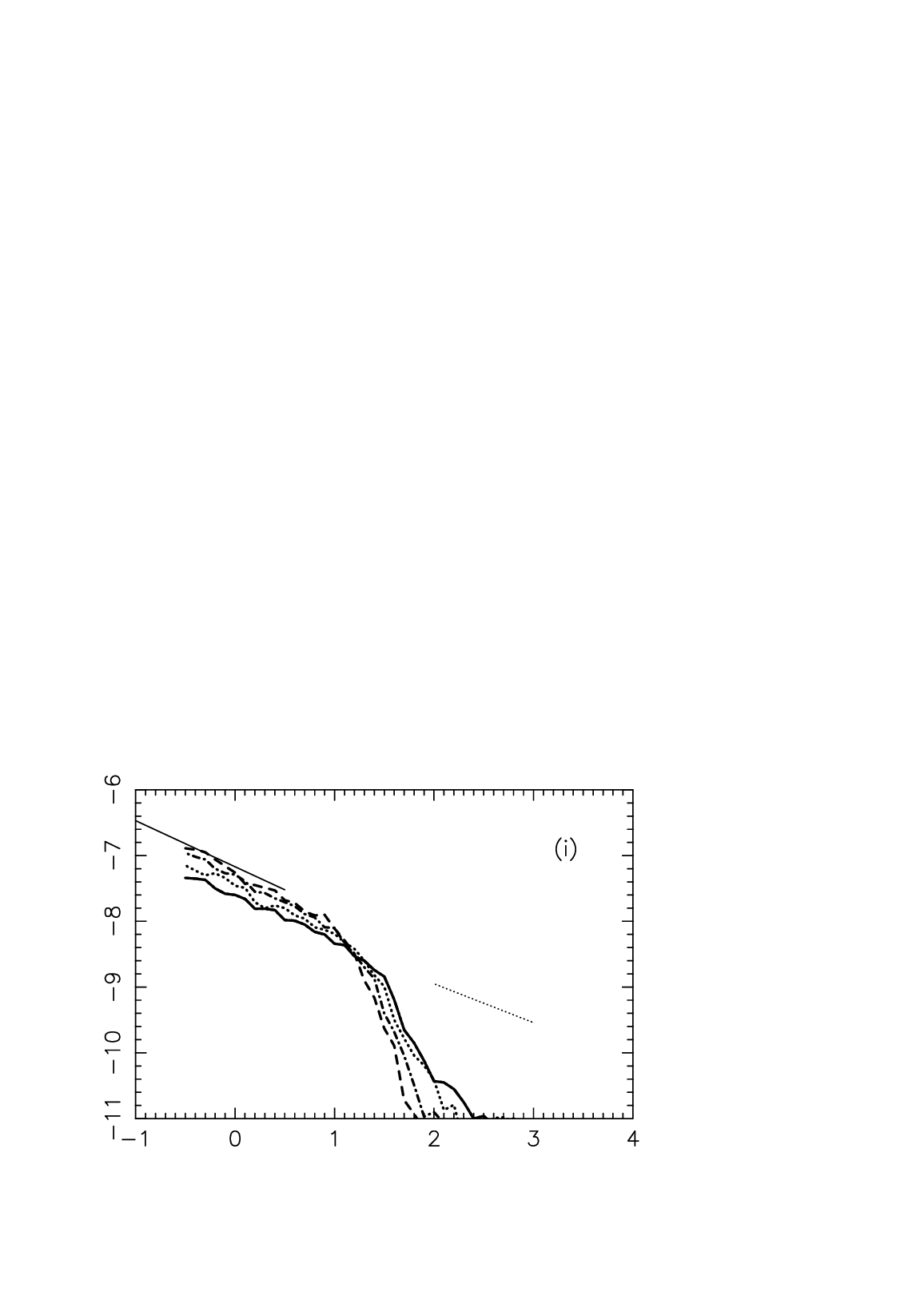}
\includegraphics{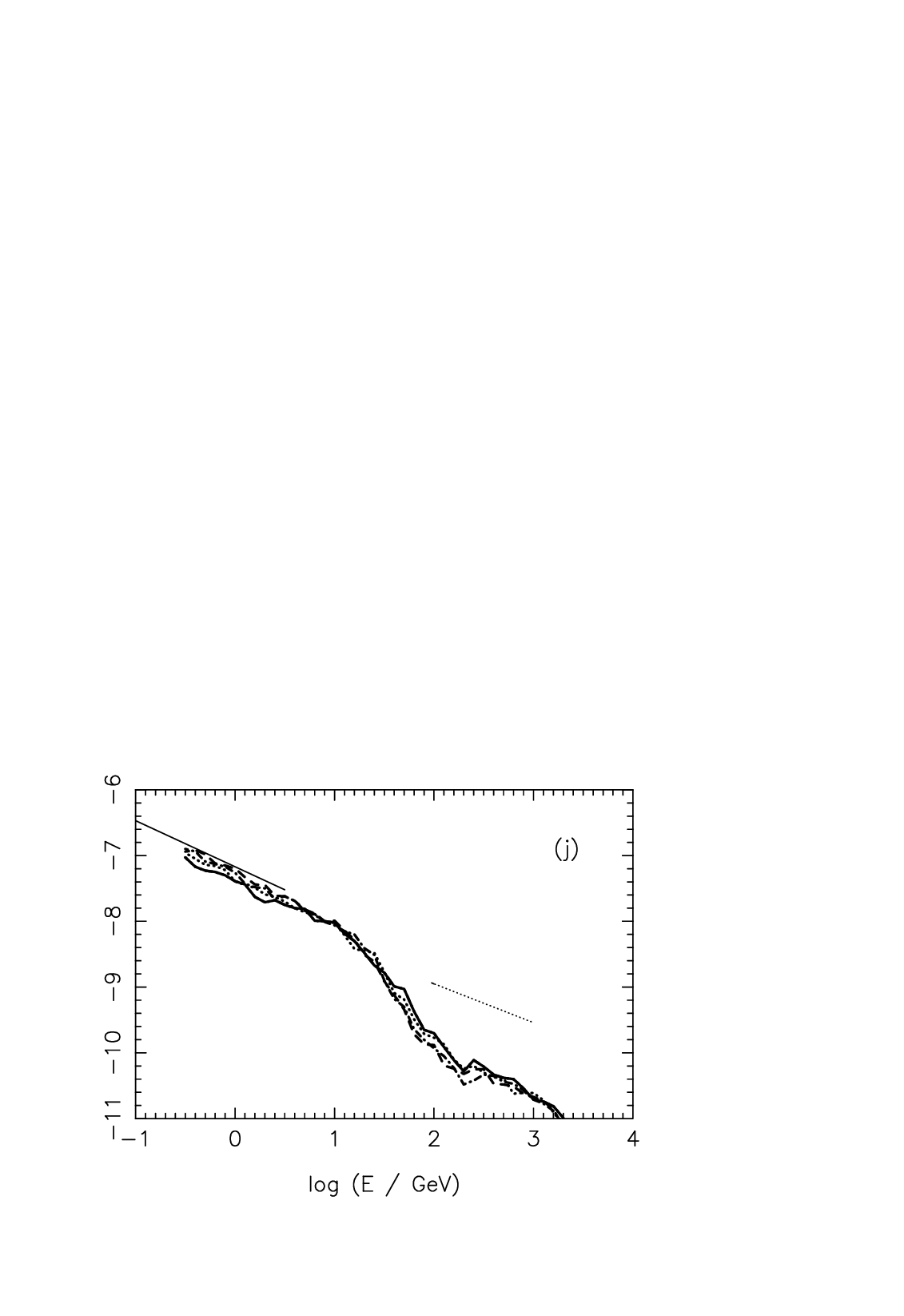}
\caption{As in Fig.~5 but for the parameters as in Fig.~4.}
\label{fig6}
\end{figure*}

We also show detailed $\gamma$-ray spectra at different phases of the binary system for the 
geometrical cases discussed above. The spectra are normalized to the peak emission
detected by the ${\it Fermi}$ satellite equal to $2\times 10^{-6}$ ph cm$^{-2}$ s$^{-1}$
(Abdo et al. 2009a). For comparison, we show the level of $1\%$ of the Crab Nebula emission (marked by $1\%$ C.U.) at energies above 100 GeV. It is clear that in all of the cases shown in Fig.~5 (dependence on 
the distance from the companion star) and Fig.~6 (dependence on the distance from the plane of
the binary system), the sub-TeV $\gamma$-ray emission is clearly below the sensitivity level of the present Cherenkov telescopes (e.g. the MAGIC or the VERITAS). Only the telescope array with the sensitivity 
on the level of $\sim 0.1\%$ C.U. (above 100 GeV) will have a chance to detect the $\gamma$-ray signal from Cyg X-3. On the other hand, Cyg X-3 might be detected by the on-ground telescope with sensitivity of the order of $\sim 1\%$ C.U. provided that its energy threshold is as low as 
$\sim$20-30 GeV. It is not clear at present whether the next stage of the MAGIC and HESS arrays will obtain such sensitivity limit. We conclude that detection of the sub-TeV $\gamma$-ray signal from the Cyg X-3 binary (WR type companion star within compact binary system) by the  Cherenkov telescopes has to probably wait for the construction of the next generation Cherenkov telescopes such as the Cherenkov Telescope Array (CTA).

\subsection{Circular versus eccentric orbit in Cyg X-3}

The shape of the GeV $\gamma$-ray light curve measured by the ${\it Fermi}$ satellite shows a modulation with the period of the binary system but also clear asymmetry.
The increasing part of the $\gamma$-ray light curve takes clearly longer than the
decreasing part. Also the asymmetry in respect to the phase 0.5 is visible (the phase zero is counted in the {\it Fermi} plot from the location of the compact object behind the companion star, Abdo et al.~2009a). 
Such asymmetry in the GeV $\gamma$-ray light curve may be expected in the case of an eccentric orbit of the compact object (injection source of relativistic particles) along the companion star. 

As an example, we perform calculations of the $\gamma$-ray light curves for different inclination angles of the binary system
and specific parameters of the eccentric orbit of the electron injection source located at or close to the compact object: the angle of the periastron is equal to $\omega = 60^o$
(measured in respect to phase zero), the eccentricity $e = 0.3$ and the semimajor axis $a = 2.25 R_\star$. In our modelling, we assumed that:
(1) electrons are injected inside a point like, isotropic source and interact only with the anisotropic radiation of the companion star; (2) the injection rate of electrons does not depend on the distance between the injection source and the companion star. 
The expected $\gamma$-ray light curves above three different energies are shown in Fig.~7. Interestingly, for the above parameters, we obtain the GeV $\gamma$-ray light curves which general shapes strongly 
resamble the $\gamma$-ray light curve measured by the ${\it Fermi}$. Note, that the level of the modulation of the $\gamma$-ray flux with the orbital period increases with the inclination angle, changing by a factor of $\sim$5 for the inclination $i = 30^{\rm o}$ and up to a factor of $\sim$100 for the inclination angle $i = 60^{\rm o}$. At present, it is difficult to decide which inclination angle describes the observations better due to the lack of knowledge on the precise location of the  
baseline emission in the {\it Fermi} light curve.

We also show the expected $\gamma$-ray light curves at sub-TeV energies for three inclination angles of the binary system.
A clear unticorrelation of the GeV and sub-TeV emission is also expected in the case of Cyg X-3 with the maximum emission at phases $\sim$0.3-0.4. The largest sub-TeV $\gamma$-ray fluxes are expected for large inclination angles. We also show the $\gamma$-ray spectra for specific phases of the binary system and compare them
with the sensitivities of the present Cherenkov telescopes (around $\sim 1\%$ Crab Unit). The spectra have been normalized to the GeV peak emission reported by the ${\it Fermi}$.  
A clear break in the $\gamma$-ray spectra is observed between 10-100 GeV in most of these spectra. Its precise location depends on the phase of the binary system. However, these spectra are not able to reach the level of the $1\%$ C.U. So then, they are undetectable by the present Cherenkov telescopes.

Note that these calculations have been performed under the assumption that the injection source is in the plane of the binary system (identified with the compact object inside the binary).
However, as we have shown above (for the case of a circular orbit), the source of relativistic electrons can not be located far away from the plane of the binary system since in such a case the modulation of the GeV $\gamma$-ray signal drops significantly being in contradiction with the measurements by the ${\it Fermi}$-LAT telescope.

\begin{figure}
\vskip 11.5truecm
\includegraphics{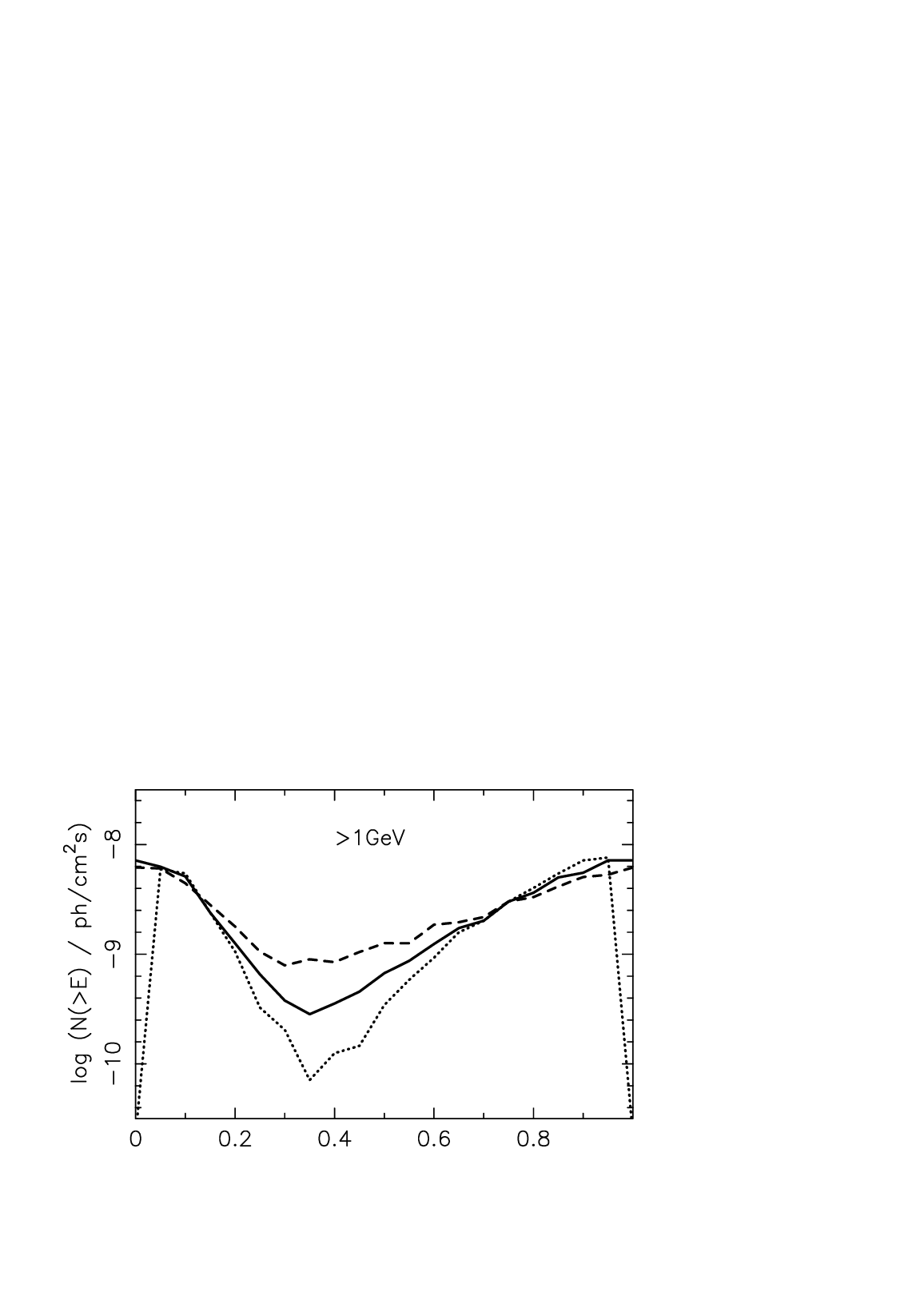}
\includegraphics{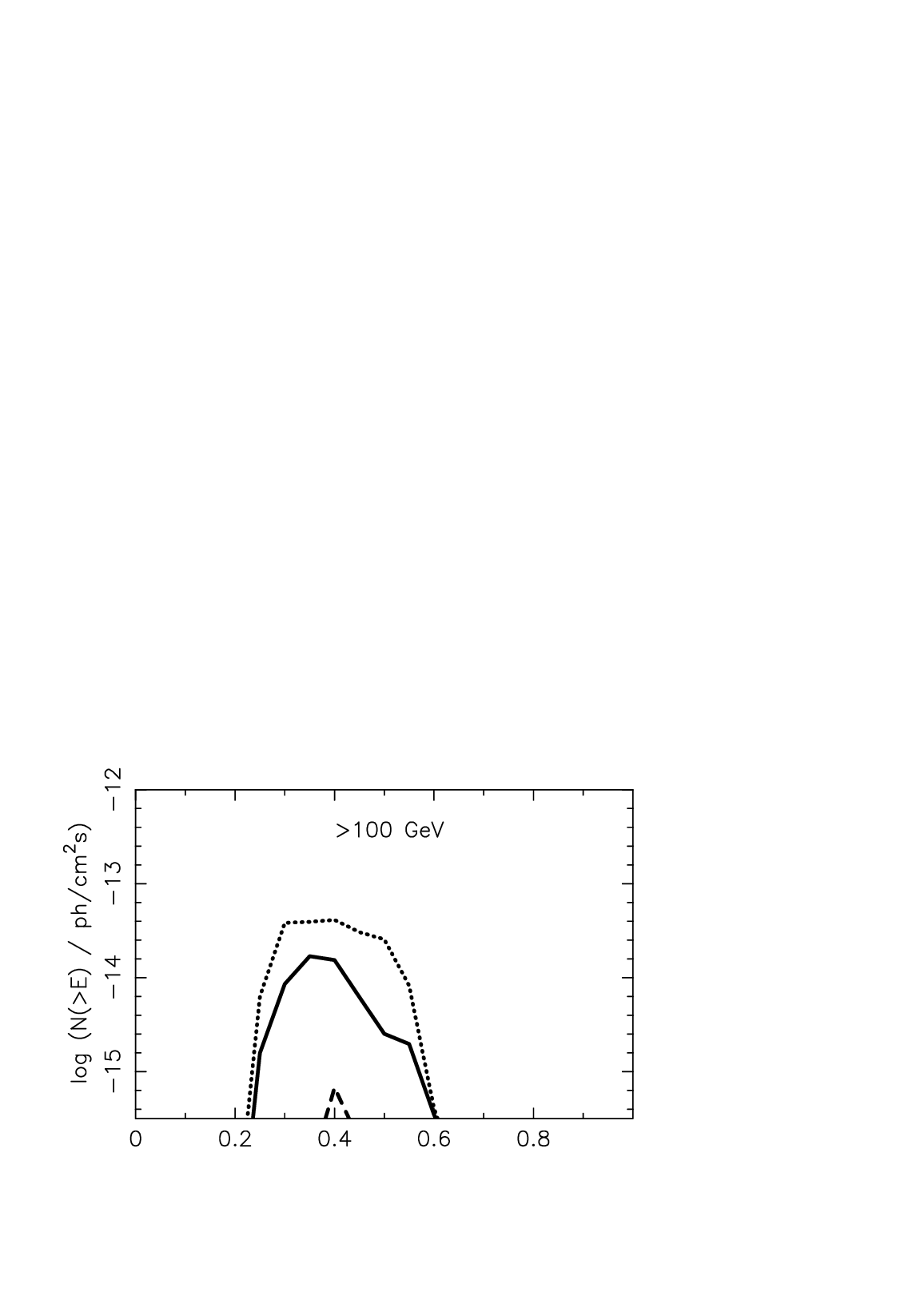}
\includegraphics{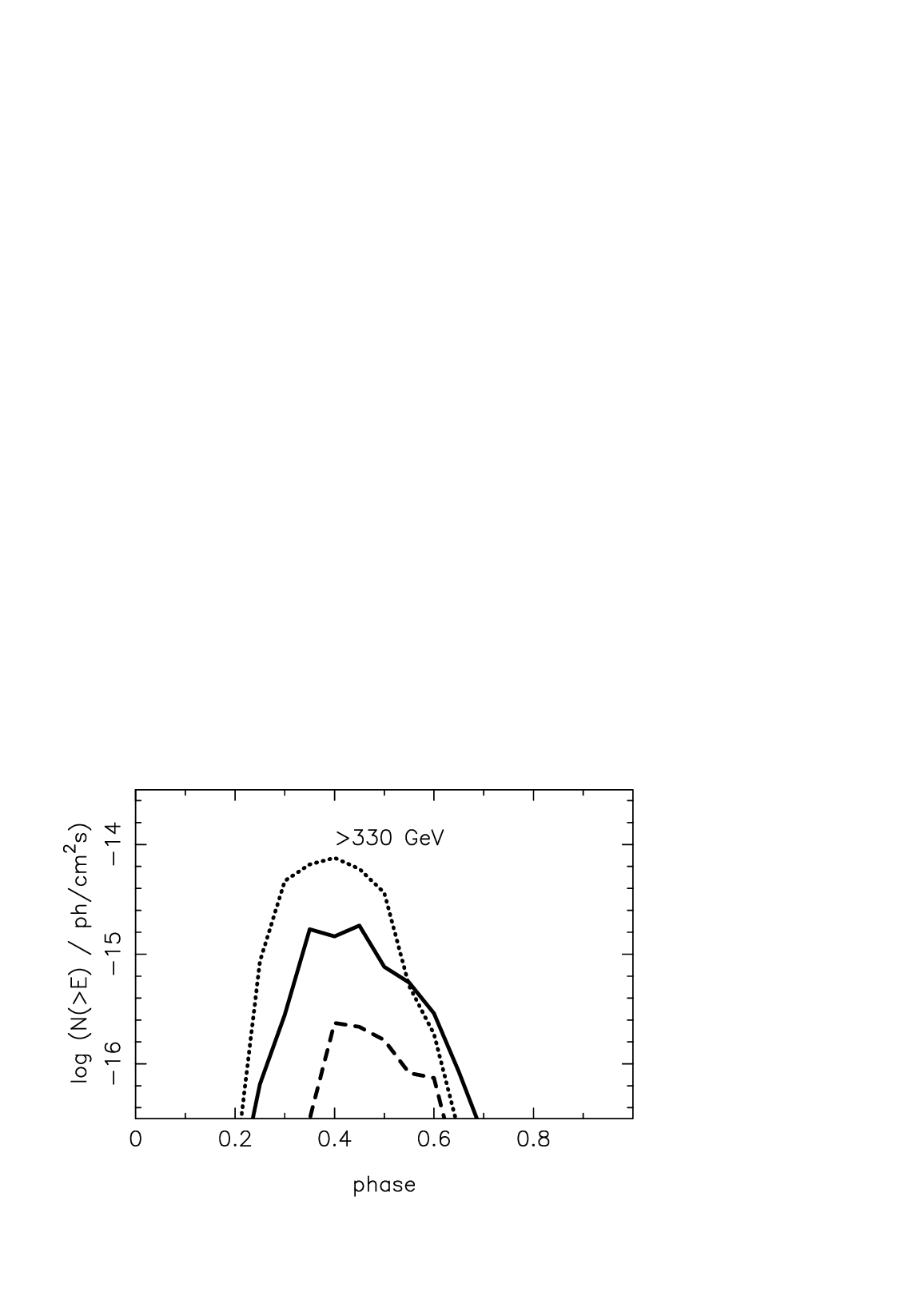}
\caption{The example $\gamma$-ray light curves for the eccentric orbit of the injection source of electrons with the angle of 
the periastron passage $\omega = 60^o$, the eccentricity $e = 0.3$, and the semimajor axis $a = 2.25R_\star$. At phase zero, the injection source (a compact object) is behind the companion star.
$\gamma$-rays are produced in the IC $e^\pm$ pair cascade process initiated by 
primary electrons (with the differential power law spectrum $\propto E^{-3.4}$ extending up to 
3 TeV) injected from a point like source. The light curves are shown for $\gamma$-rays with 
energies above 1 GeV (top panel), 100 GeV (middle), and 330 GeV (bottom).
The inclination angle of the binary system is equal to:
$i = 30^{\rm o}$ (dashed curve), $45^{\rm o}$ (solid), $60^{\rm o}$ (dotted).}
\label{fig7}
\end{figure}
\begin{figure}
\vskip 12.truecm
\includegraphics{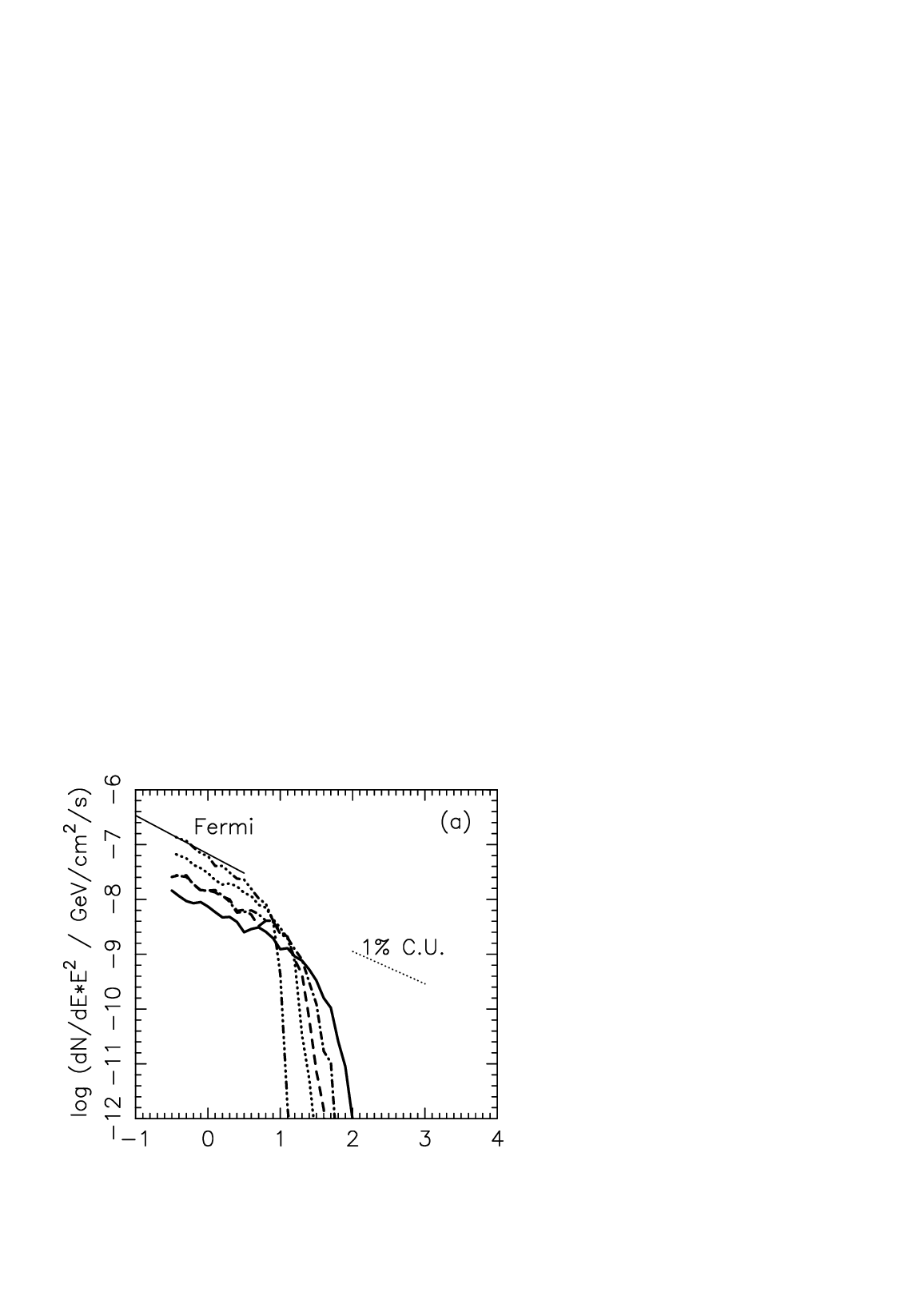}
\includegraphics{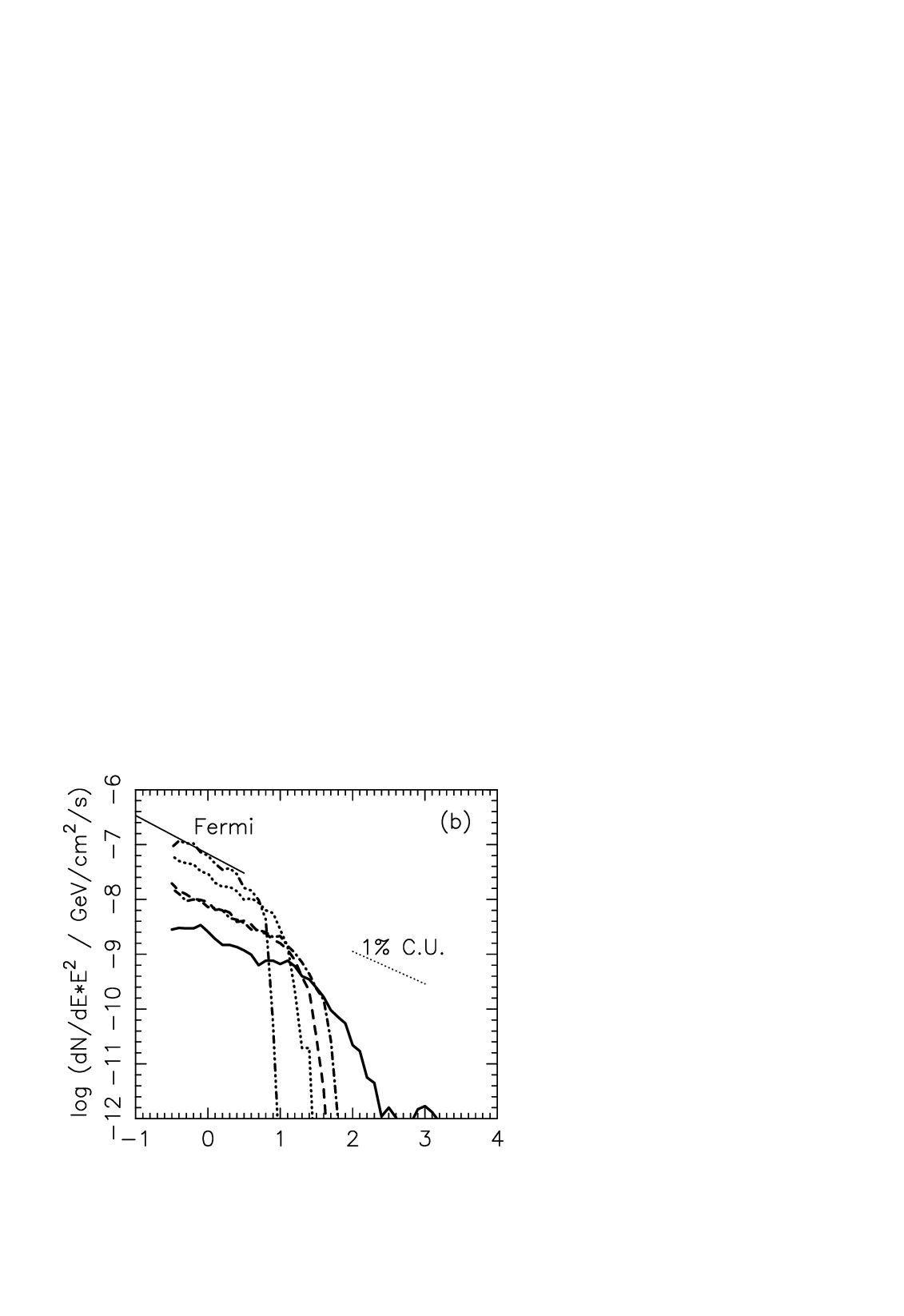}
\includegraphics{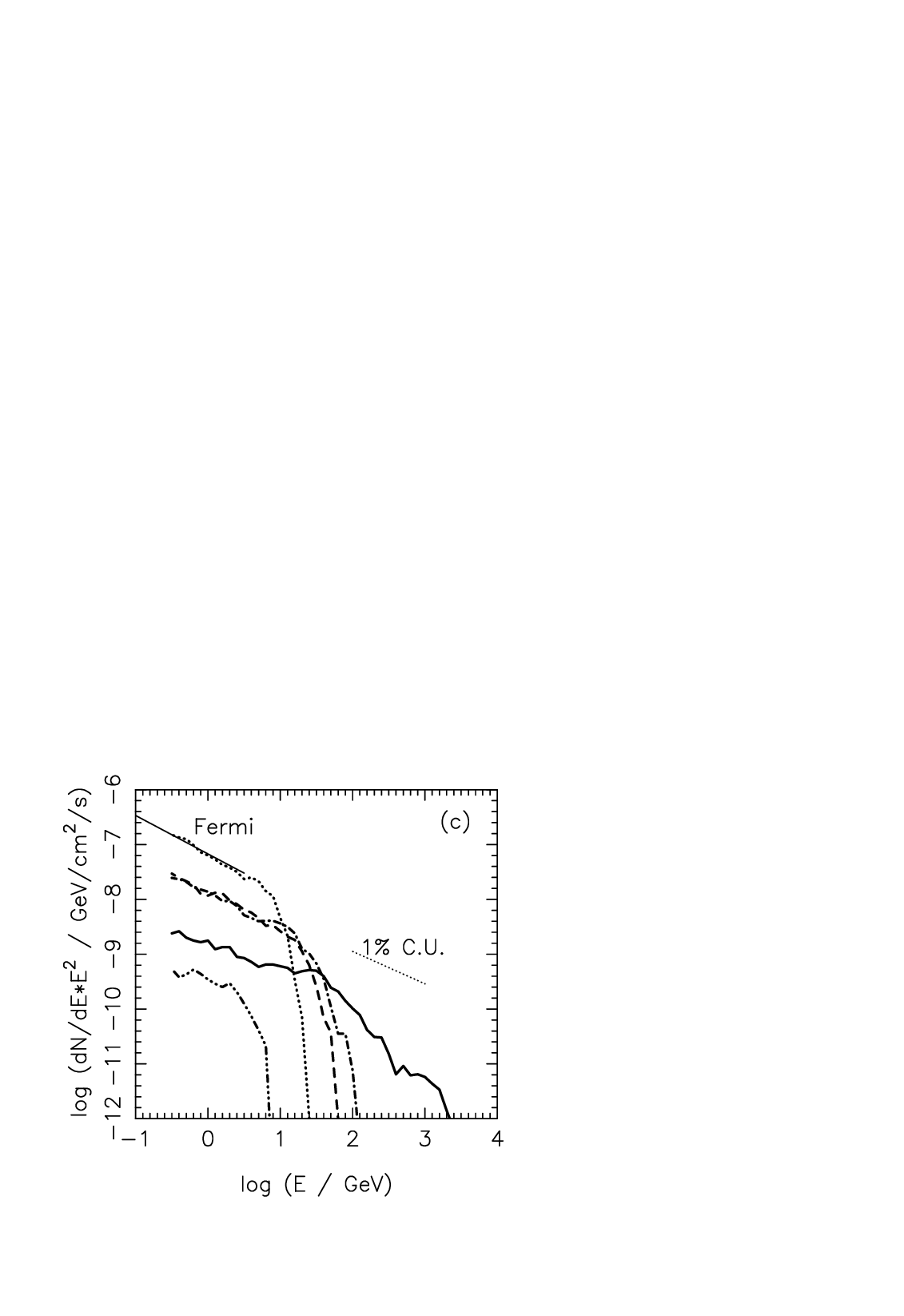}
\caption{The $\gamma$-ray spectra produced at different phases of the injection source of relativistic electrons in the binary system Cyg X-3. The parameters of the orbit and the spectrum of primary electrons are as in Fig.~7. The inclination angle of the binary system is fixed on: $i = 30^{\rm o}$ (a), $45^{\rm o}$ (b), $60^{\rm o}$ (c). The spectra for phases $\varphi = 0.$ (triple-dot-dashed curve), 0.2 (dashed), 0.4 (solid), 0.6 (dot-dashed), 0.8 (dotted).}
\label{fig8}
\end{figure}
\section{Discussion and Conclusion}

We have applied the anisotropic IC $e^\pm$ pair cascade model, developed for the $\gamma$-ray production in the TeV $\gamma$-ray binaries,
to the X-ray binary system Cyg X-3, recently discovered as a source of variable and modulated GeV $\gamma$-ray emission. The $\gamma$-ray light curves and spectra are calculated for different locations of the injection source of relativistic electrons inside the binary system following two basic models: (1) injection of $\gamma$-rays mainly within (or close to) the plane of the binary system as expected in the accreting neutron star model (Bednarek~2009); 
(2) injection of relativistic electrons from the place above the plane of the binary system as expected e.g., in the microquasar model in which particles are accelerated in the jet launched from a compact object (e.g. Levinson \& Blandford~1996, Georganopoulos et al.~2002, Romero et al.~2002). At first,  calculations are performed for circular orbits of the injection source in order to investigate their basic features. For reasonable parameters of the Cyg X-3 binary system, we obtained the GeV $\gamma$-ray light curves which have general features consistent with the observations of this source by the ${\it Fermi}$-LAT telescope. 
Moreover, it is found that the sub-TeV $\gamma$-ray emission is not expected to be produced on the level allowing its detectability by the present Cherenkov telescopes. This conclusion has been confirmed by some example calculations for a more complite scenario in which the injection source of electrons is located on an eccentric orbit around the companion star. The lack of detectable sub-TeV $\gamma$-ray emission from Cyg X-3 is a consequence of huge optical depths for $\gamma$-ray photons propagating within the binary system. Such large optical depths are  the consequence of a small compactness of the binary system and high surface temperature of the companion star
($\sim 10^5$ K). Detection of the $\gamma$-ray signal from sources similar to Cyg X-3 with  on-ground telescopes will be only possible with the next generation of Cherenkov telescopes which are expected to have sensitivity on the level of $\sim 0.1\%$ of Crab Units or by the telescopes which will be able to perform observations with the sensitivity of $\sim 1\%$ C.U. in the 20-30 GeV energy range.
Such sensitivities are planned for the future Cherenkov Telescope Array (CTA).
Note however, that in the case of Cyg X-3 another complication is introduced by the transient nature of the GeV $\gamma$-ray emission (a few day outbursts, Tavani~2009). This will additionally lower the chances for detection of Cyg X-3
in the TeV energies even with the CTA telescopes. The chances for detection can significantly rise if the $\gamma$-ray spectrum  in the GeV energies is flatter during some outbursts than reported recently (differential spectral index -2.7).
Therefore, investigation of a larger number of GeV $\gamma$-ray outbursts from Cyg X-3 by the AGILE and the ${\it Fermi}$ telescopes will  be of great importance for planning future sub-TeV $\gamma$-ray observations with the Cherenkov telescopes.

In the case of the microquasar model for the $\gamma$-ray production in binary systems two jets, propagating above and below the plane of the binary system, are expected. 
The calculations shown here concern only the $\gamma$-ray production from one of the jets (i.e. the jet propagating into the hemisphere containing the observer). For large inclination angles of the binary system, the contribution of the jet and counter-jet becomes comparable but for small inclination angles the counter-jet contribute mainly to the GeV energy range due to large angle between the injection source of electrons and the observer (measured from the companion star). 
As a result, normalization to the ${\it Fermi}$ spectrum at the GeV
energies to the sum of the $\gamma$-ray emission from the jet and the counter-jet in the case of small inclination angles of the Cyg X-3 binary system additionally decreases estimated in this paper $\gamma$-ray emission at sub-TeV energies.

Finally, we would like to comment of the possible influence of the assumptions in considered model on the predicted $\gamma$-ray light curves and spectra from Cyg X-3. 

\begin{enumerate}

\item {\it Synchrotron energy losses of primary and secondary leptons:} This process will additionally extract energy from leptons in respect to considered pure IC cascade process (e.g. Bosch Ramon et al.~2008). It is expected that with important synchrotron energy losses the calculated $\gamma$-ray fluxes specially at larger energies $>100$ GeV becomes even lower due to the dominance of the synchrotron energy losses of electrons over their IC energy losses in the Klein-Nishina regime.

\item {\it Incomplete cooling of leptons and their escape from the production place:} First process will result in  deficiency of $\gamma$-ray flux at lower energies (GeV range). In order to be consistent with the spectral index of the $\gamma$-ray spectrum even a steeper spectrum of primary electrons should be applied. However, this will result in lowering also the $\gamma$-ray flux above $\sim 100$ GeV. The influence of the second process is difficult to estimate since it is even not clear in which direction secondary leptons can be driven from the their site of origin. The effects of advection in the fast stellar wind seems to be negligible due to relatively large advection time, $\tau_{\rm adv}\approx R_\gamma/v_{\rm w}$ (where $v_{\rm w}$ is the stellar wind velocity which in the case of WR stars is of the order of $\sim (2 - 3)\times 10^3$ km s$^{-1}$), in respect to the cooling time of electrons on the IC process. The IC cooling time is $\tau^{\rm T}_{\rm IC}\approx 40r^2/(T_5^4\gamma)$ s, where the dilution factor of radiation from the stellar surface can be approximated by $r = R_\gamma/R_\star$. Even for electrons with energies 500 MeV, injected at the distance, $R_\gamma = 10R_\star$, from the star with surface temperature $T = 10^5$ K, $\tau_{\rm adv}\approx 10^4$ s and 
$\tau^{\rm T}_{\rm IC}\approx 4$ s. So then, $\tau_{\rm IC} << \tau_{\rm adv}$ even for electrons scattering soft radiation in the Klein-Nishina regime with the largest considered energies equal to 3 TeV. 

\item {\it Izotropization of cascade leptons versus their propagation along the local magnetic field lines:} The linear cascade through the radiation field of the companion star has been also considered as the limitting case when considering the importance of the IC $e^\pm$ pair cascade contribution to the $\gamma$-ray spectrum escaping from the binary system (e.g. Aharonian et al.~2006, Cerutti, Dubus \& Henri 2009). We discussed here scenario in which electrons are locally isotropized. The general conditions for local izotropization of cascade leptons has been discussed in Sect.~3. The isotropization process occurs efficiently (based on the comparison of the Larmor radius with the mean free path for IC process) even in a relatively low magnetic field. 
However, exact propagation of leptons in the ordered component of the 
magnetic field will depend on their injection angles in respect to the magnetic field direction.
Magnetic focussing may result in production of strong enhanced $\gamma$-ray emission at specific directions on the sky directly unrelated to directions of the first generation of cascade $\gamma$-rays. These effects have been studied in detail by Sierpowska \& Bednarek~(2005).

\item {\it Phase dependent injection rate of electrons:} The influence of this effect  is also very difficult to consider without detailed knowledge on the mechanism of acceleration process and generation of energy. It might be expected that both these processes could depend on the distance between the compact object and the companion star. However,  whether the higher accretion rate (or stronger stellar wind) will 
increase or decrease power converted to relativistic electrons or how this will influence their maximum energies is at present not clear.
Therefore, the only reasonable approach seems to keep these basic processes as independent on 
the distance between the stars (i.e. independent on the phase of the binary system). These effects can be studied when detailed multiwavelength observations of $\gamma$-ray binary systems become available.
 
\item {\it Possible absorption of $\gamma$-rays in the X-rays:} We postulate that $\gamma$-rays,
produced in terms of the jet model, originate at distances from its base which are comparable to
the dimensions of the binary system. The radiation from the accretion disk and the inner part of the jet can produce strong target even for hadrons (see e.g. models by Levinson \& Waxman~2001, Bednarek~2005, or more recently by Romero \& Vila~2008). So then, also $\gamma$-rays produced in the inner region of the jet are strongly absorbed.
However, $\gamma$-rays, produced at large distances from the base of the jet, suffer the radiation form the inner disk and/or jet significantly diluted.  At distances of the order of thousands larger than the inner disk radius, this radiation field is diluted by a factor proportional to the square of the distance, i.e. by a factor of $\sim 10^6$. We conclude that such strongly diluted radiation field from the inner disk and/or jet will not be able to efficiently absorb the high energy $\gamma$-rays produced farther away from the base of the jet.

\end{enumerate}

\section*{Acknowledgments}
This work is supported by the Polish MNiSzW grant N N203 390834 and 
the grant from the NCBiR.


\end{document}